\RequirePackage{ifpdf}
\documentclass[12pt,letterpaper]{article}
\pdfoutput=1
\usepackage{jheppub}
\usepackage{epsfig}
\usepackage{enumitem}
\usepackage[latin1]{inputenc}
\usepackage{bbm,amsfonts}
\usepackage{graphicx}
\usepackage{amssymb,amsmath,amsfonts,mathtools}
\usepackage{fancybox}
\usepackage{enumerate}
\usepackage{dsfont}
\usepackage{verbatim}
\usepackage{wrapfig}
\usepackage{slashed}
\usepackage{shuffle}
\usepackage{subcaption}
\usepackage{braket}
\newcommand{\bsquare}{\mbox{\scriptsize  $\blacksquare$}}
\author[a]{Marco S. Bianchi,}
\author[b]{Luca Griguolo,}
\author[c,d]{Andrea Mauri,}
\author[c,d]{Silvia Penati,}
\author[f]{and Domenico Seminara}
 
\affiliation[a]{Niels Bohr Institute, University of Copenhagen, Blegdamsvej 17, 2100 Copenhagen $\emptyset$,\\
Denmark}
\affiliation[b]{Dipartimento SMFI, Universit\`a di Parma and INFN Gruppo Collegato di Parma, Viale G.P. Usberti 7/A, 43100 Parma, Italy}
\affiliation[c]{ Dipartimento di Fisica, Universit\`a degli studi di Milano--Bicocca, Piazza della Scienza 3, I-20126 Milano, Italy }
\affiliation[d]{INFN, Sezione di Milano--Bicocca, Piazza della Scienza 3, I-20126 Milano, Italy }
\affiliation[f]{Dipartimento di Fisica, Universit\`a di Firenze and INFN Sezione di Firenze, via G. Sansone 1, 50019 Sesto Fiorentino, Italy\\}

\emailAdd{marco.bianchi@nbi.ku.dk}  
\emailAdd{luca.griguolo@pr.infn.it} 
\emailAdd{andrea.mauri@mi.infn.it} 
\emailAdd{silvia.penati@mib.infn.it} 
\emailAdd{seminara@fi.infn.it}
  
\abstract{In ABJ(M) theory, we propose a matrix model for the exact evaluation of BPS Wilson loops on a latitude circular contour, so providing a new weak-strong interpolation tool. Intriguingly, the matrix model turns out to be a particular case of that computing torus knot invariants in $U(N_1|N_2)$ Chern-Simons theory.
At weak coupling we check our proposal against a three--loop computation, performed for generic framing, winding number and representation. 
The matrix model is amenable of a Fermi gas formulation, which we use to systematically compute the strong coupling and genus expansions.
For the fermionic Wilson loop the leading planar behavior agrees with a previous string theory prediction. For the bosonic operator our result provides a clue for finding the corresponding string dual configuration.
Our matrix model is consistent with recent proposals for computing Bremsstrahlung functions exactly in terms of latitude Wilson loops.
As a by-product, we extend the conjecture for the exact $B^{\theta}_{1/6}$ Bremsstrahlung function to generic representations and test it with a four--loop perturbative computation. Finally, we propose an exact prediction for $B_{1/2}$ at unequal gauge group ranks. }

\title{A matrix model for the latitude Wilson loop in ABJM theory} 

\newcommand{\be}{\begin{equation}}
\newcommand{\ee}{\end{equation}}
\newcommand{\beq}{\begin{equation}}
\newcommand{\eeq}{\end{equation}}
\newcommand{\bea}{\begin{eqnarray}}
\newcommand{\eea}{\end{eqnarray}}
\newcommand{\ena}{\end{eqnarray}}
\newcommand {\non}{\nonumber}

\renewcommand{\a}{\alpha}
\renewcommand{\b}{\beta}

\renewcommand{\d}{\delta}
\renewcommand{\th}{\theta}

\newcommand{\pa}{\partial}
\newcommand{\g}{\gamma}
\newcommand{\G}{\Gamma}

\newcommand{\e}{\epsilon}

\renewcommand{\L}{\Lambda}
\newcommand{\m}{\mu}

\newcommand{\n}{\nu}

\newcommand{\p}{\pi}

\def\Tr{\textrm{Tr}}

\numberwithin{equation}{section}

\def\clock{{\count0=\time
           \divide\count0 60
           \ifnum\count0<10 0\fi\the\count0
           \multiply\count0 -60 \advance\count0 \time
           :\ifnum\count0<10 0\fi \the\count0
         }}
\newcommand{\timestamp}{{\small\vbox{\hbox{\tt\jobname.tex}
\hbox{\the\day/\the\month/\the\year, \clock}}}}
\begin{document}

\maketitle
\allowdisplaybreaks

\section{Introduction and Conclusions}

\subsection{Generalities}

Wilson loop operators are fundamental observables in any gauge theory. While in the case of pure Yang--Mills in two dimensions a complete solution for their vacuum expectations value exists  \cite{Kazakov:1980zj,Rusakov:1990rs}, in higher dimensions only few examples of non--perturbative calculations can be found in literature. A notable exception is provided by ordinary Wilson loops in pure three--dimensional Chern--Simons theory \cite{Witten:1988hf}, where the computation is made possible thanks to the topological nature of the model. 

A more significant class of examples is represented by the so--called supersymmetric/BPS Wilson loops, appearing quite ubiquitously in gauge theories with extended supersymmetry. Their main property consists in preserving a fraction of the supersymmetry charges, depending on the shape of the contour and on the couplings to the different fields appearing in the Lagrangian. Despite their BPS nature, the dependence of their vacuum expectation values on the coupling constant  is generically non--trivial and interpolates between the weak and strong coupling regimes, thus providing a natural playground  where to test the AdS/CFT correspondence and other non--perturbative methods. 

In recent years the technique of supersymmetric localization (for recent reviews, see \cite{Pestun:2016zxk} and references therein) has allowed for the calculation of a large variety of BPS Wilson loops, in different theories and various dimensions \cite{Erickson:2000af,Zarembo:2002an,Drukker:2007qr,Kapustin:2009kz,Drukker:2009hy,Ouyang:2015iza,Ouyang:2015bmy,Cooke:2015ila}. Localization often reduces the computation of these observables to the average of suitable matrix operators, in terms of particular matrix integrals. These integrals can then be solved by applying the powerful machinery developed along the years, as for example large $N$ expansion, orthogonal polynomials, loop equations, recursion relations.
 
In this paper we discuss a new example of such matrix model computations of Wilson loops. We focus on three--dimensional ${\cal N}=6$ superconformal $U(N_1)\times U(N_2)$ Chern--Simons theory with matter, also known as the ABJ(M) model \cite{Aharony:2008ug,Aharony:2008gk}, that can be viewed as an extended supersymmetric generalization of the familiar topological Chern--Simons theory. We propose a matrix model  for calculating the exact quantum expectation value of certain $1/12$-- and $1/6$--BPS Wilson loop operators, briefly referred to as {\em bosonic }and {\em fermionic latitudes}  and parameterized by a real number $\nu\in[0,1]$\cite{Bianchi:2014laa}. For $\nu=1$ we recover the supersymmetric Wilson loops on the great circle \cite{Berenstein:2008dc,Drukker:2008zx, Chen:2008bp, Rey:2008bh,Drukker:2009hy}, for which a precise localization procedure has been derived in \cite{Kapustin:2009kz}, leading to the ABJ(M) matrix model. Various aspects of such circular Wilson loops have been thoroughly studied in \cite{Marino:2009jd,Drukker:2010nc}. 

For generic $\nu$ we do not possess a direct derivation of our matrix model from supersymmetric localization, since these loop operators preserve supercharges that are different from the ones used in \cite{Kapustin:2009kz}, and cannot be embedded in a natural way in the ${\cal N}=2$ superspace language employed there for the localization procedure. 
Rather, we formulate an ansatz for the matrix integral from symmetry considerations and show its non--trivial consistency with a three--loop perturbative calculation of the same observables. At equal gauge group ranks, we also prove that, without the operator insertion, the matrix model reproduces the ABJM partition function. This fact supports a possible interpretation of our matrix model as the result of localizing with the $\nu$--dependent supercharge preserved by latitude operators. Encouraged by this fact, we perform a strong coupling analysis using the Fermi gas approach. In the fermionic case we obtain an important agreement with the existing string theory computation \cite{Correa:2014aga}.  Instead, in the case of unequal ranks, $N_1 \neq N_2$,  we find a residual $\nu$--dependence in the matrix model partition function. This would in principle prevent us form directly interpreting it as the result of localizing the ABJ theory on the sphere. Still, such a dependence is confined to a simple phase factor, which might stem from a framing anomaly. Therefore, the latter case requires a deeper analysis.

Latitude Wilson loops are important in order to study non--BPS observables, the Bremsstrahlung functions \cite{Bianchi:2014laa,Bianchi:2017ozk}. These are functions of the coupling constant, controlling the small angle limit of the anomalous dimension of a generalized cusp, constructed from supersymmetric Wilson lines. In particular, depending on the degree of supersymmetry, the various Bremsstrahlung functions can be obtained taking a suitable derivative of the latitude with respect to the $\nu$ parameter, at $\nu=1$ \cite{Correa:2014aga,Bianchi:2014laa,Bianchi:2017ozk}. 
The fermionic and bosonic latitudes satisfy a cohomological equivalence \cite{Bianchi:2014laa}. Using the latter and under a suitable assumption (automatically satisfied by our matrix integral), remarkable relations can be derived for the different Bremsstrahlung functions at equal ranks $N_1=N_2$. Eventually, they are all related among themselves and expressible only in terms of the phase of the undeformed 1/6--BPS Wilson loop.

Besides the derivation and the strong coupling solution of our matrix model, the paper provides details of the perturbative three--loop computation of the bosonic latitude Wilson loop, for generic representation and winding number. This general result allows for comparisons in various limits. We also present a careful discussion of the framing dependence \cite{Bianchi:2016yzj} at perturbative level. In particular, the perturbative result at generic framing is crucial for finding consistency with the prediction from our matrix model. The remarkable output is that the agreement between the perturbative and the matrix model calculations works upon the identification of framing with the effective parameter $\nu$.

\subsection{Summary of the results}

\begin{itemize}[leftmargin=*]
\item[\bsquare] The main focus of the paper is on the proposal of a matrix model computing the latitude expectation values at all orders. Based on symmetry considerations and the consistency with the perturbative results, we conjecture that the exact expectation value of the multiply wound bosonic latitude Wilson loop in the fundamental representation is given by 
\begin{align} \label{eqI:WL}
\langle W_B^m(\nu) \rangle  &= \left\langle \frac{1}{N_1} \sum_{1\leq i\leq N_1} e^{2\pi\, m\, \sqrt{\nu}\, \lambda_{i}} \right\rangle
\end{align}
where $\langle \cdots\rangle$ stands for the normalized expectation value performed using the following matrix integral 
\begin{align} 
\label{ZZnu}
&  Z (\nu)= \int \prod_{a=1}^{N_1}d\lambda _{a} \ e^{i\pi k\lambda_{a}^{2}}\prod_{b=1}^{N_2}d\mu_{b} \ e^{-i\pi k\mu_{b}^{2}}  \\
&~~ \times 
 \frac{\displaystyle\prod_{a<b}^{N_1}\sinh \sqrt{\nu} \pi (\lambda_{a}-\lambda _{b})\sinh\frac{ \pi (\lambda_{a}-\lambda _{b})}{\sqrt{\nu}} \prod_{a<b}^{N_2}\sinh\sqrt{\nu} \pi (\mu_{a}-\mu_{b}) \sinh\frac{\pi (\mu_{a}-\mu_{b})}{\sqrt{\nu}} }{\displaystyle\prod_{a=1}^{N_1}\prod_{b=1}^{N_2}\cosh\sqrt{\nu} \pi (\lambda _{a}-\mu_{b}) \cosh\frac{\pi (\lambda _{a}-\mu_{b})}{\sqrt{\nu}} } \nonumber
\end{align}
The expression for the fermionic 1/6--BPS latitude Wilson loop is obtained  by computing a suitable linear combination of bosonic latitudes, as suggested by the cohomological equivalence discussed in \cite{Bianchi:2014laa} and reviewed in Section \ref{review}. 

We also conjecture that the expectation values of latitude Wilson loops in higher dimensional representations can be obtained from this matrix model by the same generalization of the operator insertions as in the undeformed ($\nu =1$) case.

A crucial point for the consistency of the above proposal is that for $N_1=N_2$  the  partition function $Z(\nu)$ defined
in \eqref{ZZnu}
is independent of the  $\nu$ parameter and coincides with the usual partition function of ABJM on $S^3$. This important property is proved explicitly in Section \ref{fermi}. The ABJ case  ({\it i.e.}~$N_1\neq N_2$) is subtler but still the full dependence on $\nu$ is confined to a simple phase factor (see Section  \ref{fermi}). This points towards the possibility that a localization procedure performed with one of the $\nu$--dependent supersymmetric charges preserved by the latitude could produce the above matrix integral.  

\item[\bsquare]  Supported by this first evidence on the correctness of our proposal, in Section \ref{sec:Fermi} we perform a careful analysis of the large $N$, strong coupling limit, of the expectation values of the bosonic and fermionic latitudes, as defined through our conjectured matrix model. We perform a Fermi gas analysis along the lines of \cite{Klemm:2012ii} and obtain an explicit result re--summing the genus expansions both in the 1/12--BPS and 1/6--BPS cases. In particular, for the fermionic loop we obtain
\begin{equation}
\langle W_F(\nu) \rangle_\nu = -\frac{\nu\,  \Gamma \left(-\frac{\nu }{2}\right)\, \csc \left(\frac{2 \pi  \nu }{k}\right)\, \text{Ai}\left(\left(\frac{2}{\pi^2k} \right)^{-1/3} \left(N-\frac{k}{24}-\frac{6 \nu +1}{3 k}\right)\right)}{2^{\nu+2}\, \sqrt{\pi }\, \Gamma \left(\frac{3-\nu }{2}\right)\, \text{Ai}\left(\left(\frac{2}{\pi^2k} \right)^{-1/3} \left(N-\frac{k}{24}-\frac{1}{3 k}\right)\right)}
\end{equation}
From this expression we can read the leading genus--zero term that consistently coincides with the semi--classical string computation of the 1/6--BPS loop performed in \cite{Correa:2014aga}. Recovering the string result is non--trivial and depends on a delicate cancellation of various contributions appearing in the intermediate steps of the calculation.

\item[\bsquare]  The matrix model proposal must also be consistent with explicit weak--coupling calculations.  Therefore, we have performed a perturbative three--loop computation of the bosonic latitude, at generic framing $f$, representation $R$  and winding number $m$. This Feynman diagram tour de force agrees with the third--order expansion of (\ref{eqI:WL}), provided the framing number is {\it formally} identified with $\nu$.
The agreement holds for generic winding and, in addition, we have explicitly verified the consistency of our matrix model proposal with the perturbative result for a few higher dimensional representations of the gauge group.
The requirement of a specific choice of framing does not come as a surprise, in fact localization for Wilson loops on the great circle in ABJ(M) theory produces results at non--trivial framing (see \cite{Kapustin:2009kz}). 
Hence, it is reasonable to expect that our matrix model computation also yields results at a fixed framing number. 
The particular choice $f=\nu$ is natural, as it corresponds to the value at which the bosonic and fermionic latitude expectation values are related by the cohomological equivalence at the quantum level, as pointed out in \cite{Bianchi:2014laa}.
This requires an analytic continuation of the framing parameter from an integer to a real number, which is perfectly legitimate at the matrix model level. 

\item[\bsquare] We discuss the different Bremsstrahlung functions that can be defined in ABJ(M) theories, and in particular we review how they are connected to suitable derivatives of the {\it modulus} of the latitude expectation value.  The introduction of the modulus  slightly modifies the prescription  
considered previously \cite{Bianchi:2014laa, Correa:2014aga,Bianchi:2017ozk}, allowing to eliminate some {\it unexpected} imaginary contributions to the Bremsstrahlung functions, appearing at three loops from the perturbative expansion of the latitudes at framing zero. The origin of these imaginary terms is quite peculiar and is rooted into an anomalous behavior of some correlation function of scalar composite operators, as we discuss carefully in subsection \ref{sec:modulus}.  Once  we take into account this effect, it is  straightforward to understand why the correct prescription for the Bremsstrahlung functions must contain the modulus, analogously to what was originally argued in \cite{Lewkowycz:2013laa}.
We test the matrix model based expression for the Bremsstrahlung function associated to the internal angle  of the bosonic cusp, $B_{1/6}^{\theta}$, against its four--loop perturbative computation for a generic representation, which is presented in \eqref{eq:B16resultmain}. 

\item[\bsquare]  In the  case of the ABJM theories, the following relation,
\beq\label{eq:identitya}
\pa_\nu \log{\left( \langle W_B(\nu) \rangle_\nu + \langle \hat W_B(\nu) \rangle_\nu \right)}\,\, \Big|_{\nu=1} = 0 \qquad\qquad \text{for } N_1=N_2,
\eeq
was originally conjectured  in \cite{Bianchi:2014laa} and is at the core of the relation between Wilson loops and Bremsstrahlung functions proposed there. This identity holds 
for the bosonic latitudes $W_B(\nu), \hat W_B(\nu)$ with gauge group $U(N)_k\times U(N)_{-k}$ evaluated at framing $\nu$, where $\hat W_B(\nu)$ is the Wilson loop associated to the second gauge group. It is related to the first one by complex conjugation. Equation \eqref{eq:identitya} entails that the two Bremsstrahlung functions associated to the bosonic cusp and 
to the 1/2--BPS line can be expressed  only in terms of the phase $\Phi_B(\nu)$ of the bosonic latitude 
\begin{equation}\label{eqI:exactB12}
B_{1/2} = \frac{1}{8\pi}\tan\Phi_B(1)\ \ \ \ \ \ \ \ \ \ \ \ \ \ \ \ \ \ B_{1/6}^{\varphi}=2\, B_{1/6}^{\theta}=-\frac{1}{2\pi^2}\partial_\nu\log(\cos\Phi_B(\nu)) \Big|_{\nu=1}
\end{equation}
where we have defined $\langle W_B(\nu) \rangle_\nu = e^{i \Phi_B(\nu)} |\langle W_B(\nu) \rangle_\nu|$.
We explicitly check that \eqref{eq:identitya} is obeyed by our three-loop perturbative result and, remarkably, it is an exact  consequence of the proposed matrix model.  In fact we prove that more generally our matrix model satisfies this identity not only at $\nu=1$ but for any value of $\nu$. We also reproduce the three--loop result of \cite{Bianchi:2017svd} for the fermionic cusp anomalous dimension in the near--BPS regime. This  chain of  relations among the different Bremsstrahlung functions is also discussed in \cite{satana}.

\item[\bsquare]  Finally, we briefly discuss a possible generalization of our approach for computing the fermionic  Bremsstrahlung function from the 1/6--BPS fermionic latitude in the $N_1\neq N_2$ case. We put forward an exact prediction for $B_{1/2}$, whose expansion up to five loops is provided explicitly in \eqref{prop1/2}. This result calls for future perturbative confirmation. 
\end{itemize}

The paper is structured in the following way. We start in Section \ref{review} by reviewing the construction of two general families of circular Wilson loops, whose contours are the latitudes on a sphere $S^2$.  The two families, bosonic and fermionic latitudes,  differ by the nature of the connection appearing in the  holonomy and the number of preserved supercharges. We discuss the framing dependence of Wilson loops expectation values, the cohomological relation between bosonic and fermionic latitudes and the connections among the different Bremsstrahlung functions, associated to fermionic and bosonic cusps. Subsequently, Section \ref{sec:perturbative} illustrates the three--loop perturbative calculation of the bosonic latitude in the general situation described before. In particular, we elucidate the emergence of an imaginary term at framing zero at three loops and the requirement of considering the modulus of the latitude expectation value in computing the Bremsstrahlung function. Section \ref{sec:matrixlat} and Section \ref{sec:Fermi} are the heart of the paper: here we 
propose our matrix model and study its main properties by using the Fermi gas technique. The strong coupling expansion is performed and successfully compared, in the 1/6--BPS fermionic case, with the semi--classical string computation. In Section \ref{sec:ABJ} we briefly present and discuss a conjectured form of the fermionic  Bremsstrahlung function for $N_1\neq N_2$, providing a prediction up to five loops. Six appendices complete the paper with conventions, details of the computation and further checks of our results.

\section{BPS Wilson loops, Cusps and Bremsstrahlung functions: A review}\label{review}

We begin with a general review of the most fundamental properties of circular BPS Wilson loops in ABJ(M) theories and their non--trivial connections with cusped Wilson lines and the corresponding Bremsstrahlung functions. 

In $U(N_1)_k \times U(N_2)_{-k}$ ABJ(M) theory we consider the general class of Wilson loops that preserve a certain fraction of the original ${\cal N}=6$ supersymmetry\footnote{For a brief summary of our conventions for ABJ(M) theories we refer to Appendix \ref{AppA}.}. Such
operators can be constructed by generalizing the ordinary gauge holonomy with the addition of either scalar  matter bilinears (``bosonic" Wilson loops) \cite{Berenstein:2008dc,Drukker:2008zx, Chen:2008bp, Rey:2008bh} or scalar bilinears and fermions (``fermionic'' Wilson loops) \cite{Drukker:2009hy}. For the case of straight line and maximal circular contours a general classification of such BPS operators based on the amount of preserved supercharges can be found in \cite{Ouyang:2015iza,Ouyang:2015bmy, Lietti:2017gtc}.

\vskip 10pt
\noindent
{\bf  Latitude Wilson loops.} We are primarily interested in the general class of bosonic and fermionic Wilson operators introduced in \cite{Cardinali:2012ru} (latitude Wilson loops). They feature a parametric dependence on a $\a$--angle\footnote{The $\a$--angle can be freely chosen in the interval $[0, \tfrac{\pi}{2}]$, see  \cite{Cardinali:2012ru}.} that governs the couplings to matter in the internal $R$--symmetry space and a geometric angle $\theta_0 \in [-\tfrac{\pi}{2}, \tfrac{\pi}{2}]$ that fixes the contour to be a latitude circle on the unit sphere   
\beq
\G_m: \quad x^\mu = (\sin{\th_0}, \cos{\th_0} \cos{\tau}, \cos{\th_0}\sin{\tau}) \qquad \tau \in [0, 2m\pi) ~{\rm for~winding}~ m
\eeq
Note that here we are generalizing the definitions of \cite{Cardinali:2012ru}  to Wilson loops with generic winding.  As discussed in \cite{Bianchi:2014laa}, these operators can be constructed in such a way that they depend uniquely on the effective ``latitude parameter'' 
\beq
\nu \equiv \sin{2\a} \cos{\th_0} \qquad \qquad 0 \leq \nu \leq 1
\eeq
The $m$--winding bosonic latitude Wilson loops corresponding to the two gauge groups are explicitly given by 
\begin{align}\label{eq:bosonic}
&& W_B^{m}(\nu,R) = \frac{1}{ {\rm dim}(R)}\, \Tr_R\, \mathrm{P}\, \exp \left\{-i\oint_{\Gamma_m} d\tau \left(A_{\mu} \dot x^{\mu}-\frac{2 \pi i}{k} |\dot x| M_{J}^{\ \ I} C_{I}\bar C^{J}\right) \right\}
\nonumber \\ 
&& \hat{W}_B^{m}(\nu,\hat R) = \frac{1}{{\rm dim}(\hat R)}\, \Tr_{\hat R}\, \mathrm{P}\, \exp \left\{-i\oint_{\Gamma_m} d\tau \left( \hat  A_{\mu} \dot x^{\mu}-\frac{2 \pi i}{k} |\dot x| M_{J}^{\ \ I} \bar C^{J} C_I \right)  \right\} 
\end{align}
where the matrix describing the coupling to the $(C_I, \bar{C}^I)$ scalars reads
\bea 
 \mbox{\small $\! M_{J}^{\ I}=\left(\!\!
\begin{array}{cccc}
 - \nu  & e^{-i \tau } \sqrt{1-\nu ^2} & 0 & 0 \\
e^{i \tau }  \sqrt{1-\nu ^2}  & \nu  & 0 & 0 \\
 0 & 0 & -1 & 0 \\
 0 & 0 & 0 & 1 \\
\end{array}
\right)$ }
\eea
The traces in (\ref{eq:bosonic}) are taken over generic representations $R, \hat R$ of $U(N_1)$ and $U(N_2)$, respectively. The overall constants have been purposely chosen in order to normalize the tree level expectation values $\langle W_{B}^m \rangle^{(0)}$ and $\langle \hat{W}_{B}^m \rangle^{(0)}$ to one.  

Similarly, the $m$--winding fermionic latitude Wilson loop for a generic representation $\mathbf{R}$ of the superalgebra $U(N_1|N_2)$ is defined as 
\beq\label{eq:fermionic}
W_F^{m}(\nu,\mathbf{R}) =  {\cal R} \; \mathrm{STr}_{\mathbf{R}} \left[ \mathrm{P}\,\exp\left(-i\oint_{\Gamma_m}\mathcal{L}(\tau)d\tau\right)\begin{pmatrix} e^{-\frac{i\pi m\nu }{2}}  \mathds{1}_{N_1} & 0\\ 0 & e^{\frac{i\pi m\nu}{2}}  \mathds{1}_{N_2} \end{pmatrix} \right] 
\eeq
where $\mathcal{L}$ is the $U(N_1|N_2)$ superconnection 
\bea
\label{supermatrix}
\mathcal{L} &=& \begin{pmatrix}
\mathcal{A}
&i \sqrt{\frac{2\pi}{k}}  |\dot x | \eta_{I}\bar\psi^{I}\\
-i \sqrt{\frac{2\pi}{k}}   |\dot x | \psi_{I}\bar{\eta}^{I} &
\hat{\mathcal{A}}
\end{pmatrix} \ \  \  \mathrm{with}\ \ \  \left\{\begin{matrix} \mathcal{A}\equiv A_{\mu} \dot x^{\mu}-\frac{2 \pi i}{k} |\dot x| {\cal M}_{J}^{\ \ I} C_{I}\bar C^{J}\\
\\
\hat{\mathcal{A}}\equiv\hat  A_{\mu} \dot x^{\mu}-\frac{2 \pi i}{k} |\dot x| {\cal M}_{J}^{\ \ I} \bar C^{J} C_{I}
\end{matrix}\ \right. \nonumber \\
\eea
and
\bea
\label{eq:matrixfermionic}
\mbox{\small $\!{\cal M}_{I}^{ \  J}\!=\!\!\left(\!\!
\begin{array}{cccc}
 - \nu  & e^{-i \tau } 
   \sqrt{1-\nu ^2} & 0 & 0 \\
e^{i \tau }  \sqrt{1-\nu ^2}
   & \nu  & 0 & 0 \\
 0 & 0 & 1 & 0 \\
 0 & 0 & 0 & 1 \\
\end{array}\!
\right)$} \quad &,& \quad  \mbox{\small $\begin{array}{l}\eta_I^\alpha \equiv n_I \eta^\a = \frac{e^{\frac{i\nu \tau}{2}}}{\sqrt{2}}\left(\!\!\!\begin{array}{c}\!\sqrt{1+\nu}\\ -\sqrt{1-\nu} e^{i\tau}\\0\\0 \!\end{array}\!\!\right)_{\!I} \! \! \!\!(1, -i e^{-i \tau})^\alpha
\end{array}$}\!\!
\non \\
&~& \quad \bar\eta_\alpha^I \equiv \bar{n}^I \bar{\eta}_\a = i (\eta^{\alpha}_{I})^{\dagger}
\eea
The generalized prescription (\ref{eq:fermionic}) that requires taking the supertrace of the superholonomy times a constant matrix assures invariance under super gauge transformations \cite{Cardinali:2012ru}. The overall constant in (\ref{eq:fermionic}) can be chosen so as to normalize the expectation value to 1, if possible.
In the rest of the paper we consider the fermionic operator only in the fundamental representation, for which
\beq\label{eq:R}
{\cal R} = \frac{1}{ N_1\, e^{-\frac{i\pi m\nu}{2}} -\, N_2\, e^{\frac{i\pi m\nu}{2} }}
\eeq
We note that for $N_1 = N_2$, if $\nu =1$ ($\nu =0$) and $m$ is even (odd) this normalization becomes meaningless. In those cases one can simply compute the unnormalized expectation value, choosing ${\cal R}=1$.

Whenever no confusion arises we will use $W_{B,F}$ as a shorthand for the single winding operators $W_{B,F}^1$. Moreover, if no explicit dependence on the representation is displayed, the Wilson loop is understood to be in the fundamental representation.

For generic values of the parameters, the latitude bosonic operators in (\ref{eq:bosonic}) preserve 1/12 of the original ${\cal N}=6$ supercharges, whereas the fermionic one in (\ref{eq:fermionic}) is 1/6--BPS. The supersymmetry ($\th^{IJ}_\a$) and superconformal ($\e^{IJ}_\a$) charges preserved by the fermionic latitude can be expressed in terms of four constant spinor parameters $\omega_i$ as \cite{Bianchi:2014laa}
\beq
\label{chargferm}
\begin{split}
\!\!\!\bar\theta^{13}_1=&e^{-\frac{i \theta_0 }{2}} \sqrt{1-\nu } ~\omega _1+e^{\frac{i \theta_0
   }{2}} \sqrt{1+\nu} ~\omega _2  \ \ \  \ \  \  \  \  \ ~\bar\theta^{14}_1=e^{-\frac{i \theta_0 }{2}} \sqrt{1-\nu } ~\omega _3+e^{\frac{i \theta_0
   }{2}} \sqrt{1+\nu} ~\omega _4 \\
\!\!\!\bar\theta^{23}_2=& -i e^{-\frac{i \theta_0 }{2}} \sqrt{1+\nu} ~\omega _1-i e^{\frac{i
   \theta_0 }{2}} \sqrt{1-\nu } ~\omega _2 ~   \  \  \ \bar\theta^{24}_2= -i e^{-\frac{i \theta_0 }{2}} \sqrt{1+\nu} ~\omega _3-i e^{\frac{i
   \theta_0 }{2}} \sqrt{1-\nu } ~\omega _4 \\
\!\!\!{\bar\epsilon}^{13}_1=&i e^{\frac{i \theta_0 }{2}} \sqrt{1-\nu } ~\omega _1-i e^{-\frac{i \theta_0
   }{2}} \sqrt{1+\nu} ~\omega _2 \ \ \ \ \  \  \  \  {\bar\epsilon}^{14}_1=i e^{\frac{i \theta_0 }{2}} \sqrt{1-\nu } ~\omega _3-i e^{-\frac{i \theta_0
   }{2}} \sqrt{1+\nu} ~\omega _4 \\
\!\!\!{\bar\epsilon}^{23}_2=& e^{-\frac{i \theta_0 }{2}} \sqrt{1-\nu } ~\omega _2-e^{\frac{i \theta_0
   }{2}} \sqrt{1+\nu} ~\omega _1 \ \ \  \ \ \ \  \  \  \ {\bar\epsilon}^{24}_2=e^{-\frac{i \theta_0 }{2}} \sqrt{1-\nu } ~\omega _4-e^{\frac{i \theta_0
   }{2}} \sqrt{1+\nu} ~\omega _3 
\end{split}
\eeq
The supercharges preserved by the bosonic latitude can be obtained by setting $\omega_1=\omega_4=0$. We note that in both cases the preserved supercharges carry a non--trivial dependence on the parameter $\nu$.

Enhancement of preserved supersymmetry occurs at $\nu = 1$, where $W_B(1)$ coincides with the bosonic 1/6--BPS operator introduced in \cite{Drukker:2008zx, Chen:2008bp, Rey:2008bh} and the fermionic $W_F(1)$ is the 1/2--BPS operator studied in \cite{Drukker:2009hy}. 

At classical level the fermionic latitude Wilson loop (\ref{eq:fermionic}) is cohomologically equivalent to the following linear combination of bosonic latitudes
\beq \label{cohom}
 W_F^m(\nu) =  {\cal R} \left[ N_1 \, e^{-\frac{i\pi m\nu}{2}}\, W_B^m(\nu)  - N_2 \, e^{\frac{i\pi m\nu}{2}} \, \hat W_B^m(\nu) \right]  + {\cal Q}(\nu) ({\rm something})
\eeq
where for simplicity we have restricted to Wilson loops in the fundamental representation.
In the above formula ${\cal Q}(\nu)$ is the linear combination of superpoincar\'e and superconformal charges \cite{Bianchi:2014laa}
\beq
\label{supercharge}
\begin{split}
\mathcal{Q(\nu)}=&-\sqrt{\frac{1+\nu
    }{2}}\left(e^{\frac{i \theta_0 }{2}}
  Q^{13,1}-i e^{-\frac{i \theta_0 }{2}}
  S^{13,1}+e^{-\frac{i \theta_0 }{2}}
 Q^{24,2}-i e^{\frac{i \theta_0 }{2}}
S^{24,2}\right)\\
   &\qquad +i \sqrt{\frac{1-\nu  }{2} }\left(e^{\frac{i \theta_0 }{2}}
Q^{23,2}+i e^{-\frac{i \theta_0 }{2}}
S^{23,2} -e^{-\frac{i \theta_0
   }{2}}Q^{14,1}-i e^{\frac{i \theta_0 }{2}}
 S^{14,1}\right)
 \end{split}
\eeq
preserved by both bosonic and fermionic Wilson loops. 

If this equivalence survives at quantum level it allows to compute the vacuum expectation value $\langle W_{F}  (\nu) \rangle$  of the fermionic operator as a combination of the bosonic ones. 
However, in three dimensions the problem of understanding how the classical cohomological equivalence gets implemented at quantum level is strictly interconnected with the problem of understanding framing, as we review below.

\vskip 10pt
\noindent
{\bf  Matrix models for BPS Wilson loops.}
Using the procedure of supersymmetric localization the ABJ(M) partition function on the three--sphere can be reduced to the matrix model integral \cite{Kapustin:2009kz}
\begin{align}
\label{eq:matrixABJM}
Z = &\int \prod_{a=1}^{N_1}d\lambda _{a} \ e^{i\pi k\lambda_{a}^{2}}\prod_{b=1}^{N_2}d\mu_{b} \ e^{-i\pi k\mu_{b}^{2}}\,  \frac{\displaystyle\prod_{a<b}^{N_1}\sinh ^{2}\pi (\lambda_{a}-\lambda _{b})\prod_{a<b}^{N_2}\sinh ^{2}\pi (\mu_{a}-\mu_{b})}{\displaystyle\prod_{a=1}^{N_1}\prod_{b=1}^{N_2}\cosh ^{2}\pi (\lambda _{a}-\mu_{b})}
\end{align}
where we are being cavalier on the precise normalization, which is unimportant for the computation of Wilson loops. 
At $\nu=1$ the expectation values of the 1/6-- and 1/2--BPS Wilson loops can be computed as matrix model averages.
In particular, the $m$--winding bosonic 1/6--BPS Wilson loop in the fundamental representation\footnote{ The prescription can be generalized to higher dimensional representations.} is given by
\begin{align}\label{eq:matrixmodelaverage}
\langle  W_B^{m} (1) \rangle = \frac{1}{N_1}\left\langle \sum_{i=1}^{N_1} e^{2\pi\, m\, \lambda_{i} } \right\rangle \qquad , \qquad 
\langle  \hat W_B^{m} (1) \rangle = \frac{1}{N_2}\left\langle \sum_{i=1}^{N_2} e^{2\pi\, m\, \mu_{i} } \right\rangle 
\end{align}
where the right--hand--side brackets stand for the integration using the matrix model measure defined in \eqref{eq:matrixABJM}, normalized by the partition function. In this language, the 1/2--BPS Wilson loop can be computed as the average of a supermatrix operator, or equivalently using \eqref{cohom}  \cite{Drukker:2009hy}. 

For generic $\nu$, no matrix model prescription has been found so far. In fact, even though the latitude Wilson loops are BPS operators, in this case the standard localization arguments of \cite{Kapustin:2009kz} cannot be directly applied \cite{Bianchi:2014laa}.  We aim at filling this gap in Section \ref{sec:matrixlat}, where we conjecture a matrix model for the latitude Wilson loops that turns out to be compatible with all the available data points at weak and strong coupling. 
 
\vskip 10pt
\noindent
{\bf  Framing.} In three--dimensional Chern--Simons theories the computation of Wilson loop expectation values is affected by finite regularization ambiguities associated to singularities arising when two fields running on the same closed contour clash. 
In perturbation theory, this phenomenon is ascribable to the use of point--splitting regularization to define propagators at coincident points.
For ordinary Wilson loops, following e.g.~the prescription of \cite{Guadagnini:1989am}, one allows one endpoint of the gluon propagator to run on the original closed path $\Gamma$ on which the Wilson loop is evaluated, and the other to run on a framing contour $\Gamma_f$. This is infinitesimally displaced from $\Gamma$ and defined by the choice of a vector field on it.
Then the one--loop Chern--Simons contribution is proportional to the Gauss linking integral 
\begin{equation}\label{eq:gauss}
\frac{1}{4\pi} \oint_{\G} dx^\m \oint_{\G_f} dy^\n \; \varepsilon_{\m\n\rho} \frac{(x-y)^\rho}{|x-y|^3} \, \equiv \, f
\end{equation}
which evaluates to an integer $f$ (the framing number).
This is a topological invariant that counts the number of times the additional closed contour $\G_f$ introduced by the framing procedure winds around the original one $\G$. 

This phenomenon has been first discovered and extensively discussed for pure $U(N)$ Chern--Simons theory \cite{Witten:1988hf}, in connection to knot theory. In this case, the total effect of framing amounts to a phase, exponentiating the one--loop result (from now on we use the subscript on the expectation value to indicate a certain choice of framing)
\beq \label{framing}
\langle W_{CS}  \rangle_f = e^{i \pi \lambda\, f } \langle W_{CS} \rangle_0 
\eeq
where $\lambda$ is the 't Hooft Chern--Simons coupling shifted by the quadratic Casimir of the gauge group.
 
More recently, the same kind of framing dependence has been discussed  also for non--topological Chern--Simons theories coupled to matter, in particular ABJ(M) theories \cite{Bianchi:2016yzj, Bianchi:2016vvm}. 
In order to review framing effects in this context it is convenient to split the Wilson loop expectation value into its phase and its modulus. For the most general case of $m$--winding operators with a non--trivial latitude we set
\begin{equation}\label{eq:phase}
\langle W_{B}^m(  \nu) \rangle_f  = e^{i \Phi_{B}(f,m,\nu)}  \, \Big|\,\langle W_{B}^m(  \nu) \rangle_f\, \Big| \quad , \quad \langle \hat W_B^m(  \nu) \rangle_f  = e^{i \hat \Phi_B(f,m,\nu)}  \, \left|\,\langle \hat W_B^m(  \nu) \rangle_f\, \right|
\end{equation}
and similarly for $W_F^m(\nu)$.

In the $\nu=1$ case, the bosonic 1/6--BPS operators have been computed up to three loops in the large $N_1, N_2$ limit, for generic framing number and winding $m$ \cite{Bianchi:2016yzj,Bianchi:2016gpg}. 
As an effect of framing, their expectation values acquire  imaginary contributions at odd orders, as well as additional real corrections at even orders. 
We stress that in this case the imaginary contributions are entirely due to framing. 

For single winding the framing contributions can be still captured by a phase, precisely $\Phi_{B}(f,1,1), \hat \Phi_{B}(f,1,1)$ in (\ref{eq:phase}), 
while the expectation values at framing zero are real quantities and coincide with the modulus. However, the phases are no longer a one--loop effect as in the pure Chern--Simons theory, but display non--trivial quantum corrections starting at three loops \cite{Bianchi:2016yzj}
\beq \label{eq:framingfunct}
\Phi_B(f,1,1) = \pi \frac{N_1}{k} f- \frac{\pi^3}{2} \frac{N_1N_2^2}{k^3} \, f+ O(k^{-5})  \; , \; \hat \Phi_B(f,1,1) = -\pi \frac{N_2}{k} f + \frac{\pi^3}{2} \frac{N_1^2N_2}{k^3} \, f+ O(k^{-5}) 
\eeq

For multiple windings the effect of framing is more complicated and ceases to be encapsulated into a phase \cite{Bianchi:2016gpg}. This does not come as a surprise since the same pattern occurs also in the pure Chern--Simons theory.

For latitude Wilson loops the phases $\Phi_B, \hat \Phi_B$ will depend in general on the framing and winding numbers, and the latitude parameter $\nu$ as well. According to the discussion above, they will be non--trivial functions of the couplings that reduce to the expansions (\ref{eq:framingfunct}) for $\nu=1$ and single winding. Having in mind the most general scenario, we may expect them
not to necessarily account for all the framing effects (as in the multiple winding situation). Hence, in general the modulus in \eqref{eq:phase} does not coincide with the expectation value at framing zero. Moreover, for latitude operators the phase might not even be entirely produced by framing, as further framing independent imaginary contributions could arise. 
Checking if this is the case and better understanding the framing origin of $\Phi_B, \hat \Phi_B$ is one of the goals of this paper.

We conclude this short review on framing by discussing its role in localization.
It was argued in \cite{Kapustin:2009kz} that the matrix model \eqref{eq:matrixABJM} derived from localization computes the 1/6-- and 1/2--BPS Wilson loops at framing one. This is because a point--splitting regularization, implied in the derivation, is compatible with supersymmetry only if the circular path and the framing contour are two Hopf fibers in the $S^1$ fibration of the three--sphere. These in turn have linking number one, which explains the particular framing number arising in this computation. The aforementioned studies on supersymmetric Wilson loops have provided a perturbative test of such an argument.

For the purposes of this paper we therefore stress that a matrix model computing the expectation value of latitude Wilson loops is expected to imply a particular choice of framing.

\vskip 10pt
\noindent
{\bf  Cohomological equivalence and non--integer framing.} As stated above, at classical level the fermionic and bosonic latitude Wilson loop operators are related by the cohomological equivalence \eqref{cohom}.
If relation \eqref{cohom} survives at the quantum level, then $\langle W_F (\nu) \rangle$ can be obtained as a linear combination of $\langle W_B (\nu) \rangle$ and $ \langle \hat{W}_B (\nu) \rangle$. 
In particular, we expect that to be the case (namely no anomalies arise) in the localization approach, if the functional integral computing the Wilson loop expectation value is localized using its invariance under the same supercharge ${\cal Q}$ in (\ref{cohom}). 

For $\nu=1$ this reduces to the cohomological equivalence first discovered in \cite{Drukker:2009hy}.
In this case the localization computation is performed at framing 1, as recalled above, hence the equivalence is expected to hold for this particular choice of framing (in this section we restrict to the fundamental representation and set $m=1$ for simplicity)
\begin{equation}\label{eq:cohom1}
\langle W_F(1)\rangle_1 =  {\cal R} \left( N_1 \, \langle W_B(1) \rangle_1 - N_2 \, \langle \hat W_B(1) \rangle_1 \right)
\end{equation}
but could be modified if another choice of framing is taken.
In fact, this has been explicitly verified in ordinary perturbation theory at framing zero up to two loops, where the equivalent of \eqref{eq:cohom1} with $f=0$ fails.

On the other hand, from equation (\ref{eq:framingfunct}) it follows that up to two loops framing zero and one expectation values of bosonic 1/6--BPS Wilson loops are related as 
\beq
 \langle W_B(1) \rangle_1  \equiv  e^{i\pi \frac{N_1}{k}}  \, \langle W_B(1) \rangle_0  + O(k^{-3}) \quad , \quad \langle \hat{W}_B(1) \rangle_1  \equiv e^{-i\pi\frac{N_2}{k}}  \, \langle \hat{W}_B(1) \rangle_0 + O(k^{-3}) 
\eeq
Using this, and further defining 
\beq
\langle W_F (1) \rangle_1 \equiv e^{i\pi \frac{N_1-N_2}{k}} \langle W_F(1) \rangle_0 + O(k^{-3})
\eeq
identity \eqref{eq:cohom1} has been confirmed to hold, perturbatively \cite{Griguolo:2013sma, Bianchi:2013zda, Bianchi:2013rma}.

In the latitude case, the analogous two--loop calculation \cite{Bianchi:2014laa} shows that the cohomological equivalence survives at quantum level in the form   
\beq\label{eq:quantumcoho}
\langle W_F(\nu) \rangle_\nu =  {\cal R} \left[ N_1\, e^{-\frac{i\pi \nu}{2}}\, \langle W_B(\nu) \rangle_\nu \, - \, N_2 \, e^{\frac{i \pi  \nu}{2}} \, \langle \hat   W_B(\nu) \rangle_\nu \right]  
\eeq
if we define
\bea \label{framingnu}
&& \langle W_B(\nu) \rangle_\nu  \equiv  e^{i\pi \frac{N_1}{k} \nu}  \, \langle W_B( \nu) \rangle_0  + O(k^{-3}) \quad , \quad \langle \hat{W}_B( \nu) \rangle_\nu  \equiv e^{-i\pi\frac{N_2}{k}\nu}  \, \langle \hat{W}_B( \nu) \rangle_0 + O(k^{-3}) \non \\
&&  \qquad \qquad \qquad  \quad \langle W_F ( \nu) \rangle_\nu \equiv e^{i\pi \frac{N_1-N_2}{k} \nu} \langle W_F( \nu) \rangle_0 + O(k^{-3})
\eea
that is if we formally identify the framing number $f$ with the latitude parameter $\nu$. Therefore, in the general case, we allow the latitude to be evaluated at non--integer framing $\nu$.
Moreover, we expect that a matrix model computation of the latitude Wilson loop, respecting the cohomological equivalence at the quantum level, would imply framing $\nu$.

\vskip 10pt
\noindent
{\bf   Bremsstrahlung functions.} The Bremsstrahlung function $B$ is the physical quantity that measures the energy lost by a heavy quark slowly moving ($|v| \ll 1$) in a gauge background. Generalizing the well--known law of electrodynamics, it is defined as \cite{Correa:2012at} 
\beq
\Delta E = 2\pi B \int dt (\dot{v})^2
\eeq
In a  conformal field theory it also appears as the coefficient of the first non--trivial order in the small angle expansion of the cusp anomalous dimension, $\G_{cusp}(\varphi) \sim - B \varphi^2$, which governs the short distance divergences of a cusped Wilson loop.

In ABJ(M) theories, since we have bosonic and fermionic Wilson loops we can define different types of Bremsstrahlung functions \cite{Griguolo:2012iq,  Lewkowycz:2013laa}. Computing the divergent part of a fermionic, locally 1/2--BPS Wilson loop $W_F^\angle$ along a generalized cusped contour (two straight lines meeting at a point with a relative $\varphi$ angle) one finds 
\beq \label{eq:Bferm}
\langle W_F^\angle(\varphi,\theta) \rangle \sim e^{-\G^{1/2}_{cusp} (\varphi, \th) \log{\frac{\L}{\e}} } \qquad {\rm with }\quad \G^{1/2}_{cusp} (\varphi, \th) \underset{\varphi, \th \ll 1}{\sim} B_{1/2} (\th^2 - \varphi^2) 
\eeq
where $\Lambda$ and $\e$ are IR and UV cutoffs, respectively.
Here $\theta$ is the internal angle that describes possible relative rotations of the matter couplings between the Wilson loops defined on the two semi--infinite lines. $B_{1/2}$ appears as a common factor in the small angles expansion as a consequence of the fact that for $\th = \varphi$ the fermionic cusped Wilson loop is BPS and divergences no longer appear. 

Analogously, using a bosonic 1/6--BPS Wilson loop on a cusp we can define 
\beq \label{eq:Bbos}
\langle W_B^\angle(\varphi,\theta) \rangle \sim e^{-\G^{1/6}_{cusp} (\varphi, \th) \log{\frac{\L}{\e}} } \qquad {\rm with }\quad \G^{1/6}_{cusp} (\varphi, \th) \underset{\varphi, \th \ll 1}{\sim} B_{1/6}^\th \, \th^2  - B_{1/6}^\varphi \, \varphi^2 
\eeq
and similar relations for $\langle \hat W_B^\angle(\varphi,\theta) \rangle$ that give rise to $\hat B_{1/6}^\th, \hat B_{1/6}^\varphi$ associated to the second gauge group. 
In the bosonic case we have in principle two different Bremsstrahlung functions since there are no BPS conditions for cusped bosonic Wilson loops. 
   
A crucial problem consists in relating $B$ to other physical quantities that in principle can be computed exactly using localization techniques, like for instance circular BPS Wilson loops. For the ABJM theory\footnote{We postpone to Section \ref{sec:ABJ} the discussion of the more general ABJ case.}, this problem was originally addressed in \cite{Lewkowycz:2013laa}, where an exact prescription was given to compute $B_{1/6}^{\varphi}$ in terms of a $m$--winding Wilson loop
\beq\label{Bvarphi}
B_{1/6}^{\varphi} = \frac{1}{4\pi^2}\, \pa_m \log \left| \, \langle W_B^{m} \rangle\, \right|\,\, \Big|_{m=1} 
\eeq
A similar prescription has been later derived for $B_{1/2}$  and $B_{1/6}^{\th}$  in ABJM \cite{Bianchi:2014laa, Correa:2014aga}, in terms of single winding, latitude fermionic (\ref{eq:fermionic}) and bosonic (\ref{eq:bosonic}) Wilson loops, respectively 
\beq\label{Btheta}
B_{1/2} = \frac{1}{4\pi^2}\, \pa_\nu \log \left|\, \langle W_F(\nu) \rangle\, \right|\,\, \Big|_{\nu=1}     \qquad , \qquad  B_{1/6}^{\th} = \frac{1}{4\pi^2}\, \pa_\nu \log \left|\, \langle W_B(\nu) \rangle\, \right|\,\, \Big|_{\nu=1}    
\eeq
These formulae were proven in \cite{Bianchi:2017ozk} and \cite{Correa:2012at}, respectively. We note that in order to enforce the reality of the result we take the modulus of the expectation 
values\footnote{The original prescriptions in \cite{Bianchi:2014laa} were presented without the modulus, since up to two loops the expectation values at framing zero are real.}. We derive this prescription in Section \ref{sec:modulus}.
According to the previous discussion this is also supposed to remove framing ambiguities, henceforth the expectation values in \eqref{Btheta} can be computed at any convenient framing. 

These prescriptions have already passed several tests at weak and/or strong coupling. At weak coupling, the lowest order term of $B_{1/6}^{\varphi} $ computed using (\ref{Bvarphi}) agrees with the result obtained from a genuine two--loop calculation of $\G_{cusp}^{1/6}$ \cite{Griguolo:2012iq}.  $B_{1/6}^{\th} $ obtained from prescription (\ref{Btheta}) has been tested at weak coupling up to two loops \cite{Bianchi:2014laa} for Wilson loops in the fundamental representation of the gauge group. $B_{1/2}$ as computed via \eqref{Btheta} has been tested  at weak coupling up to two loops \cite{Griguolo:2012iq,Bianchi:2014laa}. Moreover, the leading term at strong coupling is successfully reproduced by the string dual configuration of $W_F(\nu)$ found in \cite{Correa:2014aga}.  

A direct perturbative calculation of $B_{1/6}^{\theta}$ at four loops has been performed in \cite{Bianchi:2017afp, Bianchi:2017ujp}, and compared with the perturbative result for $B_{1/6}^{\varphi}$ as obtained from prescription (\ref{Bvarphi}). Interestingly, it has been found that the simple relation 
\beq \label{eq:bthetaconj}
2\, B_{1/6}^{\theta}(k,N_1,N_2) =  B_{1/6}^{\varphi}(k,N_1,N_2)
\eeq
is valid up to this order and has been conjectured to be true exactly. For the ABJM case ($N_1=N_2$) this has been proved \cite{satana}.
For the more general case, taking into account prescriptions \eqref{Bvarphi} and \eqref{Btheta} it amounts to conjecturing that 
\begin{equation}\label{eq:relationbremsstrahlung}
\partial_\nu \left| \langle W_B(\nu) \rangle \right| \bigg|_{\nu=1} = \left. \frac12\, \partial_m \left| \langle W_B^m(1) \rangle \right| \right|_{m=1}
\end{equation}
We stress that this applies to generic $N_1$ and $N_2$ and no planar limit is assumed. In order to provide further support, in Appendix \ref{sect:4loops} we generalize the four--loop test to the case of Wilson loops in generic representations.
The result for the Bremsstrahlung function reads
\begin{align}\label{eq:B16resultmain}
B_{1/6}^{\theta}(R) &= \frac{N_2 C_2(R)}{4 k^2}-\frac{\pi ^2 N_2}{24 k^4} \left(\left(N_1-5 N_2\right) C_1^2(R)+\left(N_2^2+5 N_1 N_2-2\right) C_2(R)\right.\nonumber\\&\left.+2 N_1 C_3(R)-2 C_2(R) C_1(R)-2 C_4(R)\right)+O\left(k^{-6}\right)
\end{align} 
and is in agreement with \eqref{Btheta} upon using the circular Wilson loops in the appropriate representation.
We observe that the four-loop contribution exhibits an explicit dependence on higher order Casimir invariants, thereby violating quadratic Casimir scaling, as recently observed in related four--dimensional contexts \cite{Boels:2017skl,Grozin:2017css}.

Concerning $B_{1/2}$, a further point is worth mentioning separately. As discussed in \cite{Bianchi:2014laa}, 
in the $N_1=N_2$ situation one can derive from the first equation in \eqref{Btheta} an exact expression in terms of 1/6--BPS winding Wilson loops
\begin{equation}\label{eq:exactB12}
B_{1/2} = -\frac{i}{8\pi}\, \frac{\langle W_B(1)\rangle-\langle \hat W_B(1)\rangle}{\langle W_B(1)\rangle+\langle \hat W_B(1)\rangle}  \qquad\qquad \text{for } N_1=N_2
\end{equation}
This requires making use of the cohomological equivalence in (\ref{eq:quantumcoho}), assuming a certain relation between $W_B(\nu)$ and the undeformed $m$--winding Wilson loop and finally assuming the validity of the following identity 
\beq\label{eq:identity}
\pa_\nu \log{\left( \langle W_B(\nu) \rangle_\nu + \langle \hat W_B(\nu) \rangle_\nu \right)}\,\, \Big|_{\nu=1} = 0 \qquad\qquad \text{for } N_1=N_2
\eeq
As long as we do not know the exact expression for $\langle W_B(\nu) \rangle_\nu,  \langle \hat W_B(\nu) \rangle_\nu$ this identity cannot be rigorously proved. 
However, it has been indirectly verified up to three loops by testing the prediction for $B_{1/2}$ from (\ref{eq:exactB12}) against an explicit computation of $\G_{cusp}^{1/2}$ at this order \cite{Bianchi:2017svd}. In analogy with the $m$--winding case, it is likely to hold at any perturbative order and, in particular, at strong coupling. 
In fact, as an indirect check, in this regime \eqref{eq:exactB12} agrees with the explicit string theory computation of the Bremsstrahlung function performed up to the first subleading term \cite{Forini:2012bb,Aguilera-Damia:2014bqa}. Remarkably, as discussed in Section \ref{sec:matrixlat} our conjectured matrix model that computes $\langle W_B(\nu) \rangle_\nu,  \langle \hat W_B(\nu) \rangle_\nu$, for $N_1=N_2$ satisfies relation \eqref{eq:identity} not only at $\nu=1$ but as a functional identity.

Assuming (\ref{eq:identity}) to be true has far--reaching consequences, the main one being that $B_{1/6}^{\th}$ and $B_{1/2}$ can be entirely expressed in terms of the phase $\Phi_B$ introduced in (\ref{eq:phase}) for the latitude Wilson loop at framing $\nu$ and single winding (we use the shorthand notation $\Phi_B(\nu,1,\nu)\equiv \Phi(\nu)$)
\beq\label{eq:Bframing}
B_{1/6}^{\th} = \frac{1}{4\pi^2} \, \tan{ \Phi(\nu)} \; \pa_\nu \Phi(\nu) \Big|_{\nu=1}  \quad , \quad B_{1/2} = \frac{1}{8\pi} \, \tan{\Phi(1)}  \quad\quad \text{for } N_1=N_2
\eeq
In particular, it follows that a genuine perturbative computation of $B_{1/6}^{\th}$ directly from $\G_{cusp}^{1/6}$ allows us to make a prediction for the $\Phi(\nu)$ function. In fact, exploiting the four--loop result for $B_{1/6}^{\th}$ given in \cite{Bianchi:2017afp, Bianchi:2017ujp} the following prediction can be made for $N_1=N_2=N$ and in the planar limit \cite{satana} 
\beq\label{eq:prediction}
\Phi(\nu)= \pi \frac{N}{k} \nu - \frac{\pi^3}{6} \frac{N^3}{k^3} (\nu^3 + 2\nu) + O(k^{-5}) + \text{non-planar}
\eeq
Note that for $f=\nu=1$ it consistently reproduces (\ref{eq:framingfunct}).

Now, merging (\ref{eq:prediction}) with the two--loop result for $\langle W_B(\nu) \rangle_0$ \cite{Bianchi:2014laa}, from (\ref{eq:phase}) we obtain a three--loop prediction for $\langle W_B(\nu) \rangle_\nu$ in the case of ABJM theory
\beq \label{eq:prediction2}
\langle W_B(\nu) \rangle_\nu = 1 + i\pi \frac{N}{k} \nu + \frac{\pi^2}{3} \frac{N^2}{k^2} + i \frac{\pi^3}{6} \frac{N^3}{k^3} \nu^3 + O(k^{-4}) + \text{non-planar}
\eeq
In the next section we are  going to test this prediction against a perturbative three--loop calculation done at framing $\nu$. This turns out to be an indirect check of the validity of assumption (\ref{eq:identity}) and, therefore, of identities (\ref{eq:Bframing}) relating the Bremsstrahlung functions to the phases of bosonic latitude Wilson loops.

\vskip 20pt

\section{Perturbative result for the latitude Wilson loop}\label{sec:perturbative}

In this section we compute the expectation value of the bosonic latitude Wilson loop $W_B(\nu)$ at weak coupling in perturbation theory. The evaluation of $\hat W_B(\nu)$ easily follows by exchanging $N_1 \leftrightarrow N_2$ and sending $k  \to -k$. 

We consider the most general case where a non--trivial framing number $f$ is taken into account and the contour winds $m$ times around the latitude circle. We also allow for the trace in definition \eqref{eq:bosonic} to be taken in a generic representation $R$ of the $U(N_1)$ gauge group.   
The $U(N_1)$ color factors are expressed in terms of the Casimir invariants, as defined in Appendix \ref{sec:Casimir}. We work at finite $N_1$ and $N_2$, i.~e.~no planar limit is assumed.

The multiple windings and higher dimensional representations are not independent generalizations, as one can re-express the multiply wound Wilson loop as a linear combination of an alternative basis of operators in different representations \cite{Klemm:2012ii, Brini:2011wi}. Still, in perturbation theory we can treat these two properties independently and use the aforementioned relation as a consistency check of the computation.

In dealing with diagrams contributing to framing, we make use of general properties of the pure Chern--Simons perturbation theory. In particular, we apply the Alvarez--Labastida argument \cite{Alvarez:1991sx}, stating that only diagrams with collapsible propagators can contribute to framing,  to rule out their non--planar realizations.
We argue that up to three loops the whole framing dependence of the Wilson loop can be effectively ascribed to and computed from the Gauss linking integral \eqref{eq:gauss}.
We remark that although the linking number $f$ for two closed curves is naturally an integer number, we will consider its continuation to real numbers.

Throughout the computation dimensional regularization in the DRED scheme is assumed \cite{Siegel:1979wq} (see also \cite{Chen:1992ee} and \cite{Bianchi:2013rma,Griguolo:2013sma,Bianchi:2013pva,Bianchi:2014iia,Bianchi:2016yzj} for applications in perturbation theory in Chern--Simons models).

In Section \ref{singlew} we give details of the calculation for the single winding Wilson loop, whereas the generalization to multiple windings is discussed in Section \ref{sec:winding}.
For readers who want to skip technical details we summarize our results in Section \ref{3loopresult}. Finally, in Section \ref{sec:modulus} we revisit the proof of identities \eqref{Bvarphi} and \eqref{Btheta}
in light of the appearance of novel three--loop imaginary contributions not related to framing. 

\subsection{The computation}\label{singlew}

Bosonic and fermionic latitude Wilson loops in the fundamental representation and for single winding have been computed in \cite{Bianchi:2014laa}, up to two loops  and at framing zero. At non--trivial framing perturbative calculations up to three loops have been carried out in \cite{Bianchi:2016yzj} only for $W_B(1)$, in the fundamental representation and in the planar limit.

Generalizing those results to $W_B(\nu)$ in a generic representation, at one loop the only diagram contributing is a gluon exchange, which at non--trivial framing is proportional to the Gauss integral (\ref{eq:gauss})
\begin{equation}\label{eq:1loop}
\langle W_{B} (\nu) \rangle_f^{(1)} : \raisebox{-0.8cm}{\includegraphics[width=2.cm]{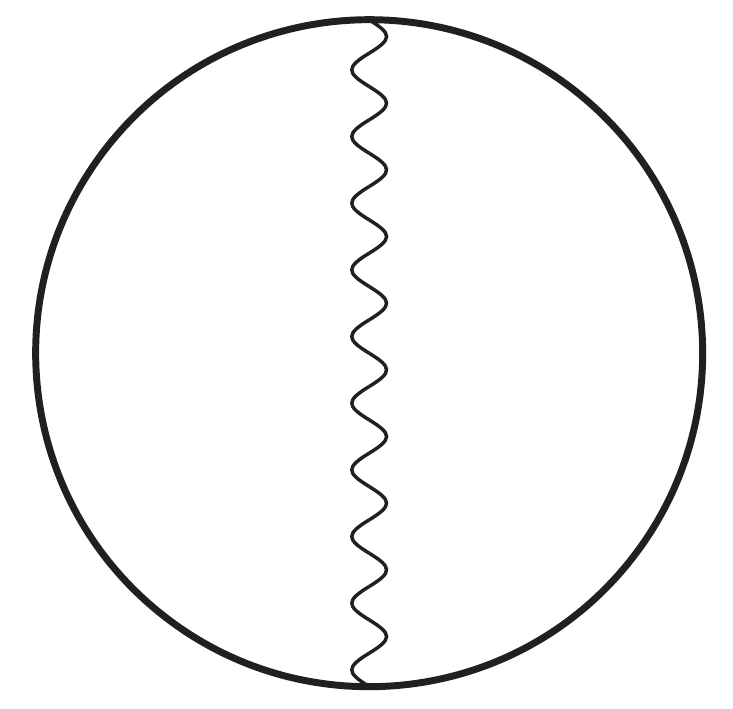}} = i\, \pi\, f\, C_2(R)
\end{equation}
where $C_2( R )$ is defined in (\ref{eq:Rcasimir}). 

In order to draw higher loop diagrams in a more concise way we find convenient to define a ``double--line exchange'' given by the combination of a bi--scalar exchange and a one--loop corrected gluon exchange evaluated in dimensional regularization 
\begin{equation}\label{eq:combined}
\raisebox{-0.325cm}{\includegraphics[width=2.cm]{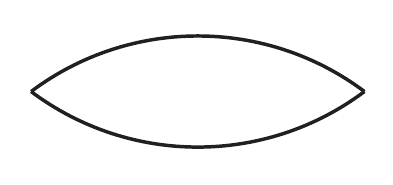}} + \raisebox{-0.32cm}{\includegraphics[width=2.cm]{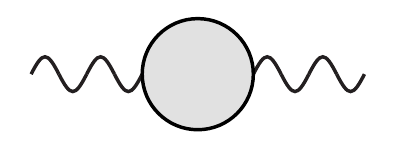}} \equiv \raisebox{0.0cm}{\includegraphics[width=2.cm]{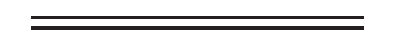}} = \frac{N_2}{k^2} \, \frac{\Gamma^2(\frac12-\epsilon)}{\pi^{1 -2\epsilon}} \; \frac{- \dot{x}_1 \cdot \dot{x}_2 + \frac{1}{4} \,  |\dot{x}_1| |\dot{x}_2| \, \Tr(M_1 M_2)}{[(x_1-x_2)^2]^{1-2\epsilon}}
\end{equation}
The dependence on the latitude parameter comes from $\Tr(M_1 M_2)$, where $M_i$ stands for the coupling matrix evaluated at point $\tau_i$ on the contour (see identity (\ref{eq:m2trace})). 

It follows that at two loops the following diagrams contribute
\begin{align}
\langle W_{B}(\nu) \rangle_f^{(2)} \, : \qquad & \raisebox{-0.8cm}{\includegraphics[width=2.cm]{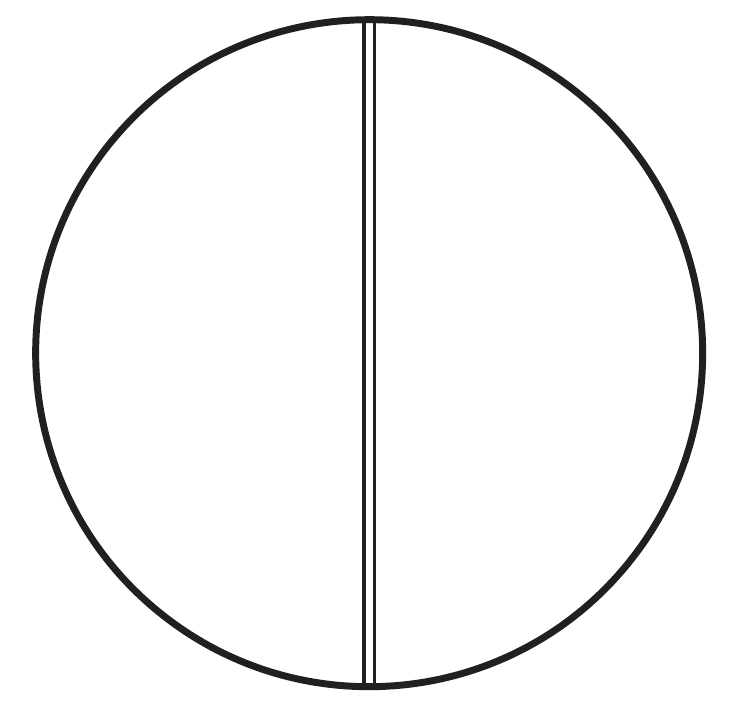}} = \pi^2\, \frac{1+\nu^2}{2}\, N_2\, C_2(R) \label{eq:matter}\\
& \raisebox{-0.8cm}{\includegraphics[width=2.cm]{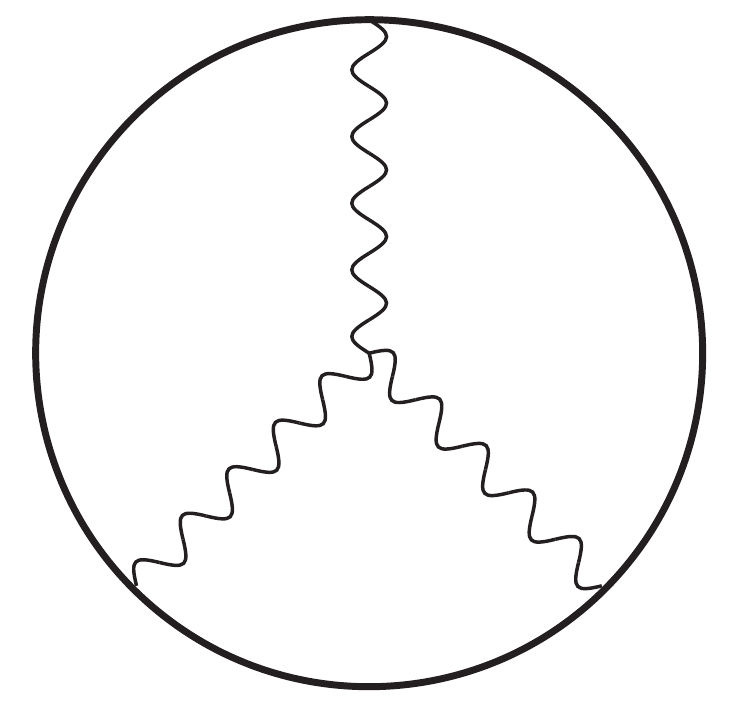}} = -\frac{\pi^2}{6}\, \left( N_1 C_2(R) - C_1^2(R) \right) \label{eq:mercedes} \\
& \raisebox{-0.8cm}{\includegraphics[width=2.cm]{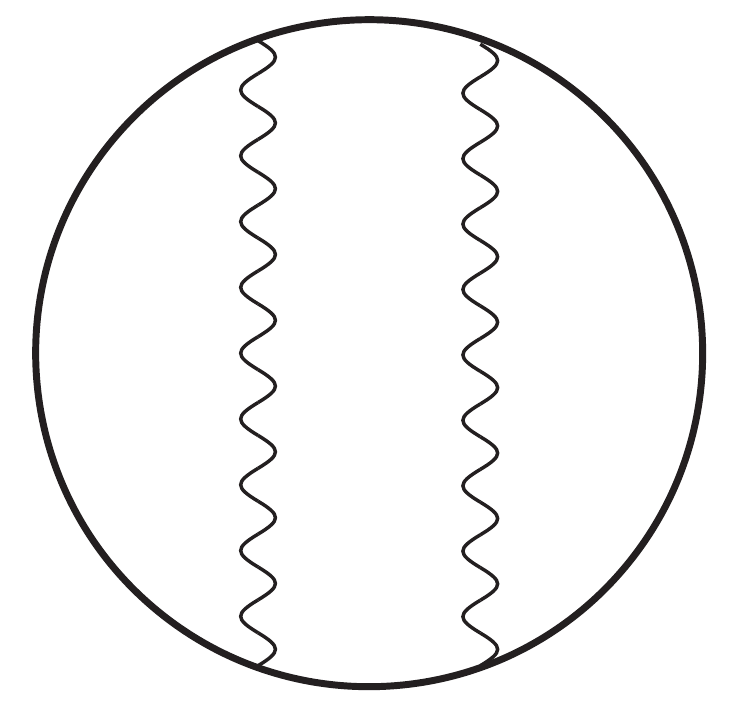}} + {\rm perms} = -\frac{\pi^2}{2}\, C_2^2(R) \, f^2 \label{eq:symm}
\end{align}
In (\ref{eq:symm}) we sum over all possible planar and non--planar permutations in order to factorize the two--loop diagram as half the squared one--loop graph (\ref{eq:1loop}). This does not contradict the Alvarez--Labastida argument, since the non--planar crossed configuration is identically vanishing.

\begin{figure}
\centering
\begin{subfigure}{2.1cm}
  \centering
  \includegraphics[width=2.cm]{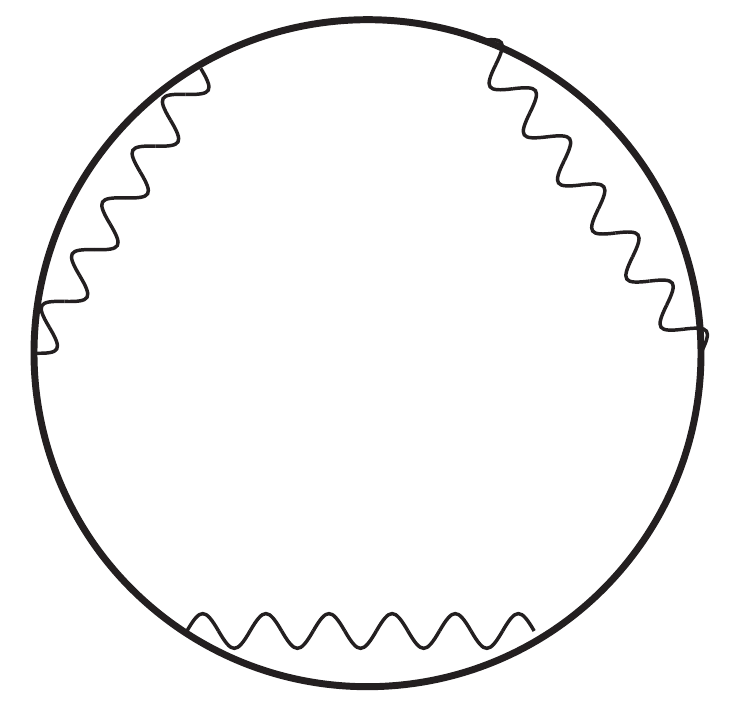}
  \caption{}
  \label{fig:3Ldiagramsa}
  \end{subfigure}
 \quad
 \begin{subfigure}{2.1cm}
  \centering
  \includegraphics[width=2.cm]{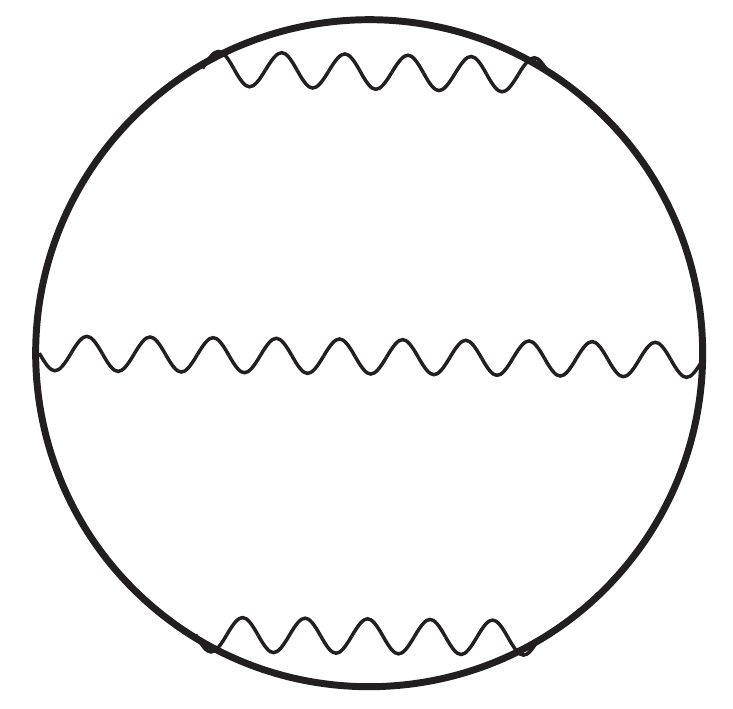}
  \caption{}
  \label{fig:3Ldiagramsb}
  \end{subfigure}  \quad
  \begin{subfigure}{2.1cm}
  \centering
  \includegraphics[width=2.cm]{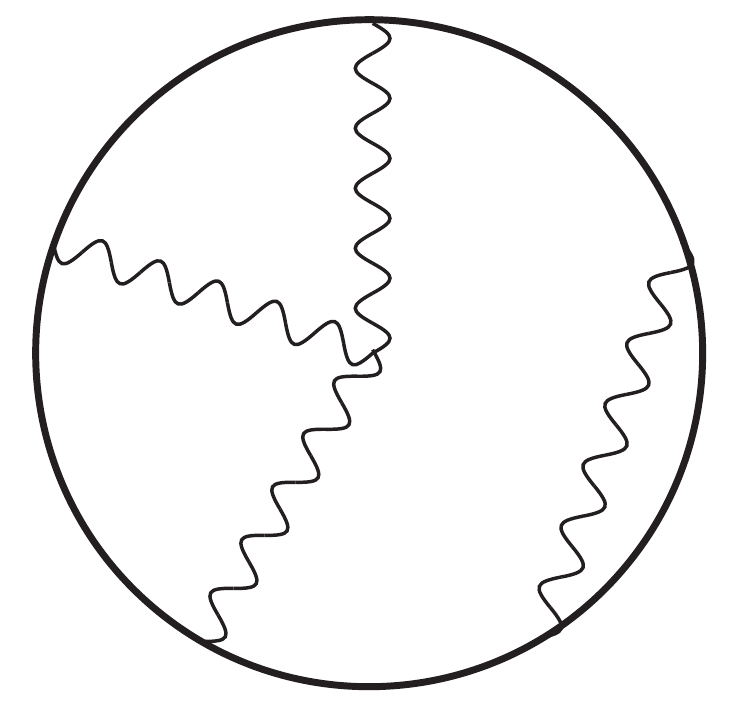}
  \caption{}
  \label{fig:3Ldiagramsc}
  \end{subfigure}  \quad
  \begin{subfigure}{2.1cm}
  \centering
  \includegraphics[width=2.cm]{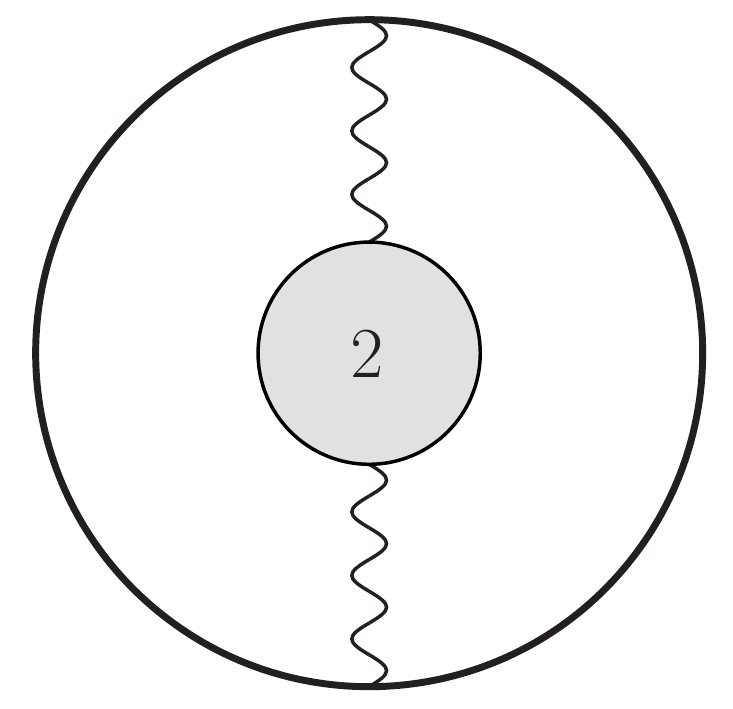}
  \caption{} 
  \label{fig:3Ldiagramsd}
  \end{subfigure} \\[3mm]
  \begin{subfigure}{2.1cm}
  \centering
  \includegraphics[width=2.cm]{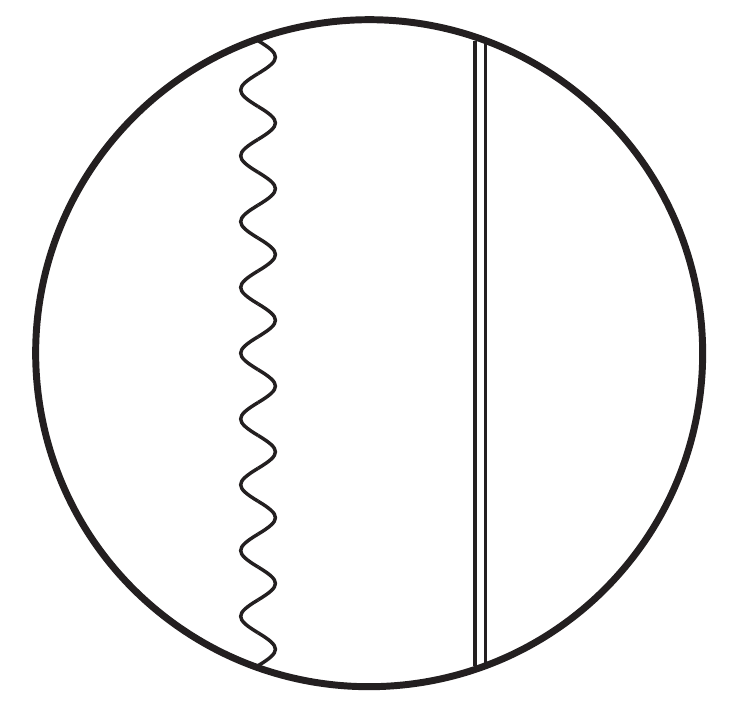}
  \caption{}
  \label{fig:3Ldiagramse}
  \end{subfigure} \quad
  \begin{subfigure}{2.1cm}
  \centering
  \includegraphics[width=2.cm]{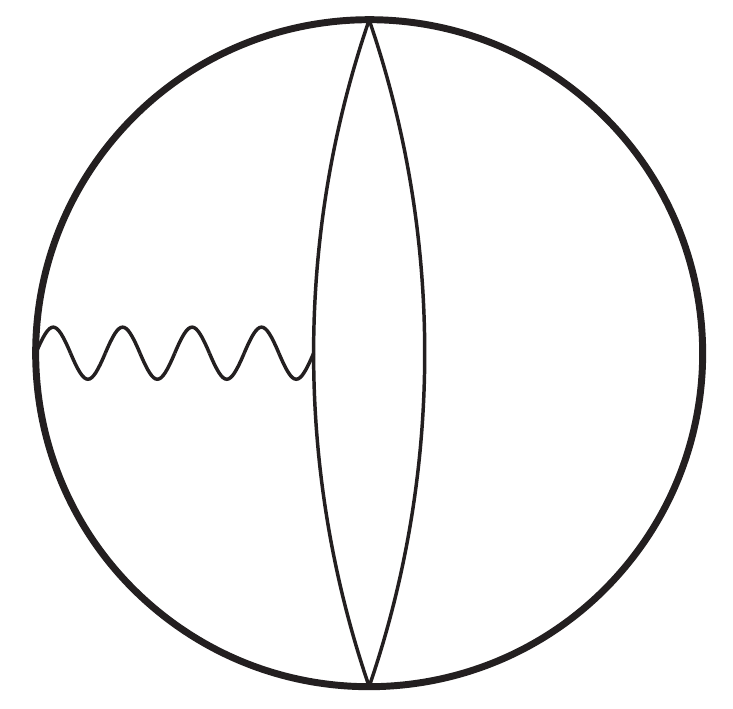}
  \caption{}
  \label{fig:3Ldiagramsf}
  \end{subfigure}  \quad
    \begin{subfigure}{2.1cm}
  \centering
  \includegraphics[width=2.cm]{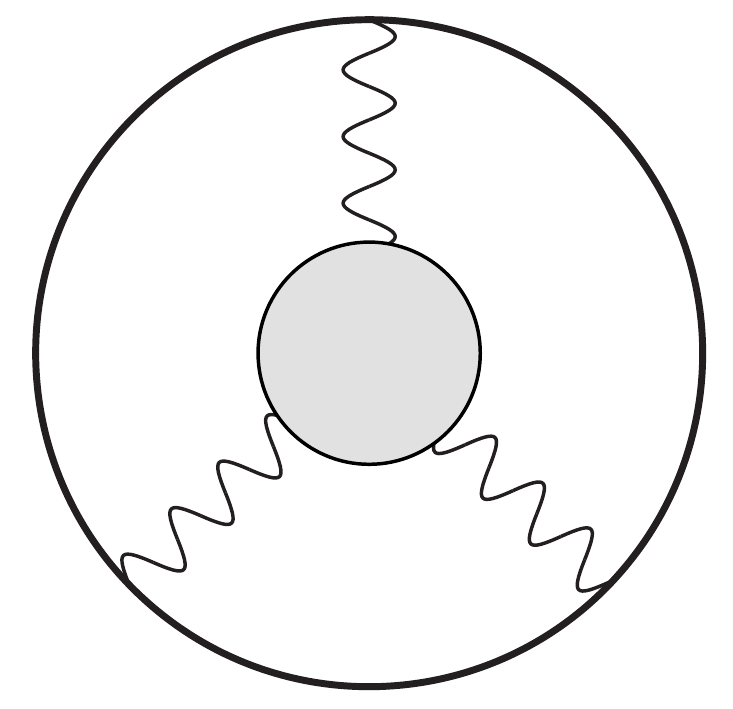}
  \caption{}
  \label{fig:3Ldiagramsg}
  \end{subfigure}  \quad
    \begin{subfigure}{2.1cm}
  \centering
  \includegraphics[width=2.cm]{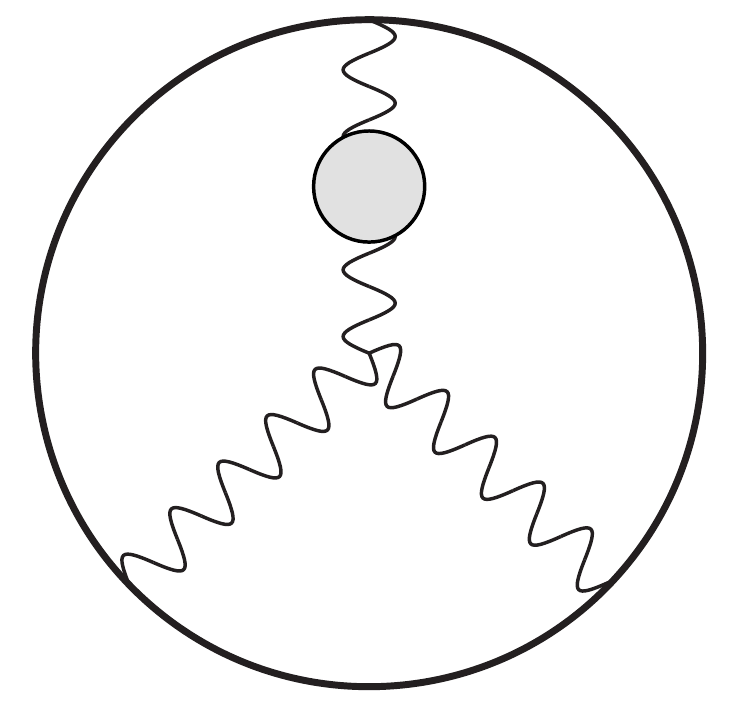}
  \caption{}
  \label{fig:3Ldiagramsh}
  \end{subfigure}  \quad
    \begin{subfigure}{2.1cm}
  \centering
  \includegraphics[width=2.cm]{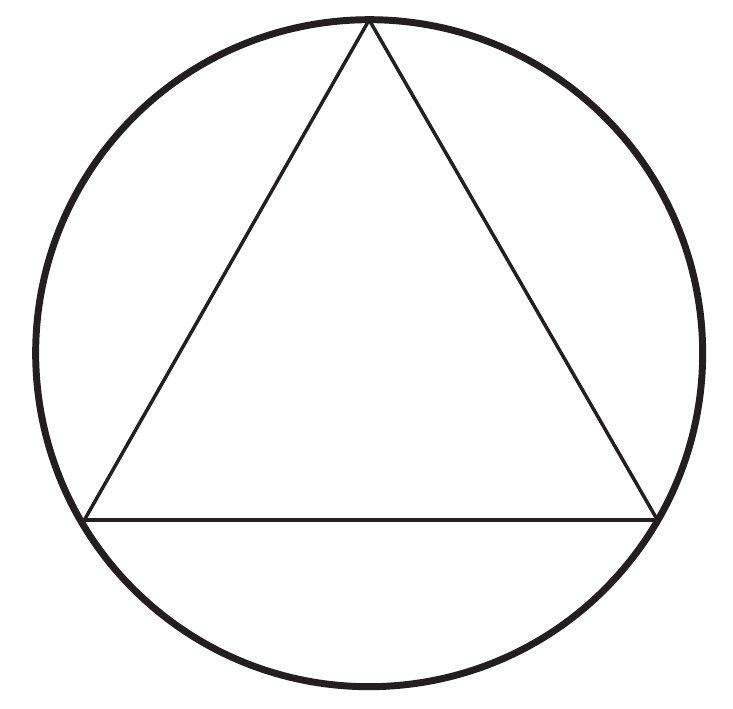}
  \caption{}
  \label{fig:3Ldiagramsi}
  \end{subfigure} 
\caption{Diagrams contributing to the bosonic latitude three--loop expectation value. Wavy and solid lines stand for gluons and scalars, respectively.}\label{fig:3Ldiagrams}
\end{figure}

The non--trivial Feynman diagrams contributing at three loops are depicted in Figure \ref{fig:3Ldiagrams}, where for diagrams with multiple insertions a sum over all planar configurations arising from permutations of contour points has to be understood. 

The triangle graph \ref{fig:3Ldiagramsi} is a new feature of the latitude Wilson loop stemming from the fact that even though $\Tr\, M$ vanishes, $\Tr\, M^3$ does not, thus allowing for a non--trivial contribution for $\nu \neq 1$ (see identity (\ref{eq:m3trace})).

Except for this graph, all the other diagrams in Figure \ref{fig:3Ldiagrams} are structurally the same contributing to the expectation value of $W_B(1)$, which were already analyzed in some detail in \cite{Bianchi:2016yzj}. We recall that all these diagrams vanish identically at framing zero \cite{Rey:2008bh}, because of the antisymmetry of the $\varepsilon$ tensors appearing ubiquitously in Chern--Simons perturbation theory, but give a non--vanishing contribution at framing $f \neq 0$. Other diagrams vanish identically even at non--trivial framing thanks to the tracelessness of the scalar coupling matrix, a property which is true in the undeformed case and remains true in the latitude case,  as well. These diagrams have not been included in Figure \ref{fig:3Ldiagrams}. Diagrams with one--loop corrections to the bi--scalar correlator have also been neglected,  as they have been argued to vanish identically \cite{Bianchi:2016yzj}, independently of framing.

In order to evaluate the diagrams in Figure \ref{fig:3Ldiagrams} we can exploit several partial results from \cite{Bianchi:2016yzj} to which we refer for more details on the computation. Those results are here generalized to include the latitude deformation, generic representations of the $U(N_1)$ gauge group, non--planar contributions and  generic framing. In particular, the latitude deformation does affect only diagrams which contain bi--scalar insertions, whereas all the others evaluate exactly as in the undeformed case.

Working with $N_1, N_2$ generically different, we can group the diagrams on the basis of their color structures. In particular we find convenient to classify them according to their leading power in $N_2$.

\vskip 5pt
\noindent 
{\bf Diagrams with no $N_2$ powers}. We start considering the subset of diagrams with no contributions from the $U(N_2)$ sector, namely pure $U(N_1)$ Chern--Simons contributions. These correspond to the class of diagrams \ref{fig:3Ldiagramsa}-\ref{fig:3Ldiagramsc}, with all possible planar permutations in the sequence of insertion points, according to the Alvarez--Labastida argument \cite{Alvarez:1991sx}. 
Having no coupling to the bi--scalar fields, these graphs do not depend on the latitude parameter $\nu$. Therefore they can be evaluated by observing that the combination of all permutations provide a factorization of the diagrams into elementary pieces involving the Gauss integral \eqref{eq:gauss}, which triggers the framing dependence. The result reads
\begin{align}\label{eq:3LCS}
& \raisebox{-0.8cm}{\includegraphics[width=2.cm]{3La5}} + \text{4 perms} = \frac{1}{6}\, i\, f\, \pi^3\, C_2(R) \left(C_1^2(R)-N_1 C_2(R)\right)\\ &
\raisebox{-0.8cm}{\includegraphics[width=2.cm]{3LA}} + \text{4 perms} = -\frac{1}{6}\, i\, f^3\, \pi ^3\, C_2^3(R)
\end{align}
where $C_1( R ), C_2( R)$ are defined in (\ref{eq:Rcasimir}). 

\vskip 5pt
\noindent 
{\bf Diagrams with $N_2^2$ powers}. A rather simple class of diagrams is the one with leading $N_2^2$ behavior, emerging from graphs \ref{fig:3Ldiagramsd} featuring gauge boson corrections at two loops \cite{Bianchi:2016yzj,Bianchi:2016rub}. Collecting the results of \cite{Bianchi:2016yzj} for the gauge two--point function insertion diagram, and extending them to the most general color structure (no new topologies arise for this case and the relevant color factors can be found in Appendix \ref{sec:two-point}) we find
\begin{equation}\label{eq:gauge2pt}
\raisebox{-0.8cm}{\includegraphics[width=2.cm]{3Lgauge}} = i\, \pi^3\, f\, \frac{N_2 \left[\left(8+5 \pi ^2\right) C_1^2(R)-\left(\left(8+\pi ^2\right) N_1+4 \pi ^2 N_2\right) C_2(R)\right]}{8 \pi ^2}
\end{equation}

\vskip 5pt
\noindent 
{\bf Diagrams linear in $N_2$}. The most complicated contribution to the three--loop expectation value comes from diagrams with leading linear $N_2$ behavior.
These include the factorized diagrams \ref{fig:3Ldiagramse}, the interaction diagrams \ref{fig:3Ldiagramsf}-\ref{fig:3Ldiagramsh} and the triangle graph \ref{fig:3Ldiagramsi}.

The most efficient way to handle factorized diagrams is to sum over all possible planar and non--planar configurations, recalling that only their contractible configurations contribute to framing, whereas the remaining ones vanish identically.
Referring to the diagram in Figure \ref{fig:3Ldiagramse}, this leads to the following factorization 
\begin{equation}\label{eq:factorization}
\raisebox{-0.8cm}{\includegraphics[width=2.cm]{mixed}} + {\rm perms} = \raisebox{-0.8cm}{\includegraphics[width=2.cm]{1Lgauge}} \times \raisebox{-0.8cm}{\includegraphics[width=2.cm]{2Lmatter}} = i\, \pi^3 f\,  \frac{1+\nu^2}{2} N_2\, C_2^2(R)  
\end{equation}
where $i \pi^3 f$ is the value of the corresponding integral, whereas the rest comes from color and combinatorics. A latitude dependent part arises from the double--line exchange \eqref{eq:combined}. The result can be correctly interpreted as emerging from the interference of the one--loop framing phase and the two--loop perturbative result from diagram \eqref{eq:matter}, reproducing the expected exponentiation of framing.

Interaction diagrams \ref{fig:3Ldiagramsf}-\ref{fig:3Ldiagramsh} are the most complicated and we do not possess an exact expression for each individual graph.
However, we can indirectly argue the value of their sum as follows.
In \cite{Bianchi:2016yzj} it was explained that for the 1/6--BPS Wilson loop $W_B(1)$ in the fundamental representation, consistency with the localization result requires the sum of \ref{fig:3Ldiagramsf}-\ref{fig:3Ldiagramsh} to cancel a suitable piece of the gauge two--point function contribution \eqref{eq:gauge2pt}.
As a first consistency check we have verified that the same reasoning holds also for a generic representation of the gauge group, as each interaction diagram has precisely the same color factor as the gauge two--point function contribution. The total sum reads
\begin{equation}
\raisebox{-0.8cm}{\includegraphics[width=2.cm]{scalar-gauge}} + \raisebox{-0.8cm}{\includegraphics[width=2.cm]{gauge-vertex-corr1}} + \raisebox{-0.8cm}{\includegraphics[width=2.cm]{gauge-vertex-corr2}} = \frac{1}{8} i \pi f \left(8+\pi ^2\right) N_2 \left(N_1 C_2(R)-C_1(R)^2\right)
\end{equation}
Turning on the latitude deformation seemingly spoils such an argument, since diagram \ref{fig:3Ldiagramsf} acquires a $\nu$--dependent factor from the trace of two $M$ matrices, equation \eqref{eq:m2trace}, that would sabotage the balance required for the aforementioned cancellation.
However, the $(\nu^2 - 1)$ term there, which would be absent in the undeformed case, is proportional to $\sin^2\frac{\tau_1-\tau_2}{2}$ and therefore vanishes for colliding insertion points. Consequently, it protects the integrand from developing the singularity that might cause a potential dependence on framing.
In fact,  the corresponding integral can be shown to be framing independent and when evaluated at framing zero it vanishes.
The remaining $\nu$--independent term in \eqref{eq:m2trace} obviously yields the same contribution as in the undeformed case.  
Altogether, the latitude deformation plays no role in the analysis of the interaction diagrams and we are led to postulate that the whole contribution in the latitude case is precisely the same as in the $\nu=1$ case, that is diagrams \ref{fig:3Ldiagramsf}-\ref{fig:3Ldiagramsh}  cancel the same part of the two--loop gauge propagator \eqref{eq:gauge2pt}.

Under this assumption, in the latitude case the only extra contribution is the triangle diagram with three bi--scalar insertions, Figure \ref{fig:3Ldiagramsi}. By direct inspection of the integrand, it is manifest that no framing dependence arises, as there are no singularities for coincident points. Therefore we can evaluate the diagram at framing zero 
and obtain
\begin{equation}\label{eq:triangle}
\raisebox{-0.8cm}{\includegraphics[width=2.cm]{triangle}} = -\frac{1}{6} i \pi ^3 \nu  \left(\nu ^2-1\right)\, N_2 \left(N_1 C_2(R)-C_1^2(R)\right)
\end{equation}
where we have used the trace of three $M$ matrices, equation \eqref{eq:m3trace}, and the integral
\begin{equation}\label{eq:triangleint}
\int_{0<\tau_3<\tau_2<\tau_1<2\pi} d\tau_1\,d\tau_2\, d\tau_3\, \frac{\sin \left(\tau _1-\tau _2\right)-\sin \left(\tau _1-\tau _3\right)+\sin \left(\tau _2-\tau _3\right)}{ \sin \frac{\tau _1-\tau _2}{2} \sin \frac{\tau _1-\tau _3}{2} \sin \frac{\tau _2-\tau _3}{2} } = \frac{16}{3}\, \pi^3
\end{equation}
can be evaluated immediately observing that the integrand is actually constant.
We remark that such a contribution is purely imaginary and therefore it mixes with the other imaginary three--loop corrections that are due to framing, though this part is framing independent. At this order this is the only imaginary contribution not arising from framing. In Section \ref{sec:modulus} we provide an additional interpretation of this term.

\vskip 10pt
\subsection{Multiple windings}\label{sec:winding}

Multiple windings introduce an overall $m$--dependent factor for each diagram, which only depends on the combinatorics of the insertions on the Wilson loops. Such factors can be computed recursively in an algorithmic manner as shown in \cite{Bianchi:2016gpg}. The strategy involves simplifying iteratively the integration contours until landing on integrals which can be immediately computed in terms of single winding ones. This translates into a system of recursion relations, which supplied with an initial condition, that is the value of the integrals at $m=1$, can be solved exactly. This procedure can be applied at any loop order and increases in complexity with the number of insertions on the Wilson contour. In some cases a computer implementation becomes necessary.

For one-- and two--loop diagrams we do not report the explicit computation, rather we state the final result in Section \ref{3loopresult}. Instead, we present some details of such a procedure for  the three--loop diagrams in Figure \ref{fig:3Ldiagrams}, in particular stressing what is new compared to the single winding case. 

The main difference that arises at multiple winding is the following. As stated in the previous section, for single winding only planar corrections contribute since the non--planar configurations, being non--contractible, vanish identically for the Alvarez--Labastida theorem.  
For multiple winding, instead, it is no longer guaranteed that non--contractible multiple winding diagrams do not contribute, since contractible single winding integrals can appear when resolving the multiply wound contours according to the procedure described above. Therefore, along the calculation we have to take into account all possible planar and non--planar configurations.

To better illustrate this point, we begin by considering the case of the factorized diagram in Figure \ref{fig:3Ldiagramse}. As in the single winding case, it is convenient to complete the sum of planar configurations with the corresponding non--planar ones in order to reduce the three--loop structure to the product of lower--order ones. This implies adding and subtracting the integrals corresponding to the crossed contributions multiplied by the same color factors of the planar ones, in such a way to symmetrize the integral. For the symmetrized part we then obtain (the double line contour stands for multiple winding)
\begin{align}\label{eq:fact}
& \phantom{+2} \raisebox{-0.8cm}{\includegraphics[width=2.cm]{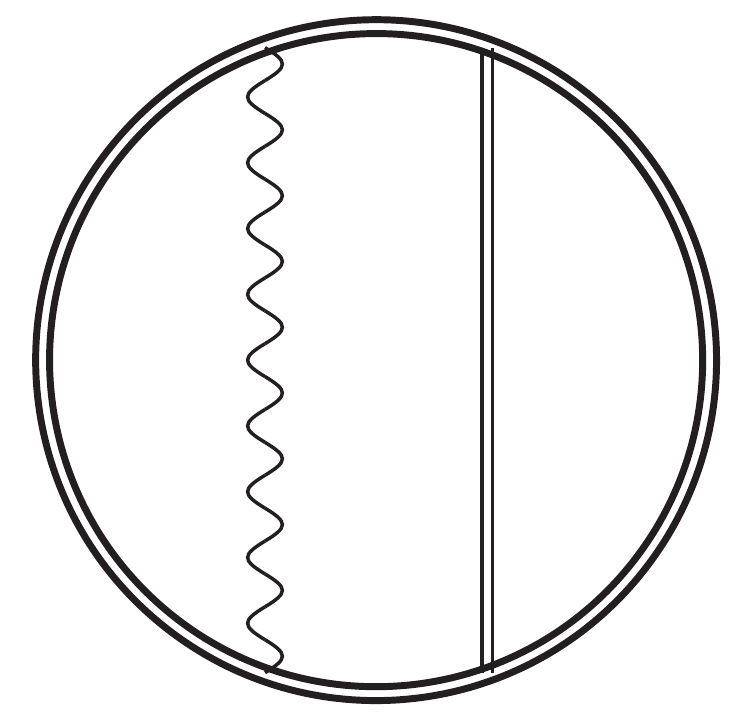}} + \raisebox{-0.8cm}{\includegraphics[width=2.cm]{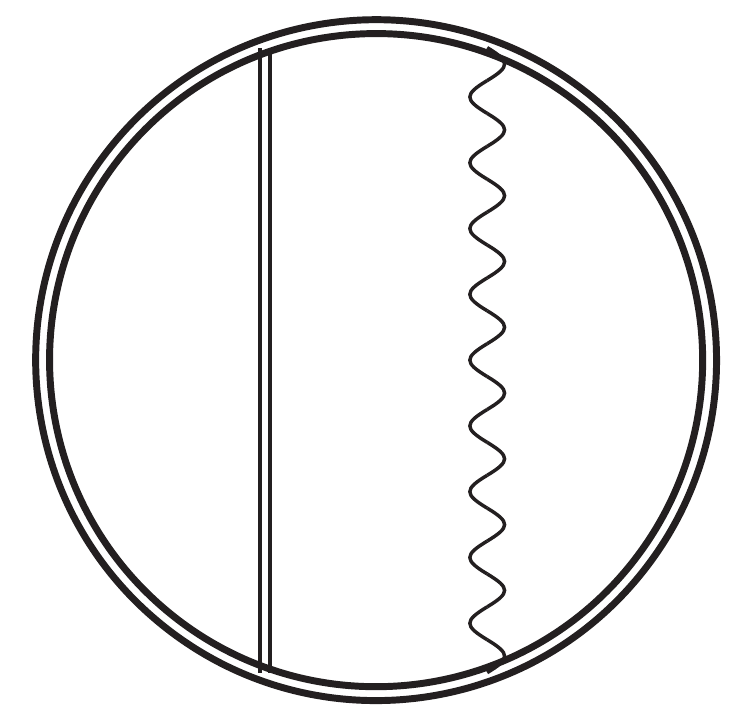}} + \raisebox{-0.8cm}{\includegraphics[width=2.cm]{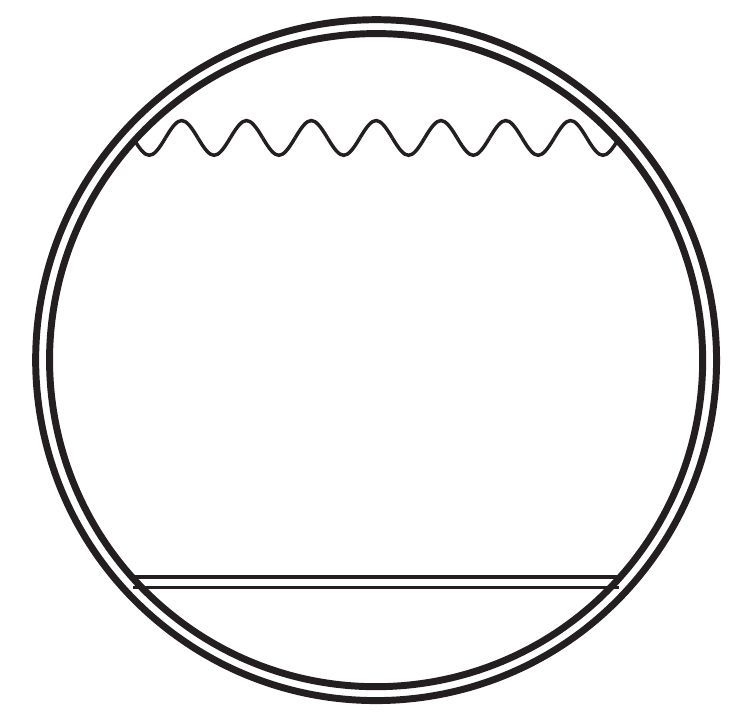}} + \raisebox{-0.8cm}{\includegraphics[width=2.cm]{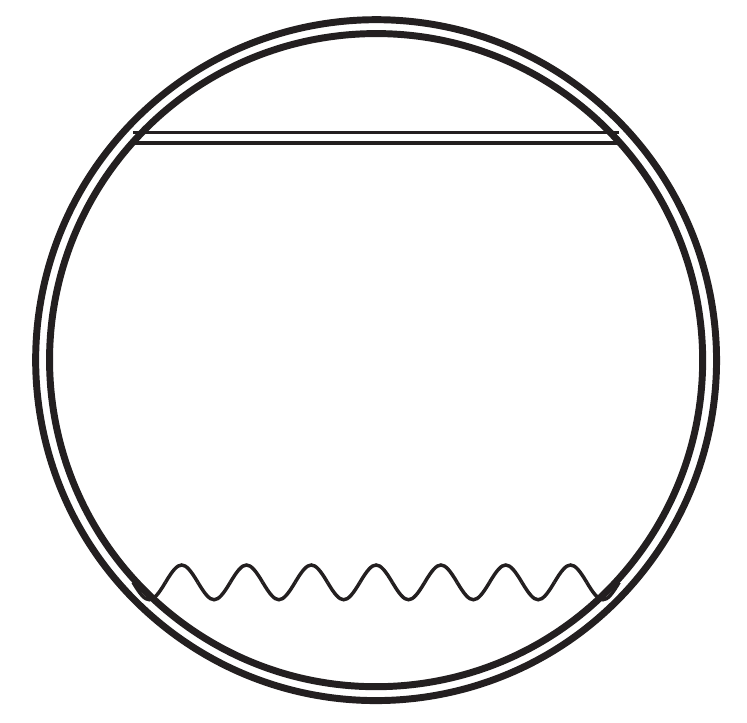}} +  \raisebox{-0.8cm}{\includegraphics[width=2.cm]{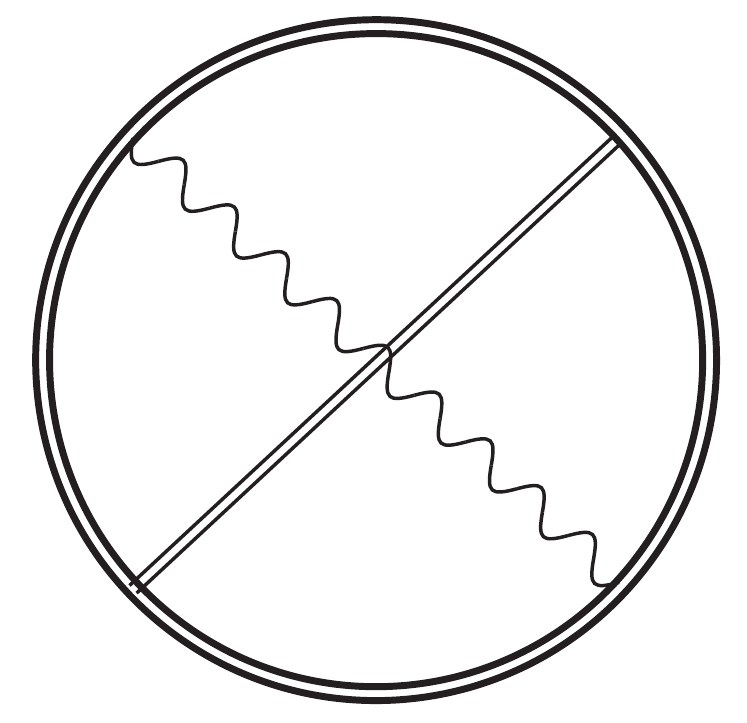}} + \raisebox{-0.8cm}{\includegraphics[width=2.cm]{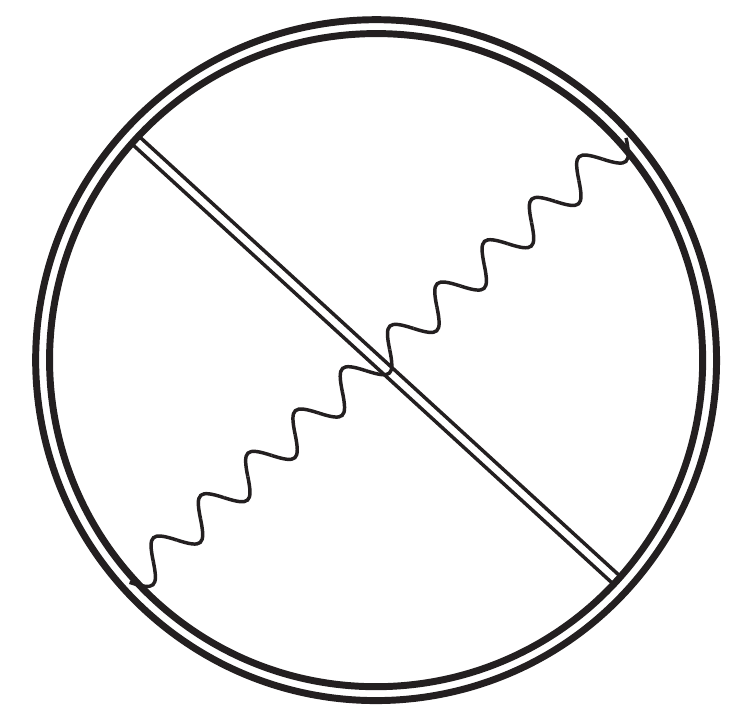}}  \nonumber\\&= \raisebox{-0.8cm}{\includegraphics[width=2.cm]{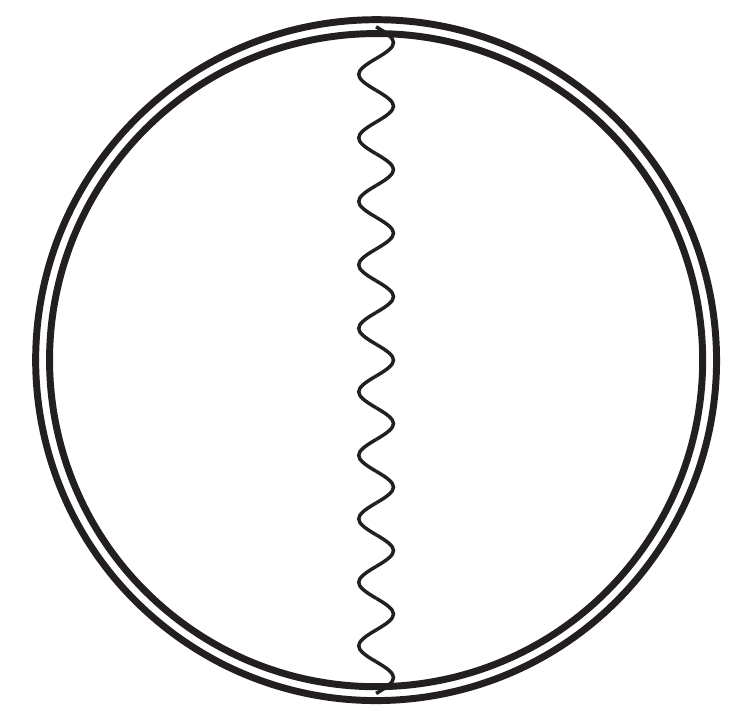}} \times \raisebox{-0.8cm}{\includegraphics[width=2.cm]{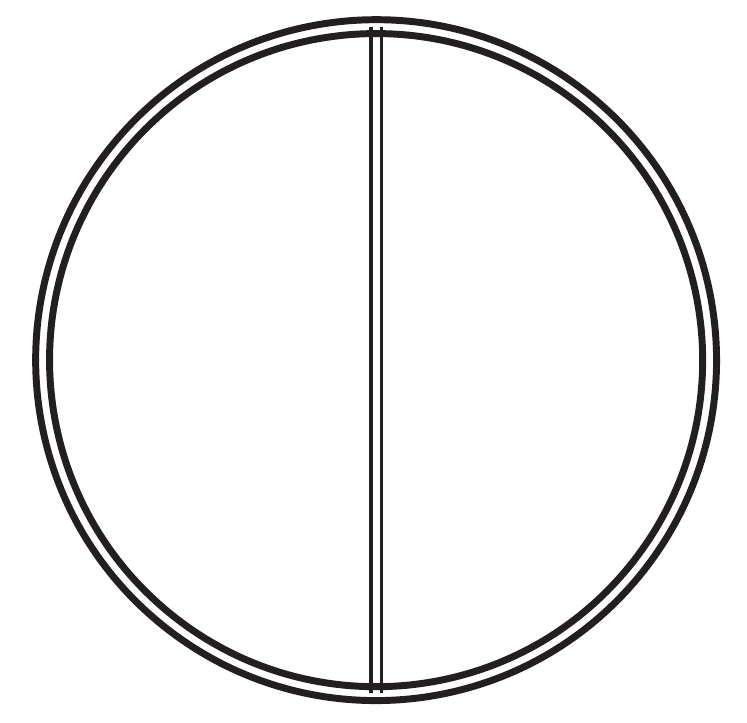}} = i\, \pi^3\, \left(m^4-\frac13 m^2(m^2-1)\right)\,f\, \frac{1+\nu^2}{2}\, N_2\, C_2^2(R)
\end{align} 
For the non--planar crossed contributions, using the algorithm of \cite{Bianchi:2016gpg}, a recursive relation yields pictorially
\bea \label{eq:crossed}
\raisebox{-0.8cm}{\includegraphics[width=2.cm]{mixed5m}} + \raisebox{-0.8cm}{\includegraphics[width=2.cm]{mixed6m}} &=& \frac13\, m^2 (m^2-1)\frac{1+\nu^2}{2}\ N_2\, \left[ C_1^2(R)-N_1 C_2(R) + C_2^2(R) \right] \non \\
& ~& \times \left( \raisebox{-0.8cm}{\includegraphics[width=2.cm]{mixed}} + \text{5 permutations} \right)
\eea
where the graphs on the right--hand--side represent integrals not diagrams, since the corresponding color and combinatorial factors have been already stripped out. We note that  the term  proportional to $C_2^2(R)$ is the relics of the symmetrization procedure in equation (\ref{eq:fact}).  Now, the combination of integrals in \eqref{eq:crossed} is exactly the one appearing in (\ref{eq:factorization}) and evaluates $i \pi^3 f$. Therefore, we find
\beq
 \raisebox{-0.8cm}{\includegraphics[width=2.cm]{mixed5m}} + \raisebox{-0.8cm}{\includegraphics[width=2.cm]{mixed6m}}  
=
i\, \frac{\pi^3}{3}\, m^2 (m^2-1)\, f\, \frac{1+\nu^2}{2}\, N_2\, \left[ C_1^2(R)-N_1 C_2(R) + C_2^2(R) \right]
\eeq
 
The evaluation of diagrams \ref{fig:3Ldiagramsd} and \ref{fig:3Ldiagramsf}-\ref{fig:3Ldiagramsi} for multiple winding is straightforward. In fact, diagrams with $n$ insertions of fields on the contour have in general a dependence on winding through a polynomial in $m^2$ of degree $\left\lfloor \frac{n}{2} \right\rfloor$. In particular, for diagrams with two and three insertions this boils down to a trivial $m^2$ factor. Consequently, at multiple winding the gauge two--point function diagram evaluates  
\begin{equation}\label{eq:gauge2ptm}
\raisebox{-0.8cm}{\includegraphics[width=2.cm]{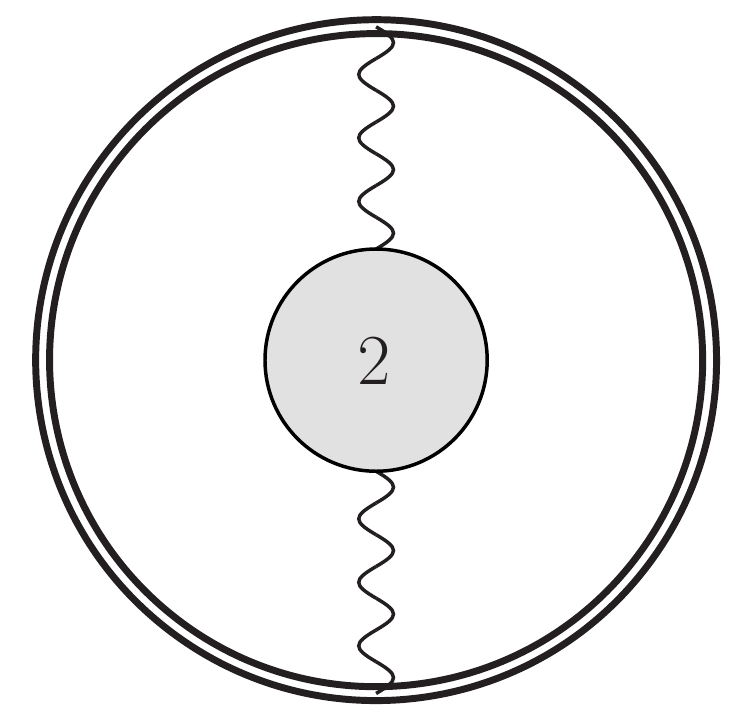}} = i\, \pi^3\, f\, \frac{m^2 N_2 \left[\left(8+5 \pi ^2\right) C_1^2(R)-\left(\left(8+\pi ^2\right) N_1+4 \pi ^2 N_2\right) C_2(R)\right]}{8 \pi ^2}
\end{equation}
The same occurs for the interaction diagrams of Figures \ref{fig:3Ldiagramsf}-\ref{fig:3Ldiagramsh} that simply acquire an overall $m^2$ factor compared to the single winding cousin.  This is important since it allows to conclude that the addition of winding does not jeopardize the argument for the cancellation of these interaction diagrams against part of \eqref{eq:gauge2ptm}.
Finally, the same $m^2$ overall factor arises for the triangle diagram as well.

Diagrams with five and six insertions of fields along the contour, see Figures \ref{fig:3Ldiagramsa}-\ref{fig:3Ldiagramsc}, are the most complicated. Their planar configurations were computed in \cite{Bianchi:2016gpg} and, extended to generic representations, they read
\begin{align}\label{eq:3LCSm}
& \raisebox{-0.8cm}{\includegraphics[width=2.cm]{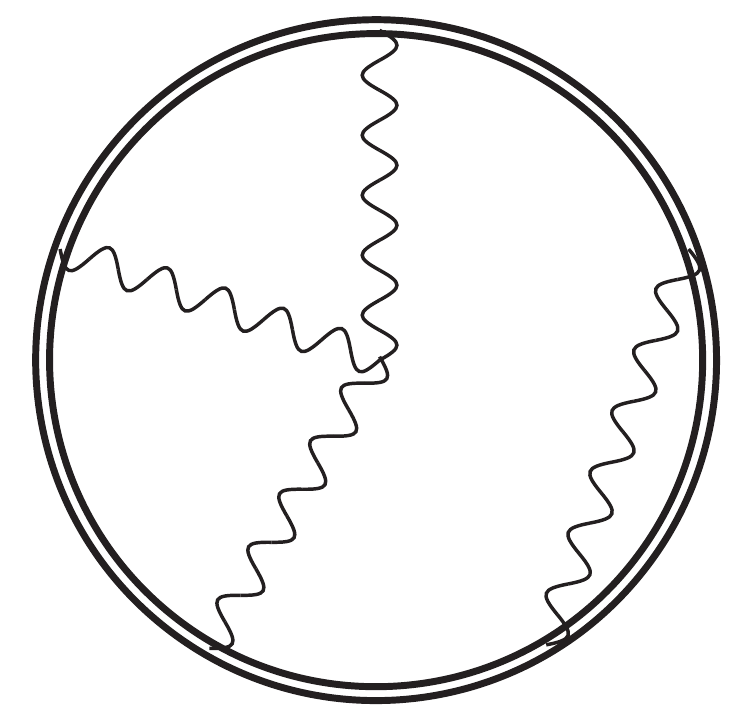}} + \text{4 perms} = \frac{1}{18} i \pi^3 f m^2 \left(2 m^2+1\right) C_2(R) \left(C_1^2(R)-N_1 C_2(R)\right)\\ &
\raisebox{-0.8cm}{\includegraphics[width=2.cm]{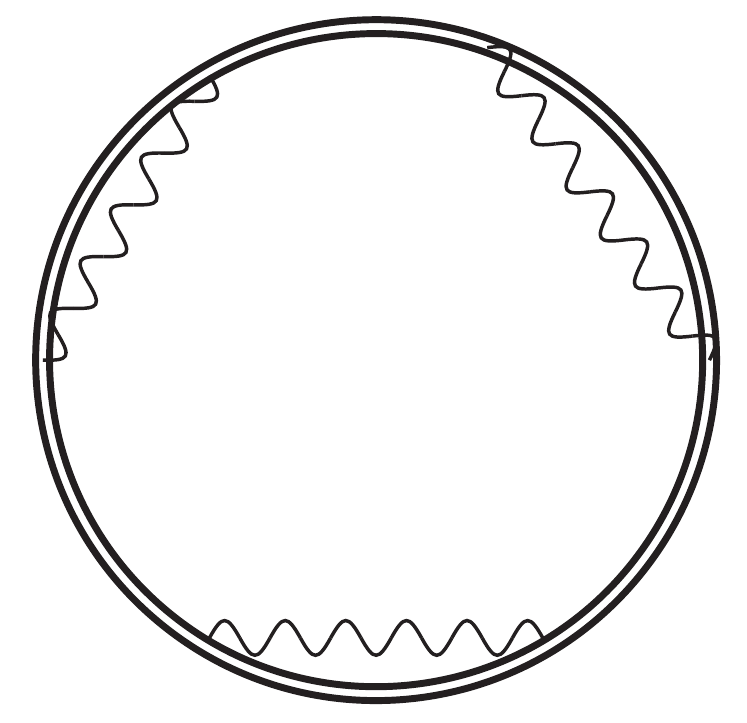}} + \text{4 perms} = -\frac{1}{18} i  \pi ^3 f^3 m^4 \left(2+m^2\right)\, C_2^3(R)\label{eq:3LCSP6}
\end{align}
where we have summed over all possible planar permutations with the same topology, and hence the same color factor.

The corresponding non--planar configurations give a non--trivial contribution since contractible planar configurations appear when decomposing the multiply wound contours.
For five insertions we obtain
\begin{equation}\label{eq:3LCSNP5}
\raisebox{-0.8cm}{\includegraphics[width=2.cm]{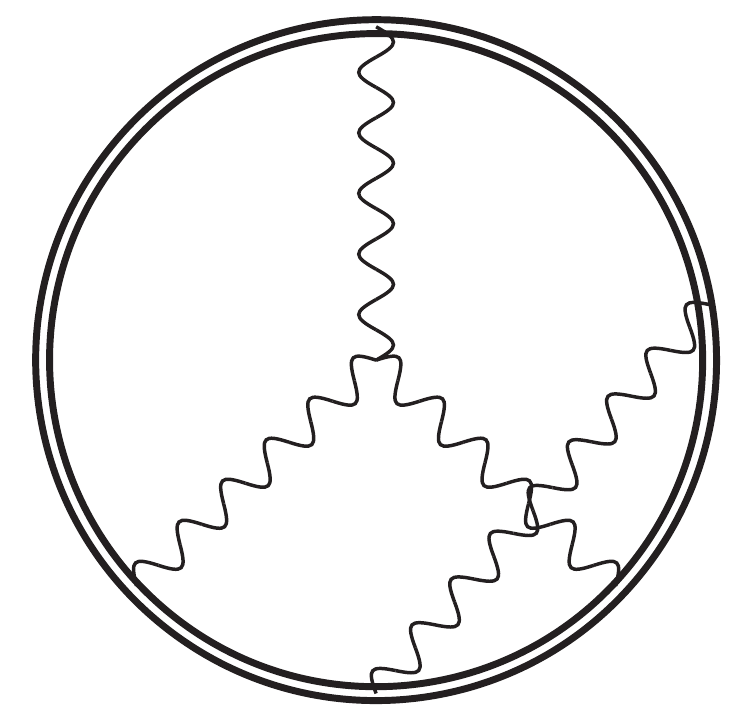}} + \text{4 perms} = \frac{1}{18} i \pi^3 f m^2 \left(m^2-1\right) \left(N_1-C_2(R)\right) \left(N_1 C_2(R)-C_1^2(R)\right)
\end{equation}
We note that in all the cases considered above the result of the recursive procedures organizes neatly in such a way that the singly wound integrals can be symmetrized and summed straightforwardly.
This is not the case for the non--planar contributions with six insertions, where the partial results for the individual topologies read
\begin{align}
& \raisebox{-0.8cm}{\includegraphics[width=2.cm]{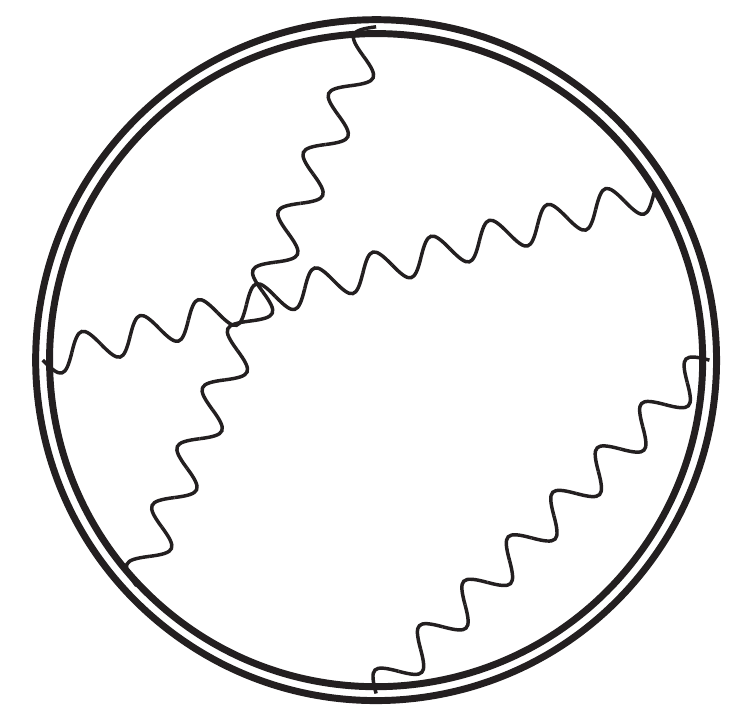}} + \text{5 perms} = \frac{2}{15} m^2 \left(m^2-1\right) C_2(R) \left(C_2(R) \left(C_2(R)-N_1\right)+C_1^2(R)\right) \label{eq:3LCSNP61}\\&
\left(3 (m^2+1) \left(\raisebox{-0.6cm}{\includegraphics[width=1.5cm]{3LA}}+\raisebox{-0.6cm}{\includegraphics[width=1.5cm]{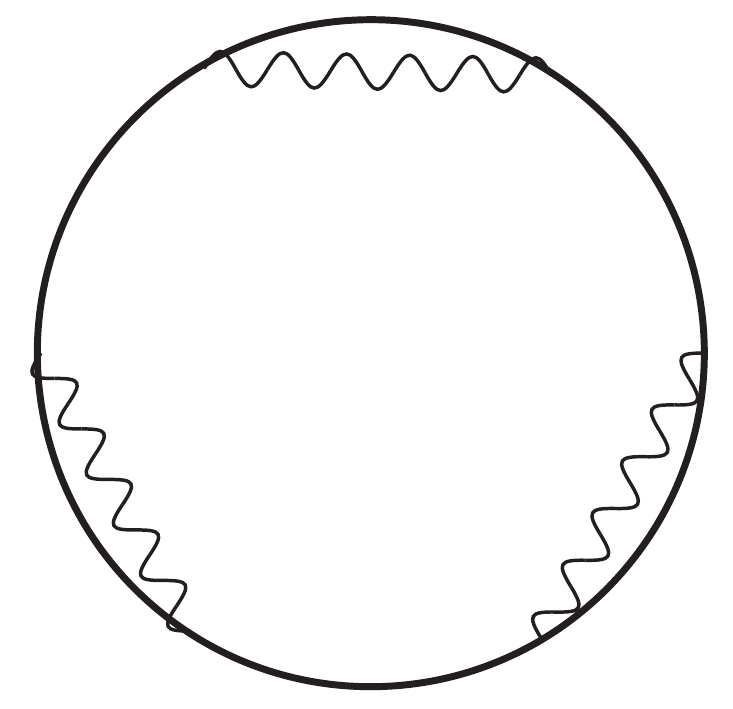}}\right)
+(3m^2-2)\left(\raisebox{-0.6cm}{\includegraphics[width=1.5cm]{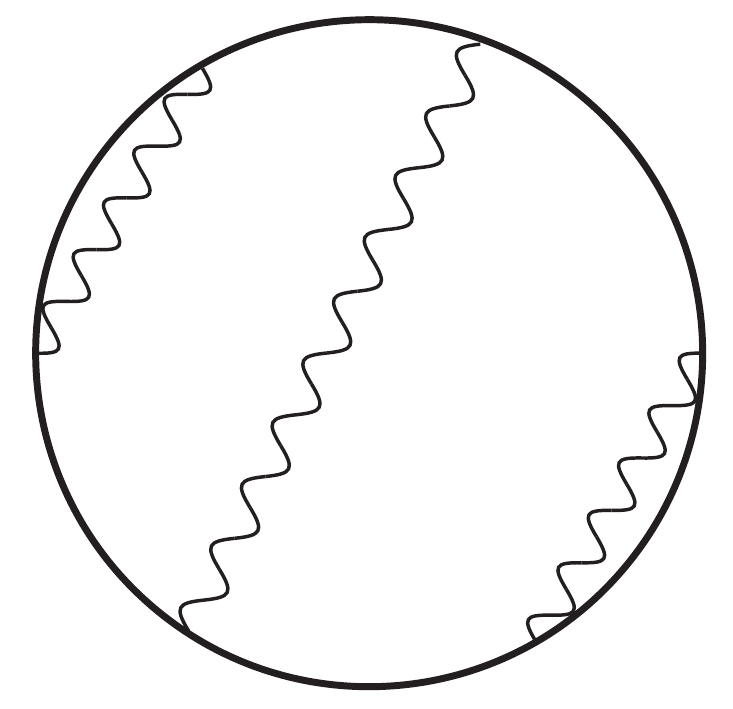}}+\raisebox{-0.6cm}{\includegraphics[width=1.5cm]{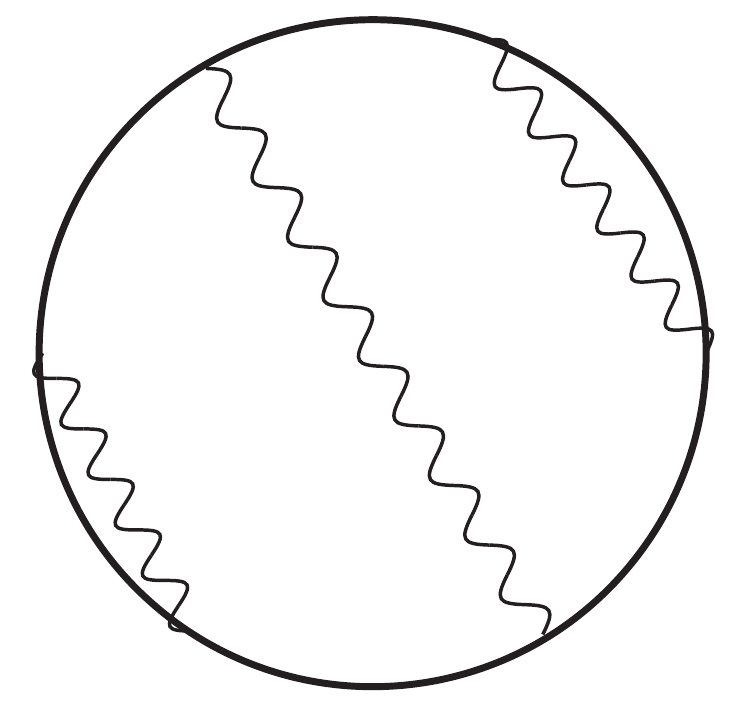}}+\raisebox{-0.6cm}{\includegraphics[width=1.5cm]{3LE}}\right)\right)\nonumber\\&
\raisebox{-0.8cm}{\includegraphics[width=2.cm]{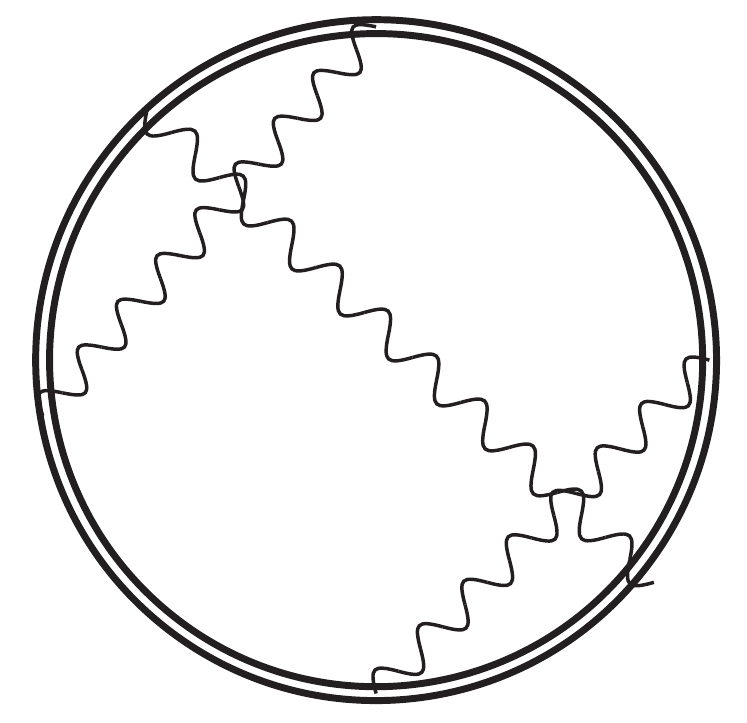}} + \text{2 perms} = \frac{1}{15} m^2 \left(m^2-1\right) \nonumber\\& ~~~~\times \left(N_1^2 C_2(R)-N_1 \left(C_1^2(R)+2 C_2^2(R)\right)+C_2^3(R)+2 C_1^2(R) C_2(R)\right) \label{eq:3LCSNP62}\\&
\left(3 (m^2-4) \left(\raisebox{-0.6cm}{\includegraphics[width=1.5cm]{3LA}}+\raisebox{-0.6cm}{\includegraphics[width=1.5cm]{3LD}}\right)
+(3m^2+8)\left(\raisebox{-0.6cm}{\includegraphics[width=1.5cm]{3LB}}+\raisebox{-0.6cm}{\includegraphics[width=1.5cm]{3LC}}+\raisebox{-0.6cm}{\includegraphics[width=1.5cm]{3LE}}\right)\right)\nonumber\\ & \label{eq:3LCSNP63} 
\raisebox{-0.8cm}{\includegraphics[width=2.cm]{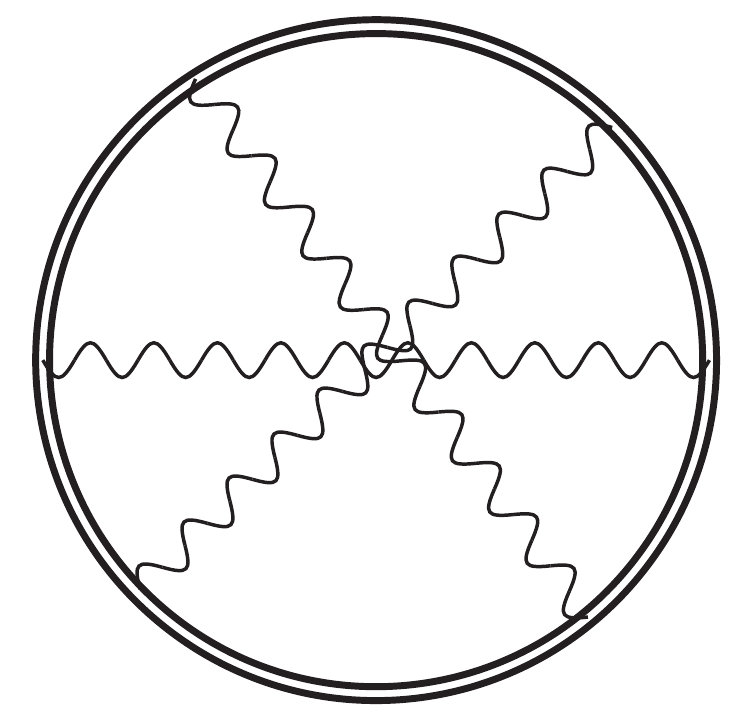}} = \frac{1}{15} m^2 \left(m^2-1\right) \nonumber\\& ~~~~\times \left(C_2(R) \left(3 C_1^2(R)+2 N_1^2\right)-3 N_1 C_2^2(R)-2 N_1 C_1^2(R)+C_2^3(R)\right)\\&
\left((m^2+6) \left(\raisebox{-0.6cm}{\includegraphics[width=1.5cm]{3LA}}+\raisebox{-0.6cm}{\includegraphics[width=1.5cm]{3LD}}\right)
+(m^2-4)\left(\raisebox{-0.6cm}{\includegraphics[width=1.5cm]{3LB}}+\raisebox{-0.6cm}{\includegraphics[width=1.5cm]{3LC}}+\raisebox{-0.6cm}{\includegraphics[width=1.5cm]{3LE}}\right)\right)\nonumber 
\end{align}
In all these formulae  the pictures on the right--hand--side stand for integrals over the contour and not diagrams, since the color factors have been already extracted.

Dealing with these integrals individually would be hard. However the non--planar sextic diagrams above combine in such a way that a symmetrized sum of integrals is reconstructed and simply evaluated. Amusingly, the final sum reads
\begin{align}
& \eqref{eq:3LCSP6}+\eqref{eq:3LCSNP61}+\eqref{eq:3LCSNP62}+\eqref{eq:3LCSNP63}=\frac{1}{3} m^4 \left(3 m^2 C_2^3(R)\right.\nonumber\\&~~~~~~~~~~~~~~~~~~~~~~~~~~\left.
+\left(m^2-1\right) \left( N_1 C_2(R) - C_1^2(R) \right) \left(N_1-3 C_2(R)\right)\right) \times \frac16\, \raisebox{-0.6cm}{\includegraphics[width=1.5cm]{1Lgaugem}} ^3
= \nonumber\\& ~~~~~~~~~=
-\frac{1}{18} i \pi^3 f^3 m^4 \left(3 m^2 C_2^3(R) +\left(m^2-1\right) \left( N_1 C_2(R) - C_1^2(R) \right) \left(N_1-3 C_2(R)\right)\right)
\end{align}

\vskip 10pt
\subsection{The general three--loop result}\label{3loopresult}

Summing all the diagrams computed in the previous sections we obtain the three--loop expectation value for the bosonic latitude Wilson loop with parameter $\nu$, framing $f$, winding number $m$ and for a generic representation $R$
\begin{align}\label{eq:Wlatgeneric}
& \langle W_B^{m}(\nu,R) \rangle_f = 1+\frac{i \pi  m^2 f  C_2(R)}{k} 
- \frac{\pi ^2 m^2}{6 k^2}\, \left(C_2(R) \left(3 f^2 m^2 C_2(R)
\right.\right.\nonumber\\&\left.\left.~~~~~~~~
+N_1 \left(f^2 \left(1-m^2\right)+1\right)-3 \left(\nu ^2+1\right) N_2\right)+C_1^2(R) \left(f^2 \left(m^2-1\right)-1\right)\right)
\nonumber\\&
+ \frac{i}{18\, k^3}\, \pi ^3\, m^2\, \left(
C_1(R)^2 \left(3 m^2 f C_2(R) \left(1-f^2 \left(m^2-1\right)\right)+f \left(m^2-1\right) N_1 \left(f^2 m^2-1\right)
\right.\right.\nonumber\\& \left.~~~~~~~~
+3 N_2 \left(f \left(\left(m^2-1\right) \nu ^2+m^2+2\right)+\nu ^3-\nu\right)\right)
\nonumber\\& ~~~~~~~~
+C_2\left(R\right) \left(-3 f m^2 C_2 \left(R\right) \left(f^2 m^2 C_2 \left(R\right)-f^2 \left(m^2-1\right) N_1-3 \left(\nu ^2+1\right) N_2+N_1\right)
\right.\nonumber\\&~~~~~~~~
-f^3 m^2 \left(m^2-1\right) N_1^2+f \left(\left(m^2-1\right) N_1^2 - 3 \left(m^2-1\right) \left(\nu ^2+1\right) N_1 N_2 - 9 N_2^2\right)
\nonumber\\&\left.\left.~~~~~~~
-3 \nu  \left(\nu ^2-1\right) N_1 N_2\right)
\right)
+O\left(k^{-4}\right)
\end{align}
where the Casimir invariants $C_1(R)$ and $C_2(R)$ for various representations are reported in \eqref{eq:casimir} and \eqref{eq:casimirh}.

As already mentioned, the multiple windings and higher dimensional representations are not independent generalizations, rather they provide two different alternative bases of operators. Using the general expression in \eqref{eq:Wlatgeneric}, we have verified explicitly that our result is in agreement with this expectation. In fact, by considering the first few windings of the Wilson loop in the fundamental representation it is easy to check that its expectation value can be obtained as a combination of single-winding operators in hook representations, according to the formula
\begin{equation}
\langle W_B^{m}(\nu,\Box) \rangle_f = \sum_{s=0}^{m-1} (-1)^s\, \bigg\langle W_B^{1}\bigg(\nu,\raisebox{-1.25cm}{\includegraphics[scale=0.2]{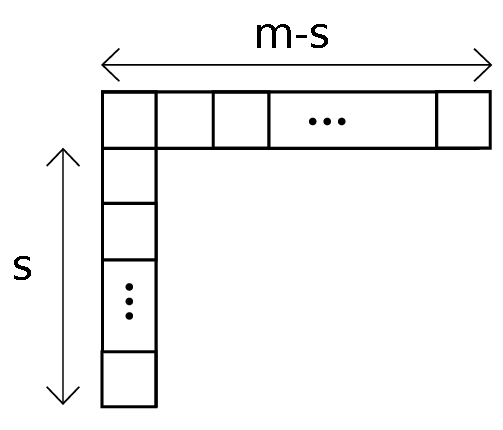}}\bigg) \bigg\rangle_f
\end{equation}
  It is important to stress that this holds for any generic framing number.

In the undeformed case ($\nu =1$) this provides a generalization of the three--loop result of \cite{Bianchi:2016yzj} to generic representations and with the inclusion of all non--planar contributions. In the case of totally symmetric and antisymmetric representations we have tested our result evaluated at $f=1$ against the weak coupling expansion of the matrix models \cite{Cookmeyer:2016dln}
(see also \cite{Hatsuda:2013yua})
\begin{align}\label{eq:higherrep}
\langle W_B^m(1) \rangle(k,S_{n}) &= \frac{1}{{\rm dim}(S_n)}\left\langle \sum_{1\leq i_1\leq\dots\leq i_n\leq N_1} e^{2 \pi \, m\, \left(\lambda_{i_1} + \dots + \lambda_{i_n}\right)} \right\rangle \non \\
\langle W_B^m(1) \rangle(k,A_{n}) &= \frac{1}{{\rm dim}(A_n)}\left\langle \sum_{1\leq i_1< \dots< i_{n}\leq N_1} e^{2\pi \, m\, \left(\lambda_{i_1} + \dots + \lambda_{i_n}\right)} \right\rangle
\end{align}
where the expectation values are defined in terms of the measure in \eqref{eq:matrixABJM}. In Appendix \ref{sec:MMexp} we supply the four--loop expansion of these matrix models  up to rank--3 representations. We find perfect agreement with the perturbative result \eqref{eq:Wlatgeneric} at $\nu=1$, thus providing a strong mutual check of the correctness of \eqref{eq:Wlatgeneric} and of the localization prediction.

Specifying result \eqref{eq:Wlatgeneric} to the fundamental representation of the gauge group $U(N_1)$ we obtain
\begin{align}\label{eq:Wlatm}
& \langle W_B^m(\nu) \rangle_f = 1+\frac{i \pi  f m^2 N_1}{k} \\&
-\frac{\pi ^2 m^2}{6 k^2} \left[ \left(f^2  (2m^2+1) +1 \right)N_1^2 -3 \left(\nu ^2+1\right) N_1 N_2  + f^2(m^2-1) - 1\right] \nonumber\\&
-\frac{i \pi ^3 m^2}{18 k^3} \left[
N_1^3 f \left(f^2 \left(m^2+2\right) m^2+2 m^2+1\right)
\right. \nonumber\\& ~~~~~~~
+N_1^2 N_2 \left(3 \nu  \left(\nu ^2-1\right)-3 f \left(2 m^2+1\right) \left(\nu ^2+1\right)\right)
+9 f N_1 N_2^2
\nonumber\\& \left.~~~~~~
+f N_1 \left(2 f^2 m^4-2 \left(f^2+1\right) m^2-1\right)
-3 N_2 \left(f \left(\left(m^2-1\right) \nu ^2+m^2+2\right)+\nu ^3-\nu \right)
\right]\nonumber\\&
+ O\left(k^{-4}\right)\nonumber
\end{align}
and for single winding it boils down to 
\begin{align}\label{eq:Wlat}
\langle W_B(\nu) \rangle_f &= 1+\frac{i \pi  f  N_1}{k}+\frac{\pi ^2 \left[-\left(3  f^2+1\right) N_1^2+3 \left(\nu ^2+1\right) N_1 N_2+1\right]}{6 k^2}
\nonumber\\&
-\frac{i \pi ^3}{6 k^3} \left[ f\left(f ^2+1\right) N_1^3 
+N_1^2 N_2 \left(\nu ^3-\nu-3 f \left(\nu ^2+1\right) \right)
\right.\nonumber\\&\left.~~~~~~~~
+3 f N_1 N_2^2 - f N_1
-N_2 \left(3 f+\nu ^3-\nu \right)
\right]+O\left(k^{-4}\right)
\end{align}

\vskip 5pt
\noindent
Finally in the ABJM theory ($N_1=N_2 \equiv N$) this reads 
\begin{align}
\langle W_B(\nu) \rangle_f &= 1+\frac{i \pi  f N}{k}+\frac{\pi ^2}{6 k^2} \left[N^2 \left(3 (\nu ^2-f^2)+2\right)+1\right] \\&
-\frac{i \pi ^3 N}{6 k^3} \left[N^2 \left(f^3+f \left(1-3 \nu ^2\right)+\left(\nu ^2-1\right) \nu \right)-4 f-\nu  \left(\nu ^2-1\right)\right]+O\left(k^{-4}\right)\nonumber
\end{align}
The crucial observation is that in the planar limit, setting $f=\nu$ this expression coincides with prediction \eqref{eq:prediction2}.
Moreover, it verifies relation \eqref{eq:identity} that was required for the consistency of expression \eqref{eq:exactB12} for the 1/2--BPS Bremsstrahlung function.
However, we stress  that in the more general case $N_1\neq N_2$ identity \eqref{eq:identity} is no longer valid, as one can easily check from \eqref{eq:Wlat}.  
Consequently, all its implications discussed in Section \ref{review} stop working. 

We conclude observing that the expectation value of the fermionic latitude can in principle be obtained from \eqref{eq:Wlat} by using the cohomological equivalence \eqref{eq:quantumcoho} conjectured in \cite{Bianchi:2016yzj}.
This applies only for the putative framing $f=\nu$, while a generic framing would require a full--fledged computation of the fermionic diagrams at three loops, which is currently not known.
At this special value of framing, using  \eqref{eq:quantumcoho} we obtain the following prediction (with single winding, for simplicity)
\begin{align}\label{eq:Wf}
\langle W_F(\nu) \rangle_\nu  
&= e^{\frac{i \pi  \nu  \left(N_1-N_2\right)}{k}} \left\{
1 + \frac{2 i \pi  \nu  N_1 N_2 {\cal R}}{k}\, \cos\frac{\pi \nu}{2}
\right.\nonumber\\&
- \frac{\pi^2 {\cal R}}{6k^2} \left[ (N_1-N_2) \left(N_1^2+N_2^2 -2 \left(3 \nu ^2+1\right) N_1 N_2-1 \right) \cos\frac{\pi \nu}{2}
\right.\nonumber\\&\left.~~~~~~~~
 - i (N_1+N_2) \left( N_1^2-4 N_2 N_1+N_2^2-1 \right) \sin\frac{\pi \nu}{2} \right]
\nonumber\\&\left. 
-\frac{i \pi ^3 \nu  N_1 N_2 {\cal R}}{3 k^3}  
 \left[\nu ^2 \left(N_1^2-3 N_2 N_1+N_2^2-1\right)-3\right] \cos\frac{\pi \nu}{2} 
+O\left(k^{-4}\right)\right\}  
\end{align}
where the normalization factor ${\cal R}$ has been defined in \eqref{eq:R}. The first two terms of this expansion reproduce the perturbative expansion of \cite{Bianchi:2014laa}\footnote{Note that there is a typo in formula (3.19) of that paper.}.

\subsection{The imaginary term at framing zero and the Bremsstrahlung functions}\label{sec:modulus}

Imaginary contributions to the expectation values of Wilson loops in ABJM are usually associated to framing. Hence, the appearance of an imaginary term at three loops in the expectation value of the latitude Wilson loop at framing zero is a bit surprising, but not inconceivable.
In fact, the operator in \eqref{eq:bosonic} that we are considering does not possess a definite hermiticity property, which would enforce the reality of its expectation value.

Still, its appearance poses a question concerning the relation between the latitude and the Bremsstrahlung function associated to the internal angle of the generalized cusp.
In the original proposal \cite{Bianchi:2014laa,Correa:2014aga}, $B_{1/6}^{\theta}$ was prescribed to be equivalent to
\begin{equation}\label{eq:Bold}
B_{1/6}^{\th} \underset{\text{naive}}{=} \frac{1}{4\pi^2}\, \pa_\nu \log \langle W_B(\nu) \rangle_0\,\, \Big|_{\nu=1}
\end{equation}
where the latitude Wilson loop at framing zero on the right--hand--side was understood to be real. According to our findings it is not. This would induce an imaginary contribution in the Bremsstrahlung function that cannot be there, as explicitly checked by the computation in \cite{Bianchi:2017afp,Bianchi:2017ujp}.

In order to resolve this tension we go back to the derivation of \eqref{eq:Bold} in \cite{Correa:2014aga} and point out where this subtlety kicks in.

Using definition \eqref{eq:Bbos}, the Bremsstrahlung function associated to $\theta$ is first determined by explicitly taking the derivatives with respect to $\theta$ of the generalized cusp Wilson loop. This is identified with the integral of two--point functions of operators constructed with the $C_{1}$, $C_2$ fields \cite{Correa:2014aga}
\begin{equation}\label{eq:twoptline}
\partial^2_\theta\, \log \langle W^{\angle}_B \rangle\,\Big|_{\theta=0}\!\! = \frac{8\pi^2}{k^2} \int_{0<t_2<t_1}\!\!\!\!\!\!\!\!\!\! dt_1\, dt_2\, \left( \langle\!\langle C_1 \bar C^2 (t_1) C_2 \bar C^1 (t_2) \rangle\!\rangle_{\rm line} + \langle\!\langle C_2 \bar C^1 (t_1) C_1 \bar C^2 (t_2) \rangle\!\rangle_{\rm line} \right)
\end{equation}
where the double bracket denotes the (normalized) correlation function of local operators at positions $t_1$ and $t_2$ on the 1/6--BPS Wilson straight line.
Such a non--local operator defines a one--dimensional superconformal defect and the first two--point function above is fixed by conformal symmetry to possess the form
\begin{equation}\label{eq:twopointlineconf}
\langle\!\langle C_1 \bar C^2 (t_1) C_2 \bar C^1 (t_2) \rangle\!\rangle_{\rm line} = \frac{\gamma }{(t_1-t_2)^2}
\end{equation}
where the coefficient $\gamma$ encapsulates the quantum corrections and is ultimately proportional to the Bremsstrahlung function.
The line configuration can be mapped to a circle via a conformal transformation, where the two--point function takes the form
\begin{equation}\label{eq:twopointcircleconf}
\langle\!\langle C_1 \bar C^2 (\tau_1) C_2 \bar C^1 (\tau_2) \rangle\!\rangle_{\rm circle} = \frac{\gamma }{2\left(1-\cos(\tau_1-\tau_2)\right)}
\end{equation}
in terms of the angles $\tau_1$ and $\tau_2$ on the circle. Conformal invariance of the theory assures that the $\g$ factors in \eqref{eq:twopointlineconf} and \eqref{eq:twopointcircleconf} are the same. 

The second correlation function in \eqref{eq:twoptline} is simply obtained from the first by exchanging $t_1$ and $t_2$ on the line or, equivalently, $\tau_1$ and $\tau_2$ on the circle. Since the operators obey bosonic statistics, we would be led to conclude that the two correlation functions are identical. However, expressions \eqref{eq:twopointlineconf} and \eqref{eq:twopointcircleconf} are correct only at non--coincident points, whereas close to the singularity at $t_1=t_2$ (or $\tau_1=\tau_2$) a suitable regularization is required (for instance by the addition of contact terms) that might introduce parity--odd corrections. Since in \eqref{eq:twoptline} we are integrating over $t_1, t_2$ this regularization can have sizable effects and, especially, lead to different results for the two integrals. Therefore, in the following we treat the two correlation functions as different objects.  

For the latitude Wilson loop at framing zero an identity analogous to \eqref{eq:twoptline} reads
\begin{align}\label{eq:twoptcircle}
\partial_\nu\, \log \langle W_B(\nu) \rangle_{0}\,\Big|_{\nu=1} &= -\frac{8\pi^2}{k^2} \int_0^{2\pi} d\tau_1 \int_0^{\tau_1} d\tau_2\, \left( e^{i(\tau_1-\tau_2)}\langle\!\langle C_1 \bar C^2 (\tau_1) C_2 \bar C^1 (\tau_2) \rangle\!\rangle_{\rm circle}\nonumber\right.\\&\left.~~~~~~~~~~~~~~~~~~~~~ + e^{-i(\tau_1-\tau_2)} \langle\!\langle C_2 \bar C^1 (\tau_1) C_1 \bar C^2 (\tau_2) \rangle\!\rangle_{\rm circle} \right)
\end{align}
where the exponentials in the integrand arise from the $e^{i\tau}$ factors in the latitude operator \eqref{eq:bosonic}. The correlation functions on the right--hand--side are on the 1/6--BPS circle, as we have set $\nu=1$ after taking the derivative.
We see that keeping the correlation functions in \eqref{eq:twoptcircle} distinct an imaginary part arises, proportional to the antisymmetric combination of the two.
For the purpose of computing the Bremsstrahlung function in terms of the latitude Wilson loop, we ascertain from \eqref{eq:twoptline} that only the symmetric combination of the correlation functions is relevant. This is equivalent to taking the real part in \eqref{eq:twoptcircle}, which on the left--hand--side amounts to enforcing the modulus of the latitude expectation value.
The precise coefficient of the relation between the Bremsstrahlung function and the latitude is computed by performing the integral in \eqref{eq:twoptcircle}, after plugging the generic form of the correlation function \eqref{eq:twopointcircleconf} on the circle and equating the parameters $\gamma$ appearing in \eqref{eq:twopointcircleconf} and \eqref{eq:twopointlineconf}. The steps are the same as in the original derivation of \cite{Correa:2014aga}. The final result reads
\begin{equation}
B_{1/6}^{\th} = \frac{1}{4\pi^2}\, \pa_\nu \log \left|\, \langle W_B(\nu) \rangle\, \right|\,\, \Big|_{\nu=1} 
\end{equation}
We can thus put such a prediction which was anticipated in \eqref{Btheta} on firmer grounds, from a first principles derivation\footnote{We acknowledge enlightening discussions with Lorenzo Bianchi for attaining this result (a similar discussion appears in \cite{satana}).}.
 
Finally, we can explicitly verify the emergence of an imaginary contribution at three loops in the latitude (which in that context comes from the triangle diagram \ref{fig:3Ldiagramsi}), as arising from the imaginary part of the right--hand--side of \eqref{eq:twoptcircle}, of the form
\begin{align}\label{eq:imint}
\partial_\nu\, \log \langle W_B(\nu) \rangle_{\nu}^{(3)}\,\Big|_{\nu=1} &= -i \, 8\pi^2 \int_0^{2\pi} d\tau_1 \int_0^{\tau_1} d\tau_2\, \sin(\tau_1-\tau_2) \\& \times \left(\langle\!\langle C_1 \bar C^2 (\tau_1) C_2 \bar C^1 (\tau_2) \rangle\!\rangle^{(1)}_{\rm circle} - \langle\!\langle C_2 \bar C^1 (\tau_1) C_1 \bar C^2 (\tau_2) \rangle\!\rangle^{(1)}_{\rm circle} \right)\nonumber
\end{align}
The only relevant diagram consists in the insertion of a bi--scalar field on the 1/6--BPS Wilson line. The matrix governing this contribution in the connection is $M=\text{diag}(-1,1,-1,1)={\cal M}\,\big|_{\nu=1}$.
Using the explicit expansion of the two--point functions, i.e. computing the whole triple integral
\begin{align}\label{eq:imint1}
& \partial_\nu\, \log \langle W_B(\nu) \rangle_{\nu}^{(3)}\,\Big|_{\nu=1} = i\, 16\pi^3 \int_0^{2\pi} d\tau_1 \int_0^{\tau_1} d\tau_2 \int_0^{\tau_2} d\tau_3\, \sin(\tau_1-\tau_2) \nonumber\\& ~~~~ \times \Tr_R \left(\langle C_1 \bar C^2 (\tau_1)\, C_2 \bar C^1 (\tau_2)\, M_I^J C_J \bar C^I (\tau_3) \rangle - \langle C_2 \bar C^1 (\tau_1)\, C_1 \bar C^2 (\tau_2)\, M_I^J C_J \bar C^I (\tau_3) \rangle \right)\nonumber\\& ~~~~ + \text{2 path ordered perms}
\end{align}
and performing exchanges in the integration variables, one can prove that this contribution precisely reconstructs the integrand of \eqref{eq:triangleint}. This automatically verifies the equality in \eqref{eq:twoptcircle} and explains the imaginary contribution of the latitude from the two--point function perspective. 

This result hints at the fact that the two--point functions in \eqref{eq:imint} are actually distinct quantum mechanically due to the necessity of regularizing them at coincident points, as already discussed.
In order to better clarify this point, we alternatively compute the one--loop two--point functions first and then plug them into \eqref{eq:imint}.
Calculating such a contribution on the circle is not straightforward, while it is immediate on the line. This computation reveals that indeed at one loop the integrands of the path--ordered correlation functions in \eqref{eq:imint} are opposite, due to R--symmetry index algebra and the properties of the matrix $M=\text{diag}(-1,1,-1,1)$. However, after integrating over the insertion of the bi--scalar field along the line, the result can be shown to vanish. On the one hand this finding is in line with the symmetry expectations on the two--point functions, on the other hand it is seemingly in contradiction with the non--vanishing result obtained above in \eqref{eq:imint1}. 
This puzzle is explained observing that the integral in \eqref{eq:imint} is actually divergent and therefore the insertion of a vanishing integrand should be handled with care.
One possibility consists in computing the one--loop two--point functions on the line in dimensional regularization
\begin{equation}\label{eq:2pt1l}
\langle\!\langle C_1 \bar C^2 (t_1) C_2 \bar C^1 (t_2) \rangle\!\rangle^{(1)}_{\rm line} = \frac{\left(N_1 C_2(R)-C_1^2(R)\right) N_2\,  \tan \pi  \epsilon\, \sec 2 \pi  \epsilon\, \Gamma^3 \left(\frac{1}{2}-\epsilon \right)}{16^{1+\epsilon} \pi ^{2-3 \epsilon} \, \Gamma (1-2 \epsilon ) \Gamma \left(2 \epsilon +\frac{1}{2}\right)\, \left((t_1-t_2)^2\right)^{1-3 \epsilon}}
\end{equation}
where consistently we obtain an order $\epsilon$ correction. The other correlation function yields the opposite result, as stated above.
We can plug this into \eqref{eq:imint}, though this requires performing a conformal map that is not justified in non--integer dimensions. Ignoring this objection and pushing the computation ahead we ascertain that the resulting integral in \eqref{eq:imint} develops a pole in the regulator $\epsilon$, thus exposing a finite contribution out of \eqref{eq:2pt1l}.
We ultimately find
\begin{align}
& -i\,8\pi^2 \int_0^{2\pi}\! d\tau_1 \int_0^{\tau_1}\! d\tau_2\, \sin(\tau_1-\tau_2) \left(\langle\!\langle C_1 \bar C^2 (\tau_1) C_2 \bar C^1 (\tau_2) \rangle\!\rangle^{(1)}_{\rm circle} \right.\nonumber\\&\left. - \langle\!\langle C_2 \bar C^1 (\tau_1) C_1 \bar C^2 (\tau_2) \rangle\!\rangle^{(1)}_{\rm circle} \right) = -\frac{i\pi^3}{3}\left(N_1 C_2(R)-C_1^2(R)\right) N_2
\end{align}
which precisely agrees with the derivative of the latitude expectation value at three loops and framing zero, that is basically the triangle diagram \eqref{eq:triangle}.
In conclusion, we have detected the interesting phenomenon that from the 1/6--BPS defect CFT perspective the triangle diagram in the latitude deformation arises from an anomalous behavior of the relevant two--point functions on the defect.

\section{The matrix model}\label{sec:matrixlat}

As reviewed in Section \ref{review},  a matrix model prescription for computing Wilson loops in ABJ(M) theory  exists for the $\nu=1$ case \cite{Kapustin:2009kz}, while it is still lacking for more general latitude operators. 
In this section we make a first attempt to fill this gap by proposing a matrix model to compute latitude Wilson loops. We then discuss some consistency checks to support our proposal.
 
The bosonic latitude Wilson loop is partially supersymmetric and its preserved supercharges are given in equation \eqref{chargferm} for $\omega_1=\omega_4=0$.
Consequently, and in parallel to the analogous situation in ${\cal N}=4$ SYM, it might be possible to compute its expectation value exactly using localization techniques. 
However, while in the four--dimensional case the latitude expectation value is obtained from the undeformed one by a simple rescaling of the coupling constant \cite{Drukker:2006ga, Drukker:2006zk,Drukker:2007dw,Drukker:2007yx,Drukker:2007qr,Giombi:2009ms,Pestun:2009nn,Giombi:2009ds,Giombi:2009ek}, this is no longer true for ABJ(M), as already appears from the perturbative results of the previous section. Therefore, we expect a significant modification of the matrix model to take place as a result of the localization process.
This program would involve localizing the ABJ(M) theory on the three--sphere using any of the supercharges in \eqref{chargferm} preserved by the bosonic latitude. In particular, since we expect the localization procedure to be consistent with the cohomological equivalence \eqref{cohom}, we should use the linear combination of supercharges \eqref{supercharge}. As already noticed in \cite{Bianchi:2014laa}, this supercharge is non--chiral and differs in nature from those considered in the original analysis of \cite{Kapustin:2009kz}. Therefore, the generalization of the localization procedure of \cite{Kapustin:2009kz} to the latitude Wilson loop is not straightforward. 

We are not going to pursue this direction here, rather we conjecture directly a matrix model that computes the latitude expectation value exactly, consistently with the perturbative results already available.

The idea is to start from the matrix model average \eqref{eq:matrixABJM} computing the expectation value of 1/6--BPS Wilson loops and try to deform it by introducing a suitable dependence on the $\nu$  parameter. As a route guidance we use the proposal \eqref{eq:bthetaconj} on the $\theta$--Bremsstrahlung function, which in turn requires the $\nu$--derivative of the matrix model to satisfy identity \eqref{eq:relationbremsstrahlung}. 

A natural way to satisfy this condition consists in requiring that the latitude Wilson loop is computed by inserting the operator $\Tr\, e^{2\pi \sqrt{\nu} \lambda}$ into a matrix model which is symmetric under the inversion $\nu\leftrightarrow 1/\nu$. In fact, in this way taking the derivative with respect to $\nu$ evaluated at $\nu= 1$, the $\nu$ dependence of the matrix model measure plays no role and the only non--trivial contribution comes from the derivative acting on the operator insertion. The result is a matrix model where the integrand, being evaluated at $\nu=1$, corresponds to the well--known ABJ(M) matrix model, except for the operator insertion $\pa_\nu \Tr\, e^{2\pi \sqrt{\nu} \lambda} |_{\nu=1}$ which can be formally identified with the $m$--derivative of a multiply wound 1/6--BPS Wilson loop evaluated at $m=1$, with $m \equiv \sqrt{\nu}$. In particular, trading $\pa_\nu$ with $\pa_m$ provides the correct 1/2 factor appearing in \eqref{eq:relationbremsstrahlung}.

Such an argument is reminiscent of the one proposed in \cite{Lewkowycz:2013laa}, but somehow with a reverse logic. 
In that case, for the ABJM theory a supersymmetric Wilson loop on a squashed sphere was considered, whose matrix model is invariant under the inversion of the squashing parameter \cite{Hama:2011ea}. This was used to argue that the derivative of this Wilson loop expectation value with respect to the squashing parameter $b$, evaluated at $b= 1$, could be traded with the derivative of the multiply wound 1/6--BPS Wilson loop with respect to the winding number $m$.  

Driven by this discussion we are led to propose the following matrix model average for the expectation value of a multiply wound latitude Wilson loop
\begin{align} \label{eq:matrixlat}
\langle W_B^m(\nu) \rangle  &= \left\langle \frac{1}{N_1} \sum_{1\leq i\leq N_1} e^{2\pi\, m\, \sqrt{\nu}\, \lambda_{i}} \right\rangle
\end{align}
where the average is evaluated and normalized using the matrix model partition function
\begin{align} \label{eq:partitionf}
&  Z = \int \prod_{a=1}^{N_1}d\lambda _{a} \ e^{i\pi k\lambda_{a}^{2}}\prod_{b=1}^{N_2}d\mu_{b} \ e^{-i\pi k\mu_{b}^{2}}  \\
&~~ \times 
 \frac{\displaystyle\prod_{a<b}^{N_1}\sinh \sqrt{\nu} \pi (\lambda_{a}-\lambda _{b})\sinh\frac{ \pi (\lambda_{a}-\lambda _{b})}{\sqrt{\nu}} \prod_{a<b}^{N_2}\sinh\sqrt{\nu} \pi (\mu_{a}-\mu_{b}) \sinh\frac{\pi (\mu_{a}-\mu_{b})}{\sqrt{\nu}} }{\displaystyle\prod_{a=1}^{N_1}\prod_{b=1}^{N_2}\cosh\sqrt{\nu} \pi (\lambda _{a}-\mu_{b}) \cosh\frac{\pi (\lambda _{a}-\mu_{b})}{\sqrt{\nu}} } \nonumber
\end{align}
Similarly, $\langle \hat W_B^m(\nu) \rangle$ corresponds to the insertion of $\frac{1}{N_2} \sum_{1\leq i\leq N_2} e^{2\pi\, m\, \sqrt{\nu}\, \mu_{i}} $. According to the discussion above, such a matrix model should arise from a suitable localization of the ABJ(M) theory.

This is the simplest non--trivial deformation of the matrix model \eqref{eq:matrixABJM} that lands back on the usual expression at $\nu=1$, and whose kernel is symmetric under $\nu\leftrightarrow 1/\nu$. 
The precise dependence on $\nu$ in the hyperbolic functions and in the operator insertion is then fixed via comparison with the perturbative results.
Indeed, we can evaluate this expression by expanding it at weak coupling. The main result of this analysis is that  we recover precisely the expectation value \eqref{eq:Wlatm} for the multiply wound latitude in the fundamental representation evaluated at framing $\nu$.
We stress that the agreement with the perturbative result holds separately for all different color structures at generically unequal and finite $N_1$ and $N_2$ and for generic winding number $m$. Moreover, in the $N_1\gg N_2$ approximation, the matrix model reconstructs the pure Chern--Simons result at framing $\nu$, and in the specular $N_2\gg N_1$ limit it reproduces the expected behavior of \cite{Bianchi:2016rub}.

 In addition, we have considered the extension of the matrix model average defined in \eqref{eq:matrixlat} to higher dimensional representations.
We have explicitly ascertained that applying the prescriptions \eqref{eq:higherrep} (with an extra $\sqrt{\nu}$ factor in the exponents) for the first few totally symmetric and antisymmetric representations, we do reproduce the corresponding perturbative results \eqref{eq:Wlatgeneric} for bosonic latitude Wilson loops.
This lead us to conjecture, that the matrix model \eqref{eq:matrixlat} computes also the expectation value of latitude Wilson loops in higher dimensional representations, upon applying the same prescriptions \eqref{eq:higherrep}, as in the undeformed case.

It is remarkable that the agreement works only at the specific choice of framing $f=\nu$. On the one hand this does not come as a surprise. In fact this occurs already for the 1/6--BPS Wilson loop where localization implies non--trivial framing $f=1$, which is the scheme compatible with the cohomological equivalence between bosonic and fermionic Wilson loops. For the latitude Wilson loop it is then highly suggestive that the agreement manifests at $f=\nu$, since this is precisely the value at which the conjectured cohomological equivalence with the fermionic Wilson loop holds (see equation  \eqref{eq:cohom1}). Since we expect a putative matrix model average to be able to compute both the bosonic and the fermionic operators, it is then natural that the matrix model indeed provides the result at framing $\nu$.

Although it might sound a bit weird to consider non--integer framing, we note that at the level of the matrix model continuing the framing from an integer value to a generically real number is perfectly legitimate and is also a common occurrence (despite usually only for rational values), for instance when computing torus knot invariants in Chern--Simons theory \cite{Brini:2011wi}.

 We can finally draw a parallel with the four--dimensional ${\cal N}=4$ SYM theory.
In that case, it is easy to realize that applying an analogous deformation procedure on the Gaussian matrix model computing the expectation value of Wilson loops in that theory, we reproduce the latitude operators. In fact, the $\nu$ dependent deformations in the matrix model measure cancel out completely and one is left with a modified operator insertion exhibiting an additional $\nu$ factor at the exponent. This eventually provides the coupling constant rescaling that characterizes the expectation value of latitude Wilson loops in ${\cal N}=4$ SYM.

\subsection{Properties and relations with other matrix models}

Supported by this first evidence on the correctness of our proposal, we devote the rest of this section to a discussion of its main properties and its possible interpretation. 

First, we note that a striking similarity exists between expression \eqref{eq:partitionf} and the kind of matrix model emerging as the result of the so--called symplectic or $SL(2,\mathbb{Z})$ transformation of the sphere partition function of pure Chern--Simons theory \cite{Brini:2011wi}. This depends on two coprime integer parameters $(P,Q)$ and has been argued to compute torus knot invariants \cite{Brini:2011wi} (see also \cite{Labastida:1990bt,Eynard:2007kz,Bouchard:2007ys} for more references on torus knot invariants), using the celebrated relation between knots and Chern--Simons theory \cite{Witten:1988hf}. 
In our case, the symplectic transformation is rather performed on the supermatrix model (or equivalently on the lens space $S^3/\mathbb{Z}_2$ partition function \cite{Stevan:2013tha,Lawrence1999}) and the result reads
\begin{align}
\label{eq:matrixPQ}
& Z^{(P,Q)}_{ABJ(M)} = \int \prod_{a=1}^{N_1}d\lambda _{a} \ e^{\frac{i\pi k}{PQ}\lambda_{a}^{2}}\prod_{b=1}^{N_2}d\mu_{b} \ e^{-\frac{i\pi k}{PQ}\mu_{b}^{2}}\, \nonumber\\& ~~ \times
 \frac{\displaystyle\prod_{a<b}^{N_1}\sinh \frac{\pi (\lambda_{a}-\lambda _{b})}{P} \sinh \frac{\pi (\lambda_{a}-\lambda _{b})}{Q} \,\prod_{a<b}^{N_2}\sinh \frac{\pi (\mu_{a}-\mu_{b})}{P} \sinh \frac{\pi (\mu_{a}-\mu_{b})}{Q} }{\displaystyle\prod_{a=1}^{N_1}\prod_{b=1}^{N_2}\cosh \frac{\pi (\lambda _{a}-\mu_{b})}{P} \cosh \frac{\pi (\lambda _{a}-\mu_{b})}{Q} }
\end{align}
Consequently, the Wilson loop averages computed with this matrix model  
\begin{equation}
\langle W_R^{(P,Q)} \rangle = \frac{\langle s_{p_R}(e^{2\pi \lambda_a},e^{2\pi \mu_b}) \rangle_{Z^{(P,Q)}_{ABJ(M)}}}{Z^{(P,Q)}_{ABJ(M)}}
\end{equation}
where $s_{p_R}$ is the supersymmetric Schur function for the partition $p_R$ associated to the representation $R$, should arguably yield torus knot invariants of $U(N_1|N_2)$ Chern--Simons theory \cite{Eynard:2014rba} (they are a generalization of colored HOMFLY polynomials \cite{freyd1985} to $U(N_1|N_2)$/lens space).
This interpretation in terms of knot invariants works only for coprime $P,Q$  integers, this condition ensuring that the contour closes on the two--torus.
The case we are considering is a (perfectly sensible) generalization of the torus knot matrix model to non--coprime integer values of the parameters, being $P=\sqrt{\nu}$ and $Q=1/\sqrt{\nu}$. Hence, in general the latitude Wilson loop is not really computing knot invariants, as the corresponding contour does not close, rather it wraps the two--torus densely. An exception arises in the degenerate situations in which the  torus reduces to one of the two cycles of length $2\pi P$ or $2\pi Q$, respectively. In these two cases  the factor in the Wilson loop operator of \eqref{eq:matrixlat} is precisely of the form required for the closure of the contour (with $P=\sqrt{\nu}$ being the length of the circle).

The matrix model in \eqref{eq:partitionf} can also be obtained by localizing the ${\cal N}=2\, $ $U(N_1+N_2)$ Chern--Simons theory on the squashed lens space $S^3_{\sqrt{\nu}}/\mathbb{Z}_2$ with squashing parameter $\sqrt{\nu}$ \cite{Hama:2011ea,Benini:2011nc,Imamura:2012rq}. Selecting the particular vacuum that breaks the gauge group to $U(N_1)\times U(N_2)$, and then continuing it to the supermatrix version as done for the ABJ(M) matrix model \cite{Marino:2002fk,Aganagic:2002wv,Brini:2008ik}, we land on \eqref{eq:partitionf}. In fact, considering Chern--Simons theory with supergroup $U(N_1|N_2)$ is (at least perturbatively) equivalent to the lens space interpretation by rewriting its matrix model in the two--cut form and taking the analytic continuation $N_2\to -N_2$. 

After integrating out the ${\cal N}=2$ auxiliary fields one expects to recover the pure Chern--Simons observables and hence possibly compute knot invariants. In \cite{Tanaka:2012nr} this procedure was indeed applied to ${\cal N}=2$ Chern--Simons theory on a squashed sphere, which was observed to yield torus knot invariants of pure Chern--Simons on the three--sphere. 
Our matrix model is formally a particular case of ${\cal N}=2$ Chern--Simons on a squashed lens space. The analysis of \cite{Tanaka:2012nr} then suggests that this matrix model should compute torus knot invariants on the lens space $\mathbb{RP}^3$.
This is indeed consistent with the identification of the latitude matrix model and that for torus knot invariants, as described above.

{\it In cauda venenum}, we conclude with few critical arguments on our proposal.
In particular we discuss whether the matrix model \eqref{eq:matrixlat} can be interpreted as the result of localizing the ABJ(M) theory on $S^3$, with the insertion of a $\nu$--dependent operator corresponding to the latitude Wilson loop. 

In principle, if this were the case one might expect the matrix model average to be computed with the ordinary $\nu$--independent measure appearing in the ABJ(M) partition function. Instead, in our proposal \eqref{eq:matrixlat} the kernel depends explicitly on $\nu$. However, as already stated, if we require the localization procedure to be compatible with the cohomological equivalence, the path integral should be localized with the supercharge \eqref{supercharge}. Since the latter exhibits an explicit dependence on $\nu$ it might reasonably lead to the conjectured  $\nu$--dependence of the matrix model in \eqref{eq:partitionf}. 

If this interpretation is correct, then \eqref{eq:partitionf} itself should, after integration, give rise to the usual, $\nu$--independent, result for the ABJ(M) partition function. 
In fact, in the $N_1=N_2$ case, expression \eqref{eq:partitionf} can be rearranged in such a way that the $\nu$--dependence disappears completely and it ends up coinciding with the ABJM partition function on $S^3$. More details on this computation can be found in the next subsection. 

Instead, for the most general $N_1\neq N_2$ case a non--trivial  $\nu$ dependence survives in the phase of the partition function (see equation \eqref{phaseABJ}), whose appearance has been ascribed to a Chern--Simons framing anomaly in literature \cite{Closset:2012vg,Closset:2012vp}.  This leads to the conclusion that the deformation affects the partition function only in its somewhat unphysical part, whereas its modulus is $\nu$ independent.  However, this is not a totally satisfactory explanation yet.  

A more general and cautious attitude could be to assume that expression \eqref{eq:matrixlat} is only a possible convenient way of rewriting the matrix model obtained properly via localization, which could arise as the result of applying some identities at the level of the matrix model average for the latitude Wilson loop.
However, while we have verified that it is possible to perform a transformation that brings us back to the ordinary ABJ(M) partition function, the form of the resulting operator insertion is not very enlightening.

In conclusion, we do not have a definite clear explanation of whether and how the matrix model \eqref{eq:matrixlat} could arise by performing an honest derivation of the latitude expectation value via localization, although there are some reassuring indications at least for the ABJM case.  
Nevertheless, supported by the striking agreement with the perturbative computation at weak coupling, we assume as a working hypothesis that the proposal 
in \eqref{eq:matrixlat} correctly reproduces the latitude expectation value. Equipped with this tool, in Section \ref{sec:Fermi} we perform a study of the matrix model average at strong coupling, where very little information is known on latitude Wilson loops \cite{Correa:2014aga}.

\subsection{Reformulation as a Fermi gas} \label{fermi}

The ABJ(M) matrix model can be reformulated in terms of a Fermi gas (see \cite{Marino:2011eh} for the original derivation in ABJM theory and 
\cite{Awata:2012jb, Matsumoto:2013nya, Honda:2013pea, Honda:2014npa} for its generalization to the ABJ model).
This perspective provides a powerful tool for expanding systematically the partition function and Wilson loop observables in powers of $1/N$ at strong coupling, by using statistical mechanics technology.
In this section we point out that the proposed matrix model for the latitude Wilson loop can also be given such an interpretation, paving the way for its study in the type IIA string and M--theory regimes. For simplicity we restrict the analysis to the ABJM slice, $N_1=N_2=N$.

\paragraph{The partition function}

The crucial property that streamlines the Fermi gas reformulation is the Cauchy identity, which we present in a form that is suitable for our purposes
\begin{equation}\label{eq:cauchy}
\frac{\displaystyle\prod_{a<b}^{N}\sinh r\, \pi\, (\lambda_{a}-\lambda _{b})\, \sinh r\, \pi\, (\mu_{a}-\mu_{b})}{\displaystyle\prod_{a=1}^{N}\prod_{b=1}^{N}\cosh\, r\, \pi\, (\lambda _{a}-\mu_{b})} = \sum_{\sigma\in S_N} (-1)^{\sigma} \prod_{a=1}^{N} \frac{1}{\cosh \,r\, \pi\, (\lambda _{a}-\mu_{\sigma(a)})}
\end{equation}
Here the final sum is over all the permutations of the $N$ eigenvalues and  $r$ is an arbitrary parameter. 

We split the integrand of \eqref{eq:partitionf} into two combinations of hyperbolic functions with arguments containing the factors $\sqrt{\nu}$ and $1/\sqrt{\nu}$ respectively, and apply the Cauchy identity separately by choosing $r$ in \eqref{eq:cauchy} appropriately. This procedure is similar for instance to the one used in \cite{Marino:2011eh} for ${\cal N}=3$ quiver models.
After few algebraic steps we end up with\footnote{Here we have rescaled the integration variables in order to obtain normalizations as in \cite{Marino:2011eh}.}
\begin{equation}\label{eq:partfuncfermi}
Z =2^{2N} N!\, \sum_{\sigma\in S_N} (-1)^{\sigma}\, \int \frac{dy^N}{(2\pi k)^N} \prod_{a=1}^N \left(2 \cosh  \frac{y_a}{2}\right)^{-1}\left( 2 \cosh \frac{y_a-y_{\sigma(a)}}{2k}\right)^{-1}
\end{equation}
We thus see that the dependence on $\nu$ drops completely and the partition function lands on the same expression as for ABJM theory.
We stress that these last steps are valid only for $N_1=N_2$. 

For different ranks of the gauge groups  the starting point \eqref{eq:cauchy} must be replaced by  a generalization of the Cauchy determinant identity  discussed in \cite{Awata:2012jb,Matsumoto:2013nya,Honda:2013pea}. Then  if we repeat the steps leading from \eqref{eq:cauchy} to \eqref{eq:partfuncfermi}, we can isolate and evaluate
the $\nu-$dependent part of the partition function. We find
\be
\label{phaseABJ}
\exp\left(\frac{\pi i}{12 k}\left(\nu+\frac{1}{\nu}\right) ((N_1-N_2)^3-(N_1-N_2))\right)
\ee
The only effect of the deformed measure consists in altering  the original phase of the ABJ partition function obtained in \cite{Drukker:2010nc} by a trivial  $\nu-$dependent multiplicative factor.

Going back to the ABJM case, it was observed in \cite{Marino:2011eh} that \eqref{eq:partfuncfermi} can be formally interpreted as the canonical partition function of an ideal Fermi gas of $N$ particles
\begin{align}\label{eq:canonicalZ}
Z_N =2^{2N} N!\, \sum_{\sigma\in S_N} (-1)^{\sigma}\, \int dy^N \prod_{a=1}^N \rho(y_a, y_{\sigma(a)}) \equiv  \Tr\,  \rho 
\end{align}
with a density matrix $\rho$ completely factorized into non--trivial one--particle density matrices 
\begin{equation} \label{eq:rho}
\rho(y_a, y_{\sigma(a)})   =   \frac{1}{2\pi k \, \big(2 \cosh  \frac{y_a}{2}\big)\,\big( 2 \cosh \frac{y_a-y_{\sigma(a)}}{2k}\big)}
\end{equation}
We can then define the corresponding one--particle quantum Hamiltonian 
\begin{equation}\label{eq:H}
\hat \rho = e^{-\hat H} \qquad {\rm such~ that} \qquad   \braket{y_1|\hat \rho|y_2} \equiv \rho(y_1,y_2) 
\end{equation}  
Using the explicit expression for $\rho$ in \eqref{eq:rho} the Hamiltonian can be written as
\begin{equation} \label{eq:H2}
e^{-\hat{H}} = e^{-U(\hat{q})} e^{-T(\hat{p})} \quad ,  \quad U(\hat{q}) = \log \left( 2\cosh\frac{\hat{q}}{2}\right) \quad , \quad T(\hat{p}) = \log \left( 2\cosh\frac{\hat{p}}{2}\right)
\end{equation}
in terms of a non--standard kinetic term $T(\hat{p})$ and a potential $ U(\hat{q})$, where $\hat q$ and $\hat p$ are canonically conjugate operators satisfying
\beq \label{eq:QM}
[\hat q,\hat p]=i\hbar \qquad , \qquad \hbar = 2\pi k
\eeq
Accordingly, the quantum average of a single--particle operator ${\hat O} \equiv O(\hat q, \hat p)$ is given by
\beq \label{eq:average}
\langle  { \hat O}   \rangle = \Tr ( \hat \rho \, \hat O )
\eeq

\vskip 10pt
\paragraph{Introducing Wilson loops}

When a Wilson loop operator of the form $\sum_{i=1}^N e^{2\pi m \sqrt{\nu} \lambda_i}$, relevant for computing the bosonic latitude, is inserted into the matrix model average we can still use the Cauchy identity and perform the same steps as those leading to \eqref{eq:partfuncfermi}. In the case of a single winding operator ($m=1$) the result we obtain reads
\begin{align}\label{eq:matrixlatcauchy}
 &\langle W_B(\nu) \rangle_\nu = \,  \frac{2^{2N}N!}{ Z} \times \\ 
 & \,  \sum_{\sigma\in S_N} (-1)^{\sigma}\, \int \frac{dy^N}{(2\pi k)^N} \, \sum_{c=1}^N\frac{ e^{i\, \frac{\pi \nu}{k}} e^{\frac{y_c}{k}}}{\displaystyle\prod_{a=1}^N 2\cosh\frac{ y_a}{2}\, \prod_{a\neq c}2\cosh \frac{y_a-y_{\sigma(a)}}{2k}\,2 \cosh\frac{y_c-y_{\sigma(c)}+2i \pi \nu}{2k}}  \non
\end{align}
As expected, the $\nu$ dependence does not drop any longer. In particular, it appears  not only in the overall phase factor, but also in an interacting piece. This can be interpreted as the statement that for this Wilson loop the framing factor is in general non--trivial and related to the $\nu$ parameter. This is a generalization of what happens already for the undeformed operator ($\nu=1$), for which the matrix model computes the average at framing one.  

Within the Fermi gas approach reviewed above, the Wilson loop expectation value \eqref{eq:matrixlatcauchy} maps to the quantum average \eqref{eq:average} of the $\nu$--dependent one--body operator 
\begin{equation}\label{eq:operator}
\hat{O}(\nu) = e^{\left( \hat{q} \, + \, \nu \, \hat{p} \right)/k}
\end{equation}
with the ABJM density operator defined in \eqref{eq:H}, \eqref{eq:H2}.

We observe that for $\nu=1$ this operator reduces to that corresponding to the undeformed Wilson loop \cite{Klemm:2012ii}. In general, the presence of the $\nu$ factor unbalances the $(\hat q,\hat p)$ symmetry, which manifests in the $\nu=1$ case.
We also point out that the insertion of operator \eqref{eq:operator} implies a normalization for the Wilson loop expectation value that differs by an overall $N$ factor from the one used at weak coupling in Section \ref{3loopresult}. Here we find convenient to use this normalization for a better comparison with the formulae of \cite{Klemm:2012ii}.
 
Equations \eqref{eq:matrixlatcauchy} and \eqref{eq:operator} can be generalized to the case of a multiply wound Wilson loop. In particular, its expectation value corresponds to the average of the single--body operator
\begin{equation}
\hat{O}^m(\nu) = e^{\frac{m}{k}\left( \hat{q} \, + \, \nu \, \hat{p} \right)}
\end{equation}

\subsection{A peculiarity of the matrix model }\label{sec:proof}

We conclude this section by highlighting a peculiar property of the matrix model average \eqref{eq:matrixlat}, valid for the ABJM theory ($N_1=N_2 \equiv N$). 
Precisely, we claim that its real part is independent of $\nu$ 
\beq\label{eq:strongeridentity}
\pa_\nu \log{\left( \langle W_B(\nu) \rangle_\nu + \langle \hat W_B(\nu) \rangle_\nu \right)}\,\, = \pa_\nu\, {\rm Re}\left(\langle W_B(\nu) \rangle_\nu\right) = 0 \qquad\qquad \text{for } N_1=N_2
\eeq
This property can be proven as follows.
We consider the Wilson loop expectation value as in \eqref{eq:matrixlatcauchy}, after the application of the Cauchy identity.
We take the real part of this expression and its derivative with respect to $\nu$ at the level of the integrand, obtaining the following expression
\begin{align}\label{eq:realder}
\pa_\nu\, {\rm Re}\left(\langle W_B(\nu) \rangle_\nu\right) & \propto \sum_{\sigma\in S_N} (-1)^{\sigma}\, \int \frac{dy^N}{\pi^{N}k^N} \frac{1}{\prod_{a=1}^N \cosh \frac{y_a}{2}\, \cosh \frac{y_a-y_{\sigma(a)}}{2k}}\, 
\nonumber\\&
\qquad \times \sum_{c=1}^N
\frac{4\pi}{k} \frac{\sin\frac{2\pi\nu}{k}\ \cosh^2 \frac{y_c-y_{\sigma(c)}}{2k}\, \sinh \frac{y_c-y_{\sigma(c)}}{2k}\, e^{\frac{y_c+y_{\sigma(c)}}{2k}}}{\left(\cos\frac{2\pi\nu}{k} + \cosh \frac{y_c-y_{\sigma(c)}}{k}\right)^2}
\end{align}
Next we work directly on the integrand and prove that \eqref{eq:strongeridentity} holds before any integration.

The permutations in the symmetric group $S_N$ can be divided into those which are idempotent $\{\sigma\in S_N | \sigma^2=1\}$ and those which are not. The former are those constructed as products of cycles of maximal length 2, whereas the latter contain at least one cycle of length greater than 2. 
For permutations belonging to the first group we find that the sum
\begin{equation}
\sum_{c=1}^N
\frac{4\pi}{k} \frac{\sin\frac{2\pi\nu}{k}\ \cosh^2 \frac{y_c-y_{\sigma(c)}}{2k}\, \sinh \frac{y_c-y_{\sigma(c)}}{2k}\, e^{\frac{y_c+y_{\sigma(c)}}{2k}}}{\left(\cos\frac{2\pi\nu}{k} + \cosh \frac{y_c-y_{\sigma(c)}}{k}\right)^2} =0  \qquad \sigma^2=1
\end{equation}
vanishes identically since the summand is antisymmetric with respect to the exchange $y_c\leftrightarrow y_{\sigma(c)}$. In fact, due to this property a given term $c=\bar c$ in the sum either vanishes if $\sigma(\bar c)=\bar c$ or, if  $\sigma(\bar c)={\bar c}'$, it will be canceled by an opposite contribution for $c={\bar c}'$.

Next we consider permutations which are not idempotent, so that they do not coincide with their inverse. This set can be divided into pairs $(\sigma, \sigma^{-1})$, which have the same signature and span the whole set. They give rise to the same factor $\prod_{a=1}^N \cosh \frac{y_a-y_{\sigma(a)}}{2k}$ in \eqref{eq:realder}.
Hence, restricting the sum in \eqref{eq:realder} to such a pair of permutations, we can focus on the two contributions 
\begin{align}\label{eq:suminverse}
\sum_{c=1}^N
\frac{\cosh^2 \frac{y_c-y_{\sigma(c)}}{2k}\, \sinh \frac{y_c-y_{\sigma(c)}}{2k}\, e^{\frac{y_c+y_{\sigma(c)}}{2k}}}{\left(\cos\frac{2\pi\nu}{k} + \cosh \frac{y_c-y_{\sigma(c)}}{k}\right)^2}
+ \sum_{c=1}^N
\frac{\cosh^2 \frac{y_c-y_{\sigma^{-1}(c)}}{2k}\, \sinh \frac{y_c-y_{\sigma^{-1}(c)}}{2k}\, e^{\frac{y_c+y_{\sigma^{-1}(c)}}{2k}}}{\left(\cos\frac{2\pi\nu}{k} + \cosh \frac{y_c-y_{\sigma^{-1}(c)}}{k}\right)^2} 
\end{align}
and choose a particular term $c=\bar c$ in the first sum for which $\sigma(\bar c)={\bar c}'$.
Then, a term exists in the second sum corresponding to the eigenvalue $y_{\bar c'}$. The sum of these two pieces vanishes
\begin{equation}
\frac{\cosh^2 \frac{y_{\bar c}-y_{{\bar c}'}}{2k}\, \sinh \frac{y_{\bar c}-y_{{\bar c}'}}{2k}\, e^{\frac{y_{\bar c}+y_{{\bar c}'}}{2k}}}{\left(\cos\frac{2\pi\nu}{k} + \cosh \frac{y_{\bar c}-y_{{\bar c}'}}{k}\right)^2}+\frac{\cosh^2 \frac{y_{{\bar c}'}-y_{\sigma^{-1}({\bar c}')}}{2k}\, \sinh \frac{y_{{\bar c}'}-y_{\sigma^{-1}({\bar c}')}}{2k}\, e^{\frac{y_{{\bar c}'}+y_{\sigma^{-1}({\bar c}')}}{2k}}}{\left(\cos\frac{2\pi\nu}{k} + \cosh \frac{y_{{\bar c}'}-y_{\sigma^{-1}({\bar c}')}}{k}\right)^2} = 0
\end{equation}
since, by construction, $\sigma^{-1}({\bar c}')=\bar c$ and consequently the latter term is equal and opposite to the former.
Such a cancellation extends pairwise to all the terms in the sum \eqref{eq:suminverse}, and therefore the terms in \eqref{eq:realder} associated to a given permutation and its inverse cancel completely. This argument in turn extends to all permutations which are not idempotent. Consequently, the whole expression \eqref{eq:realder} evaluates to zero, as claimed.

As an important corollary of identities \eqref{eq:relationbremsstrahlung} and  \eqref{eq:strongeridentity} we find that
\begin{equation}
\pa_m \log{\left( \langle W_B^m(1) \rangle_1 + \langle \hat W_B^m(1) \rangle_1 \right)} \Big|_{m=1} = 2\, \pa_\nu \log{\left( \langle W_B(\nu) \rangle_\nu + \langle \hat W_B(\nu) \rangle_\nu \right)} \Big|_{\nu=1} \underset{N_1=N_2}{=} 0
\end{equation}
This proves a property of both the multiply wound 1/6--BPS average and the latitude Wilson loops, which in particular allowed the steps leading to \eqref{eq:Bframing}.

\section{The matrix model at strong coupling} \label{sec:Fermi}

For the ABJM slice $N_1=N_2=N$, in this section we provide the expansion of the matrix model average \eqref{eq:matrixlatcauchy} that computes the expectation value of the bosonic latitude Wilson loop at strong coupling. Automatically, this also provides the average of the fermionic operator at strong coupling, via the cohomological equivalence \eqref{eq:quantumcoho}.

For simplicity, we confine the treatment to single winding operators ($m=1$), and point out how to extend the calculation to the more general case of multiply wound Wilson loops, if need be.
We work at large $N$ but we do not restrict to the planar limit. We adopt the Fermi gas approach reviewed in the previous section, as this method has the virtue of granting a systematic control on both quantum corrections in the coupling and on the genus expansion around the large $N$ limit \footnote{We will stricktly follow the procedure of \cite{Klemm:2012ii}, although it has been argued later \cite{Okuyama:2016deu} that results obtained there are correct only for winding--one 1/6 BPS Wilson loop, whereas for general winding the authors of  \cite{Klemm:2012ii} missed higher order corrections in $1/k$.}.

As recalled above, the matrix model \eqref{eq:matrixlatcauchy} can be reformulated in terms of a one--dimensional ideal (non--interacting) quantum gas of particles with Fermi statistics.
In this setting the Chern--Simons level $k$ plays the role of the Planck constant, equation \eqref{eq:QM}, and the number of colors $N$ corresponds to the number of particles.
Therefore, large $N$ is equivalent to the thermodynamic limit of the gas, whereas the  $\hbar$ expansion encoding quantum corrections corresponds to an expansion at small $k$.
Consequently, the Fermi gas approach is suitable for studying the latitude expectation value in the M--theory regime, $N\to\infty$ and $k$ fixed.
We will limit our discussion to perturbative corrections, neglecting exponentially small contributions, stemming from world--sheet and membrane instantons.

\subsection{Thermodynamic limit and quantum corrections}

When the number of particles becomes large the canonical partition function \eqref{eq:canonicalZ} is hard to deal with. We then resort to the grand--canonical ensemble with the grand--canonical partition function defined as
\begin{equation}
\Xi = 1+ \sum_{N=1}^{\infty} Z_N\, z^N
\end{equation}
where $z=e^{\mu}$ is the fugacity and $\mu$ the chemical potential. Accordingly, the canonical average for the one--body operator \eqref{eq:operator} is substituted by the grand--canonical average with Fermi statistics 
\begin{equation}\label{eq:grandaverage}
\frac{1}{\Xi}\, \langle \hat O \rangle^{GC} = \frac{1}{\Xi}\, \sum_{N=1}^{\infty} \langle \hat O \rangle\, z^N
= \Tr\left( \frac{\hat O}{e^{\hat H-\mu}+1}\right)
\end{equation}
The canonical average is then retrieved by an inverse transform in $\mu$.

The large $N$ expansion of the ABJM model can be argued to translate into a large chemical potential and energy expansion \cite{Marino:2011eh,Klemm:2012ii}. Resumming the power series from expanding \eqref{eq:grandaverage} in this limit, one obtains \cite{Klemm:2012ii}
\begin{equation} \label{eq:grandaverage2}
\frac{1}{\Xi}\, \langle \hat O \rangle^{GC} = \pi\partial_{\mu}\, \csc{\pi\partial_{\mu}}\, n_O(\mu)
\end{equation}
where $n_O(\mu)$ is the distribution
\begin{equation} \label{eq:distribution}
n_O(\mu) = \Tr \left( \Theta(\mu - \hat H) \, \hat O \right)
\end{equation}
with $\Theta$ being the Heaviside function that defines the Fermi surface.

\vskip 10pt

In order to keep track of quantum corrections for the operator average we use a semi--classical approach and look for a convenient way to expand \eqref{eq:distribution} in powers of $\hbar$. As discussed in \cite{Klemm:2012ii}, this is easily accomplished using the phase space formalism of Quantum Mechanics. This amounts to trading operators with their Wigner transform, according to
\begin{equation}
A_{W}(q,p)=\int d q' \left\langle q-\frac{q'}{2}\right|\hat A \left| q+\frac{q'}{2}\right\rangle e^{i p q'/\hbar}
\end{equation}
and operator products with $\star$--products
 \begin{equation}
\left(\hat A \hat B\right)_{W}= A_{W}  \exp\left[ \frac{i \hbar}{2} \left( {\overleftarrow{\partial}}_q {\overrightarrow{\partial}}_p  - {\overleftarrow{\partial}}_p {\overrightarrow{\partial}}_q\right) \right]    B_W \equiv A_{W}\star B_W
\end{equation}
In this formalism the trace is rewritten as an integral over the $(p,q)$ coordinates
\begin{equation}
\Tr \hat{A}=\int\frac{d p\, d q }{2\pi\hbar}\,A_W(q,p)
\end{equation}

Applying the Wigner transform to the one--particle Hamiltonian $\hat H$ defined in \eqref{eq:H}, we can expand any operator $f(\hat H)$ in powers of $(\hat H - H_W(q,p))$. Its semi--classical expansion is then obtained by taking the Wigner transform  
\begin{equation}\label{eq:kirkwood}
f(\hat H)_W = \sum_{r\geq 0} \frac{1}{r!}\, f^{(r)}(H_W)\, G_r \qquad , \qquad G_r = (\hat H-H_W(q,p))^r_W
\end{equation}
where the so--called Wigner--Kirkwood functions $G_r$ have an $\hbar$ expansion of the form 
\beq
G_0 = 1 \quad , \quad G_1 = 0 \qquad , \qquad G_r = \sum_{n\geq \left[ \frac{r+2}{3} \right]} \hbar^{2n} \, G_r^{(n)} \quad r \geq 2
\eeq

Applying this formalism to the distribution in \eqref{eq:distribution}, its semi--classical expansion reads
\begin{equation} \label{eq:distribution2}
n_{O}(\mu) = \int \frac{dp dq}{2\pi \hbar}\, \left(\Theta\left( \mu-H_W \right) O_W + \sum_{r > 1} \frac{(-1)^r}{r!}\, \delta^{(r-1)}\left( \mu-H_W \right)\, G_r\, O_W\right)
\end{equation}
The value $H_W(q,p) = \mu$ defines the Fermi surface.

In our case, from the explicit expression  \eqref{eq:H2} of the one--body Hamiltonian we realize that  $H_W$ is explicitly given by
\begin{equation} \label{eq:H3}
e^{-H_W}_{\star} = e^{-U(q)}\star e^{-T(p)} \quad ,  \quad U(q) = \log \left( 2\cosh\frac{q}{2}\right) \quad , \quad T(p) = \log \left( 2\cosh\frac{p}{2}\right)
\end{equation}
whereas for a generic operator of the form  $\hat O_{a,b} \equiv e^{(a \hat q + b \hat p)/k}$ we find
\begin{equation}\label{eq:operatorW}
(O_{a,b})_W (q,p) = e^{\frac{aq +b p}{k}}
\end{equation}

Equation \eqref{eq:distribution2} with entries \eqref{eq:H3} and \eqref{eq:operatorW} with $a=1, b=\nu$, are the key ingredients to obtain the expectation value $\langle W_B(\nu) \rangle_\nu$ at strong coupling from prescription \eqref{eq:grandaverage2}. We devote the rest of this section to its explicit evaluation.

\subsection{Expansion of Fermi gas at strong coupling}

In order to evaluate \eqref{eq:distribution2} we closely follow the treatment of \cite{Klemm:2012ii}, generalizing the form of the operator's Wigner transform as in \eqref{eq:operatorW}.
 
\begin{figure}
\centering
\begin{subfigure}{6.5cm}
  \centering
\includegraphics[scale=0.5]{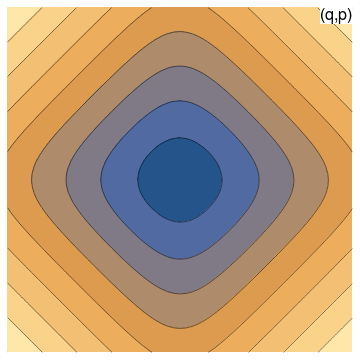}
\caption{}
\label{fig:fermi}
\end{subfigure}
\qquad\qquad
\begin{subfigure}{6.5cm}
  \centering
\includegraphics[scale=0.5]{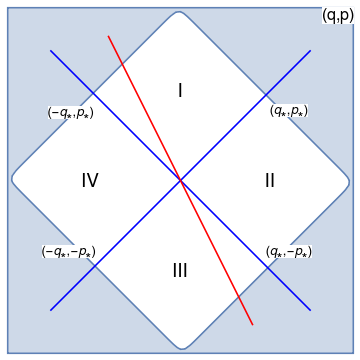}
\caption{}
\label{fig:regions}
\end{subfigure}
\caption{(a) shows the lines of constant $H_W$ in $(q,p)$ space, defining the Fermi surface. In (b) the Fermi region is divided into 4 areas. The red line has equation $p = -\frac{a}{b} q$ and separates the regions where the exponent of the operator \eqref{eq:operator} is positive or negative, for a given $0<b \equiv \nu<1$ and $a=1$.}
\end{figure}

The first step requires deriving the expression for the grand--canonical partition function of the Fermi gas. Since by construction this coincides with that of the undeformed case, we can read it 
from \cite{Klemm:2012ii}  
\begin{equation}\label{eq:grandpartition}
\Xi = \exp{\left(\frac{2\mu^3}{3\pi^2 k}+\frac{\mu}{3k}+\frac{\mu k}{24}\right)}
\end{equation}
Next, we concentrate on evaluating  the occupation number distribution  
\begin{align}\label{eq:density}
n_{O_{a,b}}(\mu) &= \int \frac{dp dq}{2\pi \hbar}\, \Theta\left( \mu-H_W \right) e^{\frac{aq+bp}{k}} + \sum_{r\geq 1} \frac{(-1)^r}{r!}\, \frac{d^{r-1}}{d\mu^{r-1}} \delta\left( \mu-H_W \right)\, G_r\, e^{\frac{aq+bp}{k}} 
\nonumber\\&
\equiv n_{O_{a,b}}^{(1)}(\mu) + n_{O_{a,b}}^{(2)}(\mu)
\end{align}
The integrals are over the Fermi region and surface, whose picture is given in Figure \ref{fig:fermi}.

As suggested in \cite{Klemm:2012ii}, it is convenient to divide the Fermi surface into four sectors.
The points where the separation occurs are 
\begin{equation}
p_\ast = \mu + \frac{i \hbar}{8} \qquad , \qquad  q_\ast = \mu + \frac{i \hbar}{8} + {\cal O}(e^{-\mu})
\end{equation}
where ${\cal O}(e^{-\mu})$ stands for exponentially suppressed terms.
The logic behind such a separation is that along the curve bounding the regions, we have that alternatively
\begin{equation}
|p|>\mu \qquad {\rm or} \qquad |q|>\mu
\end{equation}
This in turn means that exponentially small corrections in $p$ and $q$ are bounded by exponentially small corrections in $\mu$ and hence they can be neglected.
Precisely, 
\begin{align} \label{eq:regions}
\text{in regions I,III:}& \quad e^{-|p|}<e^{-\mu} \quad T(p) \sim \frac{p}{2} \nonumber\\
\text{in regions II,IV:}& \quad e^{-|q|}<e^{-\mu} \quad U(q) \sim \frac{q}{2}
\end{align}
where the approximations are correct up to exponentially small terms.

Due to the invariance of the Hamiltonian under $p\leftrightarrow q$, the different regions can be obtained from one another by exchanging the canonical coordinates. Upon the insertion of the $O_{a,b}$ operator this is equivalent to computing the contribution with the insertion of $O_{b,a}$.
Consequently, the idea is to explicitly compute the integration of $O_{a,b}$ along the curve in regions I and adding to it the contributions form the other regions obtained by changing signs and permuting labels. However, this procedure would overcount the contribution from the square $(|q|<q_\ast,|p|<p_\ast)$, which then needs to be subtracted. Naming this extra contribution $n_{O_{a,b}}^{(bulk)}$, we can finally write
\begin{equation}
n_{O_{a,b}} = n_{O_{a,b}}\big|_{I,III} + n_{O_{a,b}}\big|_{II,IV} - n_{O_{a,b}}^{(bulk)} = n_{O_{a,b}}\big|_{I,III} + n_{O_{b,a}}\big|_{I,III} - n_{O_{a,b}}^{(bulk)}
\end{equation}
The bulk contribution is easily computed and gives   
\begin{equation}\label{eq:bulk}
n_{O_{a,b}}^{(bulk)} =  \int_{-q_\ast}^{q_\ast} \int_{-p_\ast}^{p_\ast} \frac{dp\, dq}{2\pi \hbar}\, e^{\frac{a\,q+b\,p}{k}} = 
\frac{2\, k^2\, \sinh \frac{a\, q_\ast}{k}\, \sinh\frac{b\, q_\ast}{k}}{a\, b\, \hbar\, \pi }
\end{equation}
For the other contributions, we recall that for $a=b$, e.g.~for the 1/6--BPS Wilson loop, the domain of integration where the operator does not get exponentially suppressed coincides with regions I and II only and the computation can be limited to those.
For $a\neq b$ this is not necessarily true. In particular, choosing $a=1$, $b =\nu$ and assuming without loss of generality that $0<\nu<1$, we expect contributions from region III  (see Figure \ref{fig:regions}) not to be entirely exponentially suppressed.

Working in regions I and III we can use the approximated expression for $T( p )$ in \eqref{eq:regions}. Following the steps of \cite{Klemm:2012ii}, the first term in \eqref{eq:density} can be integrated in $p$ to get an integral along the Fermi surface
\begin{equation}\label{eq:nfirstterm}
n_{O_{a,b}}^{(1)}(\mu)\big|_{I,III} = \frac{k}{b}\, \int_{-q_{\ast}}^{q_{\ast}} \frac{dq}{2\pi \hbar}\, e^{\frac{aq}{k}}\left( e^{\frac{bp(\mu,q)}{k}} - e^{-\frac{bp(\mu,q)}{k}} \right)
\end{equation}
where
\begin{equation}
p(\mu,q) = 2\mu - (2H_W-|p|)
\end{equation}
This expression is indeed $p$--independent as a consequence  of \eqref{eq:regions}.

Solving for the $\delta$ function in the second term of \eqref{eq:density}, and adding it to \eqref{eq:nfirstterm} we obtain the resummed expression
\begin{equation}\label{eq:nO}
n_{O_{a,b}}(\mu)\big|_{I,III} = \frac{k}{2\pi \hbar\, b}\, \int_{-q_{\ast}}^{q_{\ast}} dq\, \left( e^{\frac{2b\mu}{k}} e^{\frac{aq+b|p|}{k}} e^{-\frac{2 b}{k}H_W}_{\star} -  e^{-\frac{2b\mu}{k}} e^{\frac{aq-b|p|}{k}} e^{\frac{2 b}{k}H_W}_{\star} \right)
\end{equation}
where we have used the definition of the Wigner--Kirkwood corrections \eqref{eq:kirkwood}.

This object has been computed in \cite{Klemm:2012ii} using the reduced Hamiltonian for regions I, III  
\begin{equation}\label{eq:HW}
e^{-t H_W}_{\star} = \exp\left( -\frac{t}{2}|p| + \frac{i t\, \hbar}{8} - \frac{t\, \hbar}{4\pi} \log\left(2\sinh\frac{q}{k}\right) \right)
\end{equation}
Plugging \eqref{eq:HW} into \eqref{eq:nO} the complete expression for $n_{O_{a,b}}$ in region I reads
\begin{equation}\label{eq:nOI}
n_{O_{a,b}} \big|_{I} = \frac{k^2}{2\pi \hbar\, b}\, e^{\frac{2b \mu}{k}}\, i^{b}\, I_{a,b}
\end{equation}
where
\begin{equation}
I_{a,b} \equiv \frac{1}{k}\, \int_{-q_{\ast}}^{q_\ast} dq\, \frac{e^{\frac{aq}{k}}}{(2\sinh\frac{q}{k})^b}
\end{equation}
The analogous expression for region III is obtained from \eqref{eq:nOI} by sending $b\to -b$.

One can now perform the change of variables $u = \exp{(q/k)}$ to put the integral in the form
\begin{equation}\label{eq:uint}
I_{a,b} \sim \tilde I_{a,b} = \int_0^{u_{\ast}} du\, \frac{u^{a-1}}{\left( u-u^{-1} \right)^{b}}
\end{equation}
where the lower limit of integration has been lowered down to $0$ at the affordable price of introducing spurious exponentially small corrections that we are neglecting anyway.

Finally, the integral $\tilde I_{a,b}$ can be evaluated in full generality  in terms of a hypergeometric function
\begin{equation}\label{eq:Ihypergeometric}
\tilde I_{a,b} = \frac{(-1)^{-b}\, u_\ast^{a+b} \, _2F_1\left(b,\frac{a+b}{2};\frac{1}{2} (a+b+2);u_\ast^2\right)}{a+b} \; ,  \; \qquad u_\ast \sim e^{\frac{\mu}{k}}
\end{equation}

\vskip 10pt
\paragraph{Leading exponential asymptotics}

We now perform the large $\mu$ asymptotics of \eqref{eq:Ihypergeometric}, discarding exponentially subleading contributions.

To this end we first invert the arguments of the hypergeometric function above using the following general identity
\begin{align}\label{eq:inversion}
_2F_1\left(\alpha ,\beta;\gamma;z\right) &= \frac{(-z)^{-\alpha }\, \Gamma (\gamma )\, \Gamma (\beta -\alpha ) \, _2F_1\left(\alpha ,\alpha -\gamma +1;\alpha -\beta +1;\frac{1}{z}\right)}{\Gamma (\beta )\, \Gamma (\gamma -\alpha )}
\nonumber\\&
+\frac{ (-z)^{-\beta }\, \Gamma (\gamma )\, \Gamma (\alpha -\beta ) \, _2F_1\left(\beta ,\beta -\gamma +1;-\alpha +\beta +1;\frac{1}{z}\right)}{\Gamma (\alpha )\, \Gamma (\gamma -\beta )}
\end{align}
and then expand each term in a power series. Since $u_{\ast}$ is exponentially large, the first term in the expansion is sufficient for retaining only the exponentially leading contributions.

We will restrict here to the cases of interest, which are $\tilde I_{1,\nu}$, $\tilde I_{\nu,1}$ and $\tilde I_{1,-\nu}$ with $0<\nu<1$, although the same analysis could be carried out for 
generic $a$ and $b$. In particular, this would allow to take into account the case of multiply wound latitude Wilson loops, for which we should set $a=m$ and $b=m\nu$.

Restricting to the single winding operator, the large $\mu$ asymptotics in the various regions reads
\begin{align}
\text{region I: } & n_{O_{1,\nu}} \big|_{I} \sim -\frac{(-1)^{\frac{\nu +1}{4}}\, k^2\, e^{\frac{\mu  (\nu +1)}{k}}}{2 \pi  \hbar\, (\nu -1) \nu } + \frac{i\, k^2\, \Gamma \left(\frac{\nu -1}{2}\right)\, \Gamma \left(\frac{\nu +3}{2}\right)\, e^{\frac{2 \mu  \nu }{k}}}{2 \pi  \hbar\, \Gamma (\nu +2)} \nonumber\\
\text{region II: } & n_{O_{1,\nu}} \big|_{II} \sim \frac{k^2\, e^{\frac{\mu }{k}} \left(2\, (-1)^{\frac{\nu +1}{4}} e^{\frac{\mu  \nu }{k}} + \pi\,  (\nu -1) e^{\frac{\mu }{k}} \left(1+i \tan \frac{\pi  \nu }{2}\right)\right)}{4 \pi  \hbar\, (\nu -1)} \nonumber\\
\text{region III: } & n_{O_{1,\nu}} \big|_{III} \sim -\frac{k^2\, e^{-\frac{\mu  (\nu-1) }{k}-\frac{i \pi\,  (\nu-1) }{4}}}{2 \pi  \hbar\, \nu (\nu+1)}
\end{align}
We note that the contribution from region III can be neglected, even though it is not exponentially suppressed (in the sense that it does not vanish exponentially for large $\mu$). In fact, compared to the contributions from the other regions it is subdominant and bounded from above by the subleading exponential $e^{\mu/k}$ in the whole $0< \nu < 1$ range. Moreover, it possesses a different behavior in the $\nu\to 1$ limit.

Summing up the contributions from regions I and II and subtracting the asymptotic expansion of $n_{O_{1,\nu}}^{(bulk)}$  (see equation \eqref{eq:bulk}), we obtain
\begin{equation} \label{eq:asymptoticbulk}
n_{O_{1,\nu}}^{(bulk)}  \sim \frac{k^2\, e^{\frac{\mu  (\nu +1)}{k}+\frac{i \pi  (\nu+1) }{4}}}{2 \pi  \hbar\, \nu }
\end{equation}
where we have neglected exponentially small corrections. The exponent $e^{\frac{\mu (\nu+1)}{k}}$ from $n_{O_{1,\nu}}^{(bulk)}$ cancels against a similar contribution in  $n_{O_{1,\nu}} \big|_{I}$, so removing unexpected singularities at $\nu\to 0$. The final result reads
\begin{equation}\label{eq:nOnu}
n_{O_{1,\nu}} = \frac{k^2\, \left(\pi\,  e^{\frac{2 \mu }{k}} \left(1+i\, \tan \frac{\pi  \nu }{2}\right)+\frac{i\, \Gamma \left(\frac{\nu -1}{2}\right) \Gamma \left(\frac{\nu +1}{2}\right) e^{\frac{2 \mu  \nu }{k}}}{\Gamma (\nu +1)}\right)}{4 \pi\,  \hbar}
\end{equation}
This expression is well--defined in the whole physical region $0 \leq \nu \leq 1$. In fact, for $\nu \to 0$ we explicitly obtain
\begin{equation}
n_{O_{1,\nu}} \underset{\nu\to 0}{=} \frac{k^2\, e^{\frac{2 \mu }{k}}}{4\, \hbar}
\end{equation}
after discarding a constant term in $\mu$ arising from the first piece in \eqref{eq:nOnu}.

Although the two terms in \eqref{eq:nOnu} are separately singular for $\nu\to 1$, the singularities cancel, leaving the finite remainder  
\begin{equation}
n_{O_{1,\nu}} \underset{\nu\to 1}{=} \frac{k\, e^{\frac{2 \mu }{k}} (4 i\, \mu + (\pi -2 i) k )}{4 \pi\,  \hbar}
\end{equation}
This expression coincides with the result of \cite{Klemm:2012ii} for the singly wound undeformed 1/6--BPS operator.

\paragraph{Genus expansion}

Having computed the asymptotic expansion of the $n_{O_{1,\nu}}$ distribution, using prescription \eqref{eq:grandaverage2} we can finally evaluate the expansion at strong coupling of $\langle W_B(\nu) \rangle$  at framing $\nu$
\begin{equation} \label{eq:Wgrancanonical}
\langle W_B(\nu) \rangle_\nu = \frac{1}{2\pi i \,Z} \, \int d\mu\, e^{-N\mu}\, \langle O_{1,\nu} \rangle^{GC}
\end{equation}
through the grand--canonical average (setting $\hbar = 2\pi k$)
\begin{equation}
\frac{1}{\Xi} \langle O_{1,\nu} \rangle^{GC} = \frac{1}{4 \pi} \left(\pi\, e^{\frac{2 \mu }{k}} \csc \frac{2 \pi }{k} \left(1+i\, \tan \frac{\pi\,  \nu }{2}\right) + \frac{i \,  \nu\,  \Gamma \left(\frac{\nu -1}{2}\right) \Gamma \left(\frac{\nu +1}{2}\right) e^{\frac{2 \mu  \nu }{k}} \csc \left(\frac{2 \pi\,  \nu }{k}\right)}{\Gamma (\nu +1)}\right)
\end{equation}
with $\Xi$ given in \eqref{eq:grandpartition}.
The result can be expressed in terms of Airy functions as in \cite{Klemm:2012ii} \footnote{We recall that here the normalization of the operator has been chosen as in \cite{Klemm:2012ii} and differs by a factor $N$ from that used at weak coupling in Section \ref{sec:perturbative}.}
\begin{align} \label{bosonicfinal}
\langle W_B(\nu) \rangle_\nu &= \frac{1}{8 \pi ^2} \left( 2\, \pi ^2\, \csc \frac{2 \pi }{k}\, \left(1+i \tan \frac{\pi  \nu }{2}\right)\, \frac{\text{Ai}\left(\left(\frac{2}{\pi^2k} \right)^{-1/3} \left(N-\frac{k}{24}-\frac{7}{3 k}\right)\right)}{\text{Ai}\left(\left(\frac{2}{\pi^2k} \right)^{-1/3} \left(N-\frac{k}{24}-\frac{1}{3 k}\right)\right)}
\right.\nonumber\\& \left.
+ 2\, i\, \pi\,  \nu\,  \frac{\Gamma \left(\frac{\nu -1}{2}\right)\, \Gamma \left(\frac{\nu +1}{2}\right)}{\Gamma (\nu +1)}\, \csc \frac{2 \pi  \nu }{k}\, \frac{\text{Ai}\left(\left(\frac{2}{\pi^2k} \right)^{-1/3} \left(N-\frac{k}{24}-\frac{6 \nu +1}{3 k}\right)\right)}{\text{Ai}\left(\left(\frac{2}{\pi^2k} \right)^{-1/3} \left(N-\frac{k}{24}-\frac{1}{3 k}\right)\right)} \right)
\end{align}
As in the case of the $m$--winding 1/6 BPS Wilson loop computed in \cite{Klemm:2012ii}, this result is missing $1/k$ corrections. In fact, an alternative derivation of this equation \cite{Okuyama} reveals that the coefficient  
$ 2\, i\, \pi\,  \nu\,  \frac{\Gamma \left(\frac{\nu -1}{2}\right)\, \Gamma \left(\frac{\nu +1}{2}\right)}{\Gamma (\nu +1)}$  in the second line should be modified and would contain a non--trivial dependence on $k$  \footnote{We are grateful to Kazumi Okuyama for sharing with us his results.}.

Introducing the string coupling 
\begin{equation}
g_s = \frac{2\pi i}{k}
\end{equation}
we can now expand \eqref{bosonicfinal} at strong coupling and in the genus series
\begin{equation}\label{eq:genus}
\langle W_B(\nu) \rangle_\nu = \sum_g g_s^{2g-1} \langle W_B(\nu) \rangle_{\nu}\,\big| _{g}
\end{equation}
While $g_s>0$ terms will be not reliable due to the subtlety mentioned above, we can safely compute the genus--zero term. To this end, 
it is convenient to define the new variable $\kappa$ throught the identity
\begin{equation}
\frac{N}{k} = \frac{\log^2 \kappa}{2\pi^2} + \frac{1}{24} + O\left(\kappa^{-2}\right)
\end{equation}
In terms of this variable the genus--zero contribution reads
\begin{equation} \label{eq:genus0}
\langle W_B(\nu) \rangle_\nu\,\big|_{g=0} = \frac{-\kappa ^{\nu }\, \Gamma \left(\frac{\nu -1}{2}\right) \Gamma \left(\frac{\nu +1}{2}\right)+ i\, \pi\,  \kappa  \left(1+i\, \tan \frac{\pi  \nu }{2}\right) \Gamma (\nu +1)}{4 \pi\, \Gamma (\nu +1)}
\end{equation}

To conclude, we mention that by taking the complex conjugate of these expression one obtains the genus--zero contribution  for the bosonic Wilson loop $\langle \hat W_B(\nu) \rangle_\nu$ corresponding to the second gauge group of the ABJM theory.

\paragraph{The fermionic operator}

As already discussed, once we know the expectation values for the bosonic latitude Wilson loops we can recover that of the fermionic operator thanks to the identity \eqref{eq:quantumcoho}.  At strong coupling, the  cohomological equivalence can be implemented already at the stage of the occupation number distribution. Its fermionic version then reads\begin{equation}
n^F_{O_{1,\nu}} = -\frac{k^2\, 2^{-\nu -2}\, \Gamma \left(-\frac{\nu }{2}\right)\, e^{\frac{2 \mu  \nu }{k}}}{\sqrt{\pi }\, \Gamma \left(\frac{3-\nu }{2}\right)\, \hbar}
\end{equation}
This is the only surviving $\nu$--exponential behavior, due to an unforeseen cancellation.
Following the steps described above, we obtain the expression of the expectation value in terms of Airy functions
\begin{equation} \label{fermionicWL}
\langle W_F(\nu) \rangle_\nu = -\frac{\nu\,  \Gamma \left(-\frac{\nu }{2}\right)\, \csc \left(\frac{2 \pi  \nu }{k}\right)\, \text{Ai}\left(\left(\frac{2}{\pi^2k} \right)^{-1/3} \left(N-\frac{k}{24}-\frac{6 \nu +1}{3 k}\right)\right)}{2^{\nu+2}\, \sqrt{\pi }\, \Gamma \left(\frac{3-\nu }{2}\right)\, \text{Ai}\left(\left(\frac{2}{\pi^2k} \right)^{-1/3} \left(N-\frac{k}{24}-\frac{1}{3 k}\right)\right)}
\end{equation}
and its genus expansion. The first term reads
\begin{equation}\label{eq:fermionicgenus0}
\langle W_F(\nu) \rangle_\nu\,\big|_{g=0} = -i\, \frac{2^{-\nu -2}\, \kappa ^{\nu }\, \Gamma \left(-\frac{\nu }{2}\right)}{\sqrt{\pi}\, \Gamma \left(\frac{3}{2}-\frac{\nu }{2}\right)}
\end{equation}
while higher genus contributions would be still affected by the lack of $1/k$ corrections in \eqref{fermionicWL}, as inherited from the bosonic result. 

As already mentioned in Section \ref{review}, in this case the $\nu\to 0$ limit is ill--defined, due to the normalization factor ${\cal R} = 1/(e^{-\frac{i \pi \nu}{2}}-e^{\frac{i \pi \nu}{2}})$ that drives the limit to infinity, much alike what happens for the 1/2--BPS operator at even winding numbers. A sensible result at $\nu=0$ can be obtained by removing the ${\cal R}$ factor and replacing it with $1$.

\vskip 10pt
As a last comment, we note that taking the $\nu\to 0$ limit at the level of the Airy functions allows to derive the following curious relation valid at strong coupling
\begin{equation}
\langle W_B(0) \rangle_0 = -\frac{i\, k}{4\pi} + \langle W_{F}(1) \rangle_1
\end{equation}
between the singly wound bosonic latitude  Wilson loop at $\nu=0$ and framing zero, and the fermionic 1/2--BPS Wilson loop at framing one.

\subsection{Comparison with the string prediction}

Classical string configurations that are dual to the latitude operators have been discussed in \cite{Correa:2014aga}.

The fermionic operator maps to a type IIA string configuration in the $AdS_4\times \mathbb{CP}^3$ background, whose endpoints are not fixed in the internal space, but rather move along a circle in $\mathbb{CP}^3$. This accounts for the non--trivial profile of the matter couplings \eqref{eq:matrixfermionic} arising in the field theoretical definition of the Wilson loop.
The semi--classical analysis of \cite{Correa:2014aga} reveals that the leading exponential behavior of such a configuration scales according to  
\begin{equation}
\langle W_F(\nu) \rangle \sim e^{\pi\, \nu\, \sqrt{2\,\lambda}} \qquad , \qquad \lambda =\frac{N}{k}
\end{equation}
 
Expansion \eqref{eq:fermionicgenus0} for the matrix model at strong coupling remarkably agrees with this string prediction, thus providing a further non--trivial test of the correctness of our proposal. Beyond that, result \eqref{eq:fermionicgenus0} predicts the precise normalization of this exponential as well as its quantum corrections, which call for further string theory checks.

For the bosonic latitude operator no precise dual string configuration has been determined yet and therefore our findings constitute a brand new prediction, begging for a string theory confirmation.
We remark that in the undeformed case the ratio between the bosonic and fermionic operators is simply proportional to $\sqrt{\lambda}$, which has been interpreted as the volume of $\mathbb{CP}^1$ inside $\mathbb{CP}^3$ \cite{Drukker:2010nc}, in agreement with the proposed interpretation of the 1/6--BPS Wilson loop as a string smeared over that cycle \cite{Drukker:2008zx}.
For the latitude operator, instead, the bosonic expectation value displays a more complicated structure with two exponential behaviors (see equation \eqref{eq:genus0}), potentially suggesting that the smearing over $\mathbb{CP}^1$ interpretation does not carry through this case. This is in line with the comments in \cite{Correa:2014aga}, which seem to rule out the possibility to describe the bosonic latitude through a simple geometric smearing in the internal space.

\section{A conjecture for \texorpdfstring{$B_{1/2}$}~ in the ABJ theory}  \label{sec:ABJ}

In this section we discuss a possible generalization of the Bremsstrahlung functions in (\ref{eq:Bferm}) and (\ref{eq:Bbos}) to the case of $U(N_1)_k \times U(N_2)_{-k}$ ABJ theory. 

In general, for $N_1 \neq N_2$ less is known about the $B$--functions compared to the $N_1 = N_2$ case. Perturbative results for all the Bremsstrahlung functions exist from a direct evaluation of the corresponding cusped Wilson loop. Based on the two--loop result, in \cite{Griguolo:2012iq} it was argued that in the ABJ case the cusped Wilson loop has a double--exponentiation structure. A different exponentiation structure still involving two terms has been further derived in \cite{Bonini:2016fnc} by resumming ladder diagrams to all orders. In particular, the result of \cite{Griguolo:2012iq} seems to point towards a non--trivial $B_{1/2}$ at first order, while having a two--loop vanishing contribution from both the exponents. It turns out instead, that the $B_{1/6}^{\varphi}$ and $ B_{1/6}^\th$ expansions start at order two in the couplings.   

In \cite{Bianchi:2017afp,Bianchi:2017ujp} the evaluation of $B_{1/6}^{\th}$ has been pushed to four loops by computing the corresponding cusped bosonic Wilson loop at that order. The remarkable output is that the result
\beq
B_{1/6}^\th = \frac{N_1 N_2}{4k^2} - \frac{\pi^2}{24 k^4} \left( 5 N_1^2 N_2^2 + N_1 N_2^3 \right) + O(k^{-6})
\eeq
coincides with $1/2$ a putative result for $B_{1/6}^\varphi$ obtained by generalizing prescription (\ref{Bvarphi}) to the $m$--winding Wilson loop in ABJ whose expansion up to eighth order can be found in \cite{Bianchi:2016gpg}. 
This property, although tested only at few perturbative orders, points towards the validity of identity \eqref{eq:bthetaconj} at any loop order also for $N_1 \neq N_2$, and supports the conjecture that the general prescriptions (\ref{Bvarphi}, \ref{Btheta}) hold also in the ABJ theory.

In fact, if we trust our matrix model as the correct prescription for computing latitude Wilson loops, we are guaranteed by construction that identity \eqref{eq:relationbremsstrahlung} is valid also for the ABJ theory. This identity, together with \eqref{eq:bthetaconj}, conspires to sustain the conjecture that prescriptions (\ref{Bvarphi}, \ref{Btheta}) have an obvious generalization to the ABJ case. 

Therefore, supported by these reassuring facts, we use the prescription in (\ref{Btheta}) to make a prediction for $B_{1/2}$ in the  $N_1 \neq N_2$ case. Inserting there the explicit matrix model expansion at weak coupling (here also including the fifth order term) we find
\begin{align}
\label{prop1/2}
B_{1/2} &\overset{?}{=} \frac{N_1 N_2}{4k \left(N_1+N_2\right)}-\frac{\pi ^2 N_1 N_2 \left(N_1 N_2-3\right)}{24 k^3 \left(N_1+N_2\right)}
\nonumber\\&
-\frac{\pi ^6 N_1 N_2 \left(7 N_1^3 N_2 - 62 N_2^2 N_1^2+ 7 N_1 N_2^3 + 120\left(  N_1 N_2 - 1\right)\right)}{360 k^5 \left(N_1+N_2\right)}
+O\left(k^{-7}\right)
\end{align}
as a sensible proposal for the fermionic Bremsstrahlung function in the ABJ theory. 

There are of course some non--trivial aspects in this proposal that should be better understood. First of all, while the vanishing of the two--loop contribution is consistent with the result of \cite{Griguolo:2012iq}, the one--loop term does not coincide and seems to suggest a different exponentiation structure. Moreover,  the prescription of taking the modulus seems to be crucial to recover the vanishing of the second order term, a fact that needs a deeper understanding.

Further perturbative data, obtained from generalizing the three--loop computation of \cite{Bianchi:2017svd} to the ABJ case, would surely give more insights on the exponentiation procedure of the cusped Wilson loop, leading possibly to a check of conjecture (\ref{prop1/2}). We leave this open problem for the future.

\acknowledgments

We thank Lorenzo Bianchi, Michelangelo Preti and Edoardo Vescovi for sharing with us their results prior to publication. We also thank Marcos Marino, Kazumi Okuyama and Sara Pasquetti for useful discussions. MSB is supported by DFF-FNU through grant number DFF-4002-00037.
This work has been supported in part by Italian Ministero dell'Istruzione, Universit\`a e Ricerca (MIUR), and Istituto
Nazionale di Fisica Nucleare (INFN) through the  ``Gauge and String Theory'' (GAST) and ``Gauge Theories, Strings, Supergravity'' (GSS)   research projects.

\vfill
\newpage

\appendix

\section{ABJ(M) theory} \label{AppA}

In this section we summarize basic notions on the quiver U$(N_1)_k \times$ U$(N_2)_{-k}$ ABJ(M) theory. Its field content includes two gauge fields 
$(A_\mu)_i{~}^j$  and $(\hat {A}_\mu)_{\hat i}{~}^{\hat{j}}$ belonging respectively to
the adjoint of U$(N_1)$ and U$(N_2)$, and matter scalar fields  
$(C_I)_i{~}^{\hat j}$ and $(\bar {C}^I)_{\hat{i}}{~}^j$ plus their fermionic superpartners $(\psi_I)_i{~}^{\hat j}$ and $(\bar {\psi}^I)_{\hat{i}}{~}^j$ . The fields $(C_I, \bar\psi^I)$ transform in 
the $({\bf N_1},{\bf \bar N_2})$ of the gauge group (small latin indices) while the pair $(\bar C^I,\psi_I)$ belongs to the
representation $({\bf \bar N_1},{\bf N_2})$.
Index $I = 1,2,3,4$  labels the fundamental representation of the $SU(4)$ R--symmetry group.

After gauge fixing, the action reads  
\beq
\label{action}
S = S_{\mathrm{CS}} \big|_{\mathrm{g.f.}}+ S_{\mathrm{mat}} + S_{\mathrm{pot}}^{\mathrm{bos}} + S_{\mathrm{pot}}^{\mathrm{ferm}} 
\eeq 
where
\begin{subequations}
\begin{align}
\label{action1}
S_{\mathrm{CS}}\big|_{\mathrm{g.f.}} &=\frac{k}{4\pi}\int d^3x\,\varepsilon^{\mu\nu\rho} \Big\{ i \, \Tr \!\left(\hat{A}_\mu\partial_\nu 
\hat{A}_\rho+\frac{2}{3} i \hat{A}_\mu \hat{A}_\nu \hat{A}_\rho \right) \!-\! i \, \Tr \left( A_\mu\partial_\nu A_\rho+\frac{2}{3} i A_\mu A_\nu A_\rho\! \right)\!\non \\
&~   + \, \Tr \Big[ \frac{1}{\xi}  (\pa_\mu A^\mu)^2 -\frac{1}{\xi} ( \pa_\mu \hat{A}^\mu )^2 + \pa_\mu \bar{c} D^\mu c  
  - \pa_\mu \bar{\hat{c}} D^\mu \hat{c} \Big] \Big\}
\\
\label{action2}
S_{\mathrm{mat}} =& \int d^3x \, \Tr \Big[ D_\mu C_I D^\mu \bar{C}^I - i \bar{\Psi}^I \g^\mu D_\mu \Psi_I \Big] 
\\
 S_{\mathrm{pot}}^{\mathrm{bos}} =& -\frac{4\pi^2}{3 k^2} \int d^3x \, \Tr \Big[ C_I \bar{C}^I C_J \bar{C}^J C_K \bar{C}^K + \bar{C}^I C_I \bar{C}^J C_J \bar{C}^K C_K\non
 \\
&~ \qquad \qquad \qquad \qquad + 4 C_I \bar{C}^J C_K \bar{C}^I C_J \bar{C}^K - 6 C_I \bar{C}^J C_J \bar{C}^I C_K \bar{C}^K \Big] 
\\
 S_{\mathrm{pot}}^{\mathrm{ferm}} =&  -\frac{2\pi i}{k} \int d^3x \, \Tr \Big[ \bar{C}^I C_I \Psi_J \bar{\Psi}^J - C_I \bar{C}^I \bar{\Psi}^J \Psi_J
+2 C_I \bar{C}^J \bar{\Psi}^I \Psi_J 
\non \\
&~ \qquad \qquad  - 2 \bar{C}^I C_J \Psi_I \bar{\Psi}^J - \e_{IJKL} \bar{C}^I\bar{\Psi}^J \bar{C}^K \bar{\Psi}^L + \e^{IJKL} C_I \Psi_J C_K \Psi_L \Big]
\end{align}
\end{subequations}
with $(\bar c,c)$ and $(\bar{\hat c},\hat c)$ being the ghosts. We use spinor and group conventions of \cite{Bianchi:2017svd}. The invariant $SU(4)$ $\epsilon$--tensors are defined as $\e_{1234}=\e^{1234} =1$ and the covariant derivatives are given by
\bea
\label{covariant}
D_\mu C_I &=& \pa_\mu C_I + i A_\mu C_I - i C_I \hat{A}_\mu,
\quad \quad 
D_\mu \bar{C}^I = \pa_\mu \bar{C}^I - i \bar{C}^I A_\mu + i \hat{A}_\mu \bar{C}^I
\non \\
D_\mu \bar{\Psi}^I  &=& \pa_\mu \bar{\Psi}^I + i A_\mu \bar{\Psi}^I - i \bar{\Psi}^I \hat{A}_\mu,
\quad  \quad
D_\mu \Psi_I = \pa_\mu \Psi_I - i \Psi_I A_\mu + i \hat{A}_\mu \Psi_I  
\eea

\subsection{Color conventions}\label{sec:Casimir}

The U$(N)$  generators are defined as $T^A = (T^0, T^a)$, where $T^0 =
\frac{1}{\sqrt{N}}\mathds{1}$ and $T^a$ ($a=1,\ldots, N^2-1$) are an orthonormal  set of traceless
$N\times N$ hermitian matrices.  The generators are normalized as
\be
\Tr( T^A T^B )= \delta^{AB}
\ee 
The structure constant are then defined by 
\be
[T^A,T^B]=i f^{AB}{}_C T^C
\ee
We perform computations associating to every generator $T^A$ in the given representation $R$ of $U(N_1)$ a matrix $R_{i}^{\phantom{i}j}$ with indices $i,j=1\dots N_1$ with commutation relation
\begin{equation}
[ R_{i_1}^{\phantom{i_1}i_2},\, R_{i_3}^{\phantom{i_1}i_4} ] = \delta_{i_3}^{\phantom{i_1}i_2}\, R_{i_1}^{\phantom{i_1}i_4} - \delta_{i_1}^{\phantom{i_1}i_4}\, R_{i_3}^{\phantom{i_1}i_2}
\end{equation}
in such a way that the most generic Casimir invariant reads
\begin{equation}\label{eq:Rcasimir}
R_{i_1}^{\phantom{i_1}i_2}\, R_{i_2}^{\phantom{i_1}i_3}\dots R_{i_p}^{\phantom{i_1}i_1} = C_p(R)\, 1_{{\rm dim}(R)}
\end{equation}
For the rank $n$ totally symmetric and totally antisymmetric representations they evaluate
\begin{align}\label{eq:casimir}
& C_{p}(S_n) = n (N_1+n-1)^{p-1}\nonumber\\
& C_{p}(A_n) = n (N_1-n+1)^{p-1}
\end{align}
For Hook representations with a total of $m$ boxes and $m-s$ boxes in the first row the quadratic Casimir invariants read 
\begin{equation}\label{eq:casimirh}
C_2(m,m-s) = m N_1 + m (m - 2 s - 1)
\end{equation}

\section{Feynman rules}\label{app:FeynmanRules}

We use the Fourier transform definition
\begin{equation}\label{eq:Fourier}
\int \frac{d^{3-2\e}p}{(2 \p)^{3-2\e}} \frac{p^{\m}}{(p^2)^s} e^{i p \cdot (x-y)} =  \frac{\G(\frac{3}{2}-s-\e)}{4^s \pi^{3/2-\e}\Gamma(s)} \big(-i \partial^{\m}_x \big)\frac{1}{(x-y)^{2(3/2-s-\e)}}
\end{equation}
In euclidean space we define the functional generator as $Z \sim \int e^{-S}$, with action  \eqref{action}.  This gives rise to the following Feynman rules
\begin{itemize}
\item Vector propagators in Landau gauge
\bea
\label{treevector}
 \langle (A_\mu)_i{}^j (x) (A_\nu)_{k}{}^\ell(y) \rangle^{(0)} &=&  \d^{\ell}_i \delta_k^j  \, \left( \frac{2\pi i}{k} \right) \frac{\G(\frac32-\e)}{2\pi^{\frac32 -\e}} \varepsilon_{\mu\nu\rho} \frac{(x-y)^\rho}{[(x-y)^2]^{\frac32 -\e} }
\non \\
&=&  \d^{\ell}_i \delta_k^j   \left( \frac{2\pi }{k} \right) \varepsilon_{\mu\nu\rho} \, \int \frac{d^np}{(2\pi)^n}   \frac{p^\rho}{p^2} e^{ip(x-y)}
\non \\
\non \\
 \langle (\hat{A}_\mu)_{\hat{i}}{}^{\hat{j}} (x) ( \hat{A}_\nu)_{\hat k}{}^{\hat\ell }(y) \rangle^{(0)} &=&  - \d^{\hat\ell}_{\hat i} \delta_{\hat k}^{\hat j}     \, \left( \frac{2\pi i}{k} \right) \frac{\G(\frac32-\e)}{2\pi^{\frac32 -\e}} \varepsilon_{\mu\nu\rho} \frac{(x-y)^\rho}{[(x-y)^2]^{\frac32 -\e} }
\non \\
&=& -\d^{\hat\ell}_{\hat i} \delta_{\hat k}^{\hat j} \left( \frac{2\pi }{k} \right) \varepsilon_{\mu\nu\rho} \, \int \frac{d^np}{(2\pi)^n}   \frac{p^\rho}{p^2} e^{ip(x-y)}
\eea
\item Scalar propagator
\bea
\label{scalar}
\langle (C_I)_i{}^{\hat{j}} (x) (\bar{C}^J)_{\hat{k}}{}^l(\; y) \rangle^{(0)}  &=& \d_I^J \d_i^l \d_{\hat{k}}^{\hat{j}} \, \frac{\G(\frac12 -\e)}{4\pi^{\frac32-\e}} 
\, \frac{1}{[(x-y)^2]^{\frac12 -\e}}
\non \\
&=& \d_I^J \d_i^l \d_{\hat{k}}^{\hat{j}} \, \int \frac{d^np}{(2\pi)^n}   \frac{e^{ip(x-y)} }{p^2}  
\eea
\item Fermion propagator
\bea
\label{treefermion}
\langle (\psi_I^\a)_{\hat{i}}{}^{ j}  (x) (\bar{\psi}^J_\b )_k{}^{ \hat{l}}(y) \rangle^{(0)} &=&  i \, \d_I^J \d_{\hat{i}}^{\hat{l}} \d_{k}^{j} \, 
\frac{\G(\frac32 - \e)}{2\pi^{\frac32 -\e}} \,  \frac{(\g^\mu)^\a_{\; \, \b} \,  (x-y)_\mu}{[(x-y)^2]^{\frac32 - \e}}
\non \\
&=& \d_I^J \d_{\hat{i}}^{\hat{l}} \d_{k}^{j} \, \int \frac{d^np}{(2\pi)^n}   \frac{(\g^\mu)^\a_{\; \, \b} \, p_\mu  }{p^2}  e^{ip(x-y)}
\eea
\item Gauge cubic vertex
\beq
\label{gaugecubic}
i \frac{k}{12\pi} \varepsilon^{\mu\nu\rho} \int d^3x \, f^{abc} A_\mu^a A_\nu^b A_\rho^c
\eeq
\item Gauge--fermion cubic vertex
\beq
\label{gaugefermion}
-\int d^3x \, \Tr \Big[ \bar{\Psi}^I \g^\mu \Psi_I A_\mu - \bar{\Psi}^I \g^\mu \hat{A}_\mu \Psi_I  \Big]
\eeq 
\end{itemize}

\noindent
The one loop gauge propagators are given by
\begin{subequations}
\begin{align}
\label{1vector}
 \langle (A_\mu)_i{}^j (x) &(A_\nu)_{k}{}^\ell(y) \rangle^{(1)} =  \d^{\ell}_i \delta_k^j    \left( \frac{2\pi }{k} \right)^2 N_2 \, \frac{\G^2(\frac12-\e)}{4\pi^{3 -2\e}} 
\left[ \frac{\d_{\mu\nu}}{ [(x- y)^2]^{1-2\e}} - \pa_\mu \pa_\nu \frac{[(x-y)^2]^{2\e}}{4\e(1+2\e)} \right]  
\non  \\
  =&   \d^{\ell}_i \delta_k^j    \left( \frac{2\pi }{k} \right)^2 N_2 \, \frac{\G^2(\frac12-\e)\G(\frac12 +\e)}{\G(1-2\e) 2^{1-2\e} \pi^{\frac32 -\e}} 
\, \int \frac{d^np}{(2\pi)^n}   \frac{e^{ip(x-y)}}{(p^2)^{\frac12 +\e}}  \left( \d_{\mu \n} - \frac{p_\mu p_\nu}{p^2} \right) 
\\
& \non \\
 \langle (\hat{A}_\mu)_{\hat{i}}{}^{\hat{j}} (x) &( \hat{A}_\nu)_{\hat k}{}^{\hat\ell }(y) \rangle^{(1)} = \d^{\hat\ell}_{\hat i} \delta_{\hat k}^{\hat j}     \left( \frac{2\pi }{k} \right)^2 N_1 \, \frac{\G^2(\frac12-\e)}{4\pi^{3 -2\e}} 
\left[ \frac{\d_{\mu\nu}}{ [(x- y)^2]^{1-2\e}} - \pa_\mu \pa_\nu \frac{[(x-y)^2]^{2\e}}{4\e(1+2\e)} \right] \non \\
 =& \d^{\hat\ell}_{\hat i} \delta_{\hat k}^{\hat j}   \left( \frac{2\pi }{k} \right)^2 N_1 \, \frac{\G^2(\frac12-\e)\G(\frac12 +\e)}{\G(1-2\e) 2^{1-2\e} \pi^{\frac32 -\e}} 
\, \int \frac{d^np}{(2\pi)^n}   \frac{e^{ip(x-y)}}{(p^2)^{\frac12 +\e}} \left( \d_{\mu \n} - \frac{p_\mu p_\nu}{p^2} \right) 
\end{align}
\end{subequations}
The one--loop fermion propagator reads 
\bea
\label{1fermion}
&& \langle (\psi_I^\a)_{\hat{i}}{}^{\; j}  (x) (\bar{\psi}^J_\b)_k{}^{\; \hat{l}}(y) \rangle^{(1)} =  -i \,\left( \frac{2\pi}{k} \right) \,  \d_I^J \d_{\hat{i}}^{\hat{l}} \d_{k}^{j} \,  \, \d^\a_{\; \, \b}
\, (N_1-N_2) \frac{\G^2(\frac12 - \e)}{16 \pi^{3-2\e}} \, \frac{1}{[(x-y)^2]^{1 - 2\e}}
\non \\
&~& \qquad= - \left( \frac{2\pi i}{k} \right) \, \d_I^J \d_{\hat{i}}^{\hat{l}} \d_{k}^{j} \,  \, \d^\a_{\; \, \b} \, (N_1-N_2) \frac{\G^2(\frac12 - \e) \G(\frac12 + \e)}{\G(1-2\e) 2^{3-2\e} \pi^{\frac32 -\e}} \, 
\int \frac{d^np}{(2\pi)^n}   \frac{e^{ip(x-y)} }{(p^2)^{\frac12 +\e}}
\eea 
Being proportional to the difference $(N_1-N_2)$, it vanishes in the ABJM limit.

\subsection{Gauge two-point function at two loops}\label{sec:two-point}

In Table \ref{tab} we list the non--vanishing diagrams contributing to the gluon two--loop self--energy and their properties.
The color factors are relative to the diagram already inserted into the Wilson loop.
The values up to order $\epsilon^0$ are for the two-loop diagram only and a factor $e^{2\gamma_E \epsilon}$ is understood.
\begin{table}
\begin{tabular}{cccc}
 & & color & ${\cal O}(\epsilon^0)$ \\[5mm]
(a) & \raisebox{-0.5cm}{\includegraphics[scale=0.5]{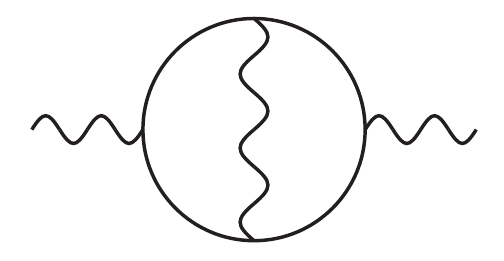}} & $N_2 \left(N_2 C_2(R)-C_1^2(R)\right)$ & $4 \pi  \left(\pi ^2-8\right)-\frac{16 \pi }{3 \epsilon }$ \\[5mm]
(b) & \raisebox{-0.5cm}{\includegraphics[scale=0.5]{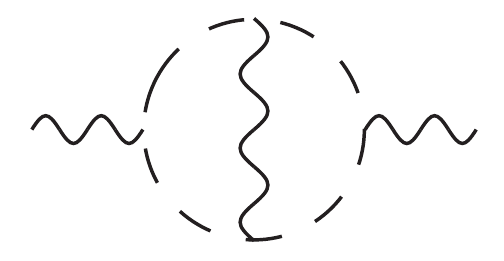}} & $N_2 \left(N_2 C_2(R)-C_1^2(R)\right)$ & $4 \pi  \left(\pi ^2-8\right)-\frac{16 \pi }{3 \epsilon }$ \\[5mm]
(c) & \raisebox{-0.5cm}{\includegraphics[scale=0.5]{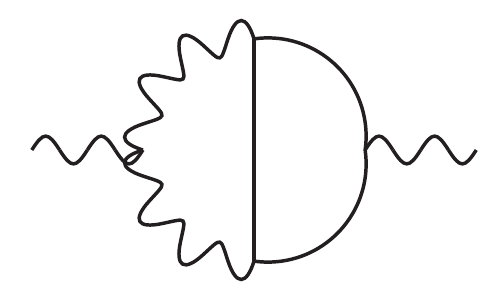}} & $N_2 \left(N_1 C_2(R)-C_1^2(R)\right)$ & $\frac{16 \pi }{3 \epsilon }-4 \pi  \left(\pi ^2-8\right)$ \\[5mm]
(d) & \raisebox{-0.5cm}{\includegraphics[scale=0.5]{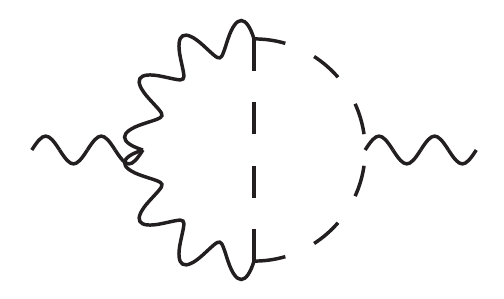}} & $N_2 \left(N_1 C_2(R)-C_1^2(R)\right)$ & $2 \pi  \left(\pi ^2-8\right)-\frac{8 \pi }{3 \epsilon }$ \\[5mm]
(e) & \raisebox{-0.5cm}{\includegraphics[scale=0.5]{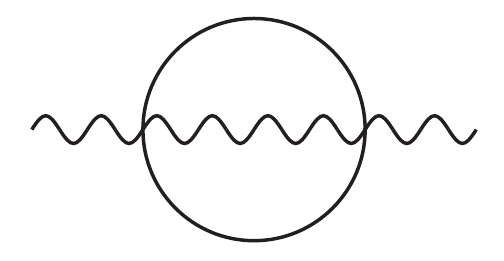}} & $2N_2 \left(-\left(N_1-2 N_2\right) C_2(R)-C_1^2(R)\right)$ & $\frac{8 \pi }{3 \epsilon }+16 \pi$  \\[5mm]
(f) & \raisebox{-0.5cm}{\includegraphics[scale=0.5]{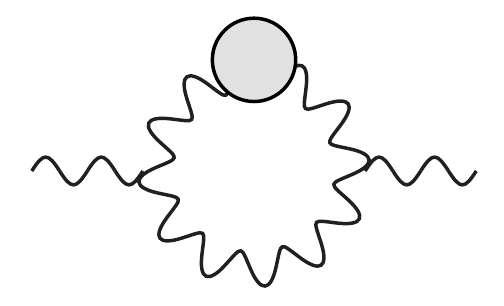}} & $N_2 \left(N_1 C_2(R)-C_1^2(R)\right)$ & $\frac{16 \pi }{3 \epsilon }+16 \pi$  \\[5mm]
(g) & \raisebox{-0.5cm}{\includegraphics[scale=0.5]{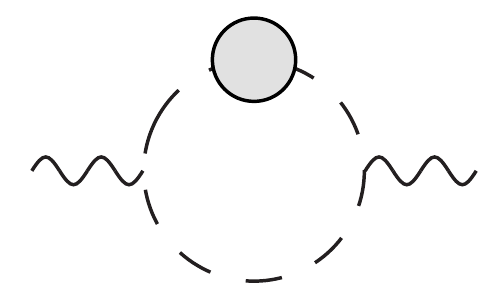}} & $N_2 \left(N_2-N_1\right) C_2(R)$ & $\frac{16 \pi }{3 \epsilon }+32 \pi$  \\[5mm]
(h) & \raisebox{-0.2cm}{\includegraphics[scale=0.5]{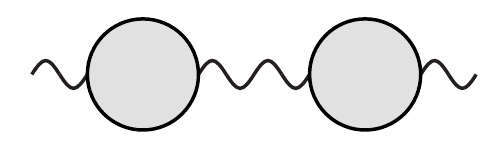}} & $N_2 \left(N_2 C_2(R)-C_1^2(R)\right)$ & $-16 \pi ^3$
\end{tabular}
\caption{Table of non--vanishing self--energy diagrams with matter contributions.}\label{tab}
\end{table}
 
\subsection{Couplings to scalars}\label{sec:traces} 

Up to three loop order the computation of the latitude expectation value involves the following traces of $M$ matrices
\begin{align}
& \Tr M(\tau) = 0\\
&\Tr\left(M(\tau_1)M(\tau_2)\right) = 4 \left(1 + (\nu ^2 -1) \sin ^2\frac{\tau _1-\tau _2}{2} \right) \label{eq:m2trace}\\
&\Tr\left(M(\tau_1)M(\tau_2)M(\tau_3)\right) = -2 i \nu  \left(\nu ^2-1\right) \left( \sin \left(\tau _1-\tau _2\right) - \sin \left(\tau _1-\tau _3\right) +  \sin\left(\tau _2-\tau _3\right) \right) \label{eq:m3trace}
\end{align}
 
\section{Weak coupling expansion of the  un--deformed matrix model}\label{sec:MMexp}

In this section we expand the ABJ(M) matrix model
\begin{align}
\label{eq:matrix}
Z = &\int \prod_{a=1}^{N_1}d\lambda _{a} \ e^{i\pi k\lambda_{a}^{2}}\prod_{b=1}^{N_2}d\mu_{b} \ e^{-i\pi k\mu_{b}^{2}}\,  \frac{\displaystyle\prod_{a<b}^{N_1}\sinh ^{2}\pi (\lambda_{a}-\lambda _{b})\prod_{a<b}^{N_2}\sinh ^{2}\pi (\mu_{a}-\mu_{b})}{\displaystyle\prod_{a=1}^{N_1}\prod_{b=1}^{N_2}\cosh ^{2} \pi (\lambda _{a}-\mu_{b})}
\end{align}
at weak coupling and compute the expectation value of 1/6--BPS Wilson loops with $m$ windings up to the fourth order, for generic $N_1$ and $N_2$.
This is performed by observing that every matrix model correlator with a total power of $2n$ eigenvalues scales as $k^{-n}$. Therefore one can expand in power series the hyperbolic functions in the integrand and the exponential accounting for the operator insertion. This boils down to computing correlators in a Gaussian matrix model and for the purpose of the present expansion those listed in the appendices of \cite{Bianchi:2017afp} are sufficient.

We evaluate the Wilson loop in higher rank totally symmetric and antisymmetric representations of $U(N_1)$ using prescriptions \eqref{eq:higherrep}.
For the three lowest rank representations, up to four loops, we find for instance
\begin{align}\label{eq:Wn}
\langle W_B^m \rangle(k,S_1) =& 1+\frac{i \pi  m^2 N_1}{k}-\frac{\pi ^2 m^2 \left(m^2 \left(2 N_1^2+1\right)+2 N_1^2-6 N_1 N_2-2\right)}{6 k^2}\nonumber\\&
-\frac{i \pi ^3 m^2}{18 k^3} \left(m^4 \left(N_1^3+2 N_1\right)+m^2 \left(4 N_1^3-12 N_2 N_1^2-4 N_1-6 N_2\right) \right.\nonumber\\&\left.+N_1^3+9 N_1 N_2^2-N_1-6 N_1^2 N_2-3 N_2\right)\nonumber\\&
+\frac{\pi ^4 m^2}{360 k^4} \left(m^6 \left(2 N_1^4+10 N_1^2+3\right)+20 m^4 \left(N_1^4-3 N_2 N_1^3-6 N_2 N_1-1\right) \right.\nonumber\\&+2 m^2 \left(13 N_1^4-75 N_2 N_1^3+5 \left(24 N_2^2-5\right) N_1^2+15 N_2 N_1 +60 N_2^2+12\right)\nonumber\\&\left.-60 N_2 \left(N_2 N_1^2+\left(N_2^2-1\right) N_1-N_2\right)\right) +{\cal O}\left(k^{-5}\right)
\end{align}
\begin{align}\label{eq:WnS2}
& \langle W_B^m \rangle(k,S_2) = 1+\frac{2 i \pi  m^2 \left(N_1+1\right)}{k}\nonumber\\&
-\frac{\pi ^2 m^2 \left(m^2 \left(5 N_1^2+11 N_1+8\right)+2 \left(N_1^2-3 N_2 N_1+N_1-3 N_2-2\right)\right)}{3 k^2}\nonumber\\&
-\frac{i \pi ^3 m^2}{9 k^3} \left(m^4 \left(7 N_1^3+25 N_1^2+40 N_1+24\right)+2 m^2 \left(5 N_1^3+\left(11-15 N_2\right) N_1^2
\right.\right.\nonumber\\&\left.
-\left(33 N_2+4\right) N_1-12 \left(2 N_2+1\right)\right)+N_1^3+N_1^2 \left(1-6 N_2\right)+3 N_2 \left(3 N_2-2\right)
\nonumber\\&\left.
+N_1 \left(9 N_2^2-6 N_2-2\right)\right)\nonumber\\&
+\frac{1}{90 k^4}\pi ^4 m^2 \left(m^6 \left(21 N_1^4+107 N_1^3+278 N_1^2+362 N_1+192\right)
\right.\nonumber\\&\left.
+10 m^4 \left(7 N_1^4+\left(25-21 N_2\right) N_1^3+\left(20-75 N_2\right) N_1^2-20 \left(6 N_2+1\right) N_1-8 \left(9 N_2+4\right)\right)
\right.\nonumber\\&
+m^2 \left(33 N_1^4+\left(71-195 N_2\right) N_1^3+\left(300 N_2^2-405 N_2-76\right) N_1^2
\right.\nonumber\\&\left.
+2 \left(330 N_2^2-45 N_2-62\right) N_1+6 \left(80 N_2^2+35 N_2+16\right)\right)
\nonumber\\&\left.
-30 N_2 \left(\left(N_1+1\right) N_2^2+\left(N_1^2+N_1-2\right) N_2-2\right)\right) +{\cal O}\left(k^{-5}\right)
\end{align}
\begin{align}\label{eq:WnS3}
&\langle W_B^m \rangle(k,S_3) = 1+\frac{3 i \pi  m^2 \left(N_1+2\right)}{k}
\nonumber\\&
-\frac{\pi ^2 m^2 \left(8 m^2 N_1^2+34 m^2 N_1+39 m^2+2 N_1^2+4 N_1-6 N_1 N_2-12 N_2-6\right)}{2 k^2}
\nonumber\\&
-\frac{1}{6 k^3}i \pi ^3 m^2 \left(m^4 \left(19 N_1^3+128 N_1^2+312 N_1+270\right)+2 m^2 \left(8 N_1^3+\left(34-24 N_2\right) N_1^2
\right.\right.\nonumber\\&\left.\left.
-6 \left(17 N_2-2\right) N_1-9 \left(13 N_2+6\right)\right)+N_1^3+N_1^2 \left(2-6 N_2\right)+9 N_2 \left(2 N_2-1\right)
\right.\nonumber\\&\left.
+3 N_1 \left(3 N_2^2-4 N_2-1\right)\right)
\nonumber\\&
+\frac{1}{120 k^4}\pi ^4 m^2 \left(m^6 \left(202 N_1^4+1908 N_1^3+7370 N_1^2+13524 N_1+9801\right)
\right.\nonumber\\&
+20 m^4 \left(19 N_1^4+\left(128-57 N_2\right) N_1^3-48 \left(8 N_2-5\right) N_1^2-36 \left(26 N_2+1\right) N_1
\right.\nonumber\\&\left.
-27 \left(30 N_2+13\right)\right)
\nonumber\\&
+2 m^2 \left(53 N_1^4+\left(222-315 N_2\right) N_1^3+5 \left(96 N_2^2-258 N_2-7\right) N_1^2
\right.\nonumber\\&\left.
+3 \left(680 N_2^2-325 N_2-188\right) N_1+12 \left(195 N_2^2+80 N_2+27\right)\right)
\nonumber\\&\left.
-60 N_2 \left(N_2 N_1^2+\left(N_2+1\right){}^2 N_1+2 N_2^2-3 N_2-4\right)\right) +{\cal O}\left(k^{-5}\right)
\end{align}
\begin{align}\label{eq:WnA2}
& \langle W_B^m \rangle(k,A_2) = 1+\frac{2 i \pi  m^2 \left(N_1-1\right)}{k}
\nonumber\\&
-\frac{\pi ^2 m^2 \left(m^2 \left(5 N_1^2-11 N_1+8\right)+2 \left(N_1^2-\left(3 N_2+1\right) N_1+3 N_2-2\right)\right)}{3 k^2}
\nonumber\\&
-\frac{i \pi ^3 m^2}{9 k^3} \left(m^4 \left(7 N_1^3-25 N_1^2+40 N_1-24\right)
\right.\nonumber\\&
+2 m^2 \left(5 N_1^3-\left(15 N_2+11\right) N_1^2+\left(33 N_2-4\right) N_1-24 N_2+12\right)+N_1^3-3 N_2 \left(3 N_2+2\right)
\nonumber\\&\left.
-N_1^2 \left(6 N_2+1\right)+N_1 \left(9 N_2^2+6 N_2-2\right)\right)
\nonumber\\&
+\frac{1}{90 k^4}\pi ^4 m^2 \left(m^6 \left(21 N_1^4-107 N_1^3+278 N_1^2-362 N_1+192\right)
\right.\nonumber\\&
+10 m^4 \left(7 N_1^4-\left(21 N_2+25\right) N_1^3+5 \left(15 N_2+4\right) N_1^2-20 \left(6 N_2-1\right) N_1+8 \left(9 N_2-4\right)\right)
\nonumber\\&
+m^2 \left(33 N_1^4-\left(195 N_2+71\right) N_1^3+\left(300 N_2^2+405 N_2-76\right) N_1^2
\right.\nonumber\\&\left.
-2 \left(330 N_2^2+45 N_2-62\right) N_1+6 \left(80 N_2^2-35 N_2+16\right)\right)
\nonumber\\&\left.
-30 N_2 \left(\left(N_1-1\right) N_2^2
+\left(N_1^2-N_1-2\right) N_2+2\right)\right) +{\cal O}\left(k^{-5}\right)
\end{align}
\begin{align}\label{eq:WnA3}
& \langle W_B^m \rangle(k,A_3) = 1+\frac{3 i \pi  m^2 \left(N_1-2\right)}{k}
\nonumber\\&
-\frac{\pi ^2 m^2 \left(m^2 \left(8 N_1^2-34 N_1+39\right)+2 \left(N_1^2-\left(3 N_2+2\right) N_1+6 N_2-3\right)\right)}{2 k^2}
\nonumber\\&
-\frac{1}{6 k^3}i \pi ^3 m^2 \left(m^4 \left(19 N_1^3-128 N_1^2+312 N_1-270\right)
\right.\nonumber\\&
+2 m^2 \left(8 N_1^3-2 \left(12 N_2+17\right) N_1^2+6 \left(17 N_2+2\right) N_1-117 N_2+54\right)
\nonumber\\&\left.
+N_1^3-9 N_2 \left(2 N_2+1\right)-2 N_1^2 \left(3 N_2+1\right)+3 N_1 \left(3 N_2^2+4 N_2-1\right)\right)
\nonumber\\&
+\frac{1}{120 k^4}\pi ^4 m^2 \left(m^6 \left(202 N_1^4-1908 N_1^3+7370 N_1^2-13524 N_1+9801\right)
\right.\nonumber\\&
+20 m^4 \left(19 N_1^4-\left(57 N_2+128\right) N_1^3+48 \left(8 N_2+5\right) N_1^2
\right.\nonumber\\&\left.
+\left(36-936 N_2\right) N_1+810 N_2-351\right)+2 m^2 \left(53 N_1^4-3 \left(105 N_2+74\right) N_1^3
\right.\nonumber\\&\left.
+5 \left(96 N_2^2+258 N_2-7\right) N_1^2-3 \left(680 N_2^2+325 N_2-188\right) N_1
\right.\nonumber\\&\left.\left.
+12 \left(195 N_2^2-80 N_2+27\right)\right)-60 N_2 \left(N_2 N_1^2+\left(N_2-1\right){}^2 N_1-2 N_2^2-3 N_2+4\right)\right)
\nonumber\\&
 +{\cal O}\left(k^{-5}\right)
\end{align}
We can compare these results with the general three--loop expression derived from \eqref{eq:Wlatgeneric} by setting $\nu=1$, and framing $f=1$ as required by comparison with localization predictions
\begin{align}\label{eq:result}
\langle W_B^{m}(1) \rangle_1 &= 1+\frac{i \pi  m^2 C_2(R)}{k}\nonumber\\&
+\frac{\pi ^2 m^2}{6 k^2} \left(C_2(R) \left(-3 m^2 C_2(R)+\left(m^2-2\right) N_1+6 N_2\right)-\left(m^2-2\right) C_1^2(R)\right)\nonumber\\&
+\frac{i \pi ^3 m^2}{18 k^3} \left[-3 m^4 C_2^3(R)+3 m^2 C_2^2(R) \left(\left(m^2-2\right) N_1+6 N_2\right)
\right.\nonumber\\&
+C_1^2(R) \left(\left(m^2-1\right)^2 N_1+3 \left(2 m^2+1\right) N_2\right) \nonumber\\&\left.-C_2(R) \left(3 m^2 \left(m^2-2\right) C_1^2(R)+\left(\left(m^2-1\right) N_1+3 N_2\right)^2\right)\right] + {\cal O}\left(k^{-4}\right)
\end{align}
Selecting $R=S_1$, $S_2$, $S_3$, $A_2$ and $A_3$ and using relations \eqref{eq:casimir} we find perfect agreement with \eqref{eq:Wn}--\eqref{eq:WnA3}.

\section{\texorpdfstring{$B^\theta_{1/6}$}~  at four loops for generic representations}\label{sect:4loops}

In this section we provide the details of the computation of the $\theta$--Bremsstrahlung function up to four loops, for generic representations of the $U(N_1)$ gauge group.
Most of the computation has already been addressed in \cite{Bianchi:2017afp,Bianchi:2017ujp}, to which we refer for a more complete discussion.  There the cusp has been evaluated for the fundamental representation of the gauge group. Here we extend it to general representations.
This is aimed at verifying that the conjecture for the exact value of $B_{1/6}^{\theta}$ (cfr.~equation \eqref{eq:relationbremsstrahlung})
\begin{equation}\label{eq:B16conjecture}
B_{1/6}^{\theta} = \frac{1}{8\pi^2}\, \partial_m\, |\langle W_B^m(1) \rangle|\, \bigg|_{m=1}
\end{equation}
put forward in \cite{Bianchi:2017afp,Bianchi:2017ujp} for operators in the fundamental representation, actually holds for any representation of the gauge group.

We compute the $\theta$--Bremsstrahlung function from a direct evaluation of the cusped 1/6--BPS Wilson loop using the definition \eqref{eq:Bbos}, setting $\varphi=0$. This  amounts to computing the operator along a straight line, but at a non--trivial internal $\th$ angle. 

\begin{figure}[h]
\centering
\includegraphics[scale=0.325]{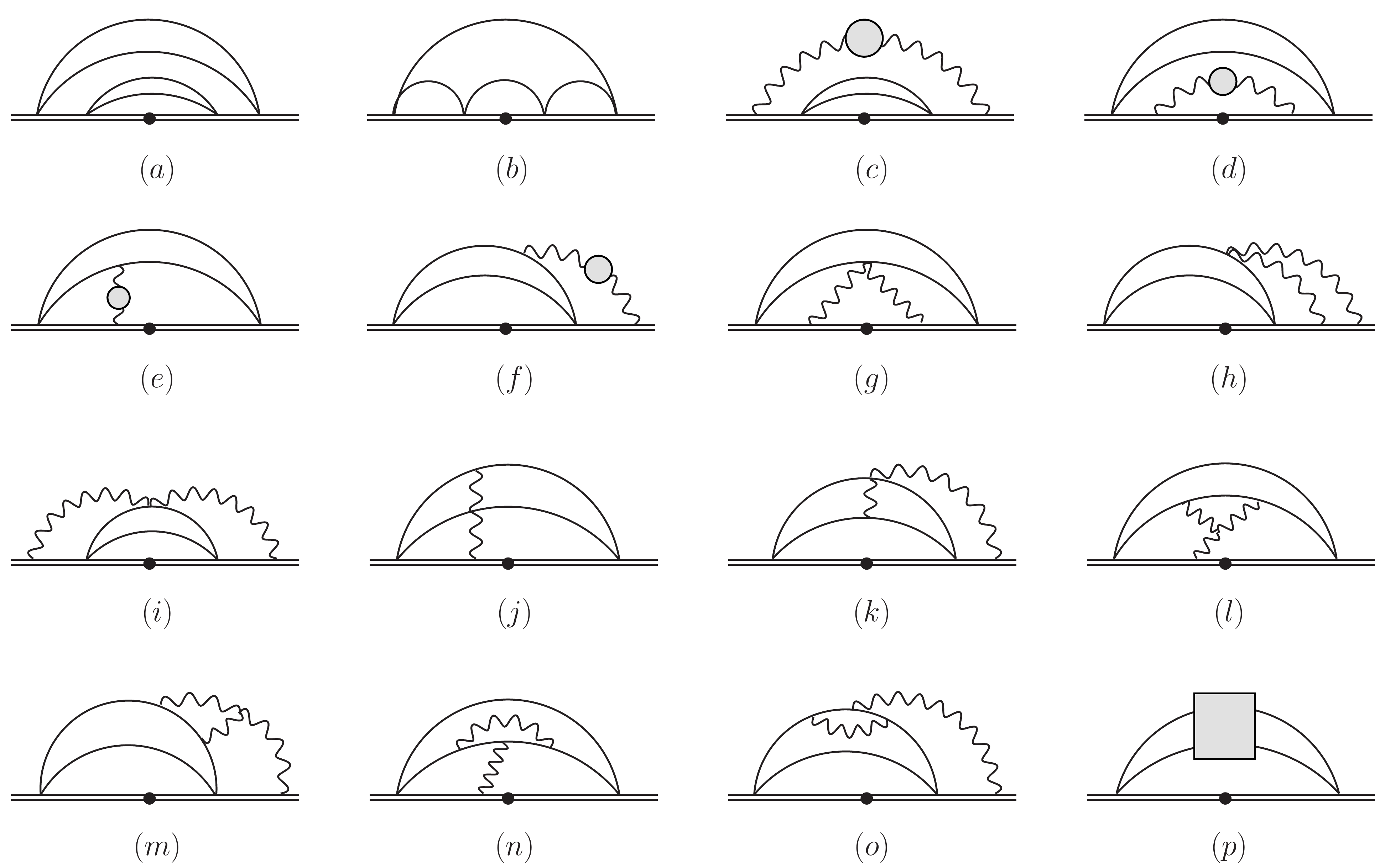}
\caption{List of planar diagrams contributing to the four--loop $\theta$--Bremsstrahlung function. Gray bullets stand for one--loop corrections to the gauge propagator. The gray box collects the two--loop corrections to the bi--scalar two--point function. Here solid lines stand for scalars and dashed lines for fermions.} \label{fig:4loops}
\end{figure}

We focus only on graphs with an explicit dependence on $\th$. These are depicted in Figures \ref{fig:4loops} and \ref{fig:4loopsNP}.
After evaluating the algebra of these diagrams using the Feynman rules in Appendix \ref{AppA} and expressing color factors in terms of Casimir invariants using \eqref{eq:Rcasimir}, we perform an integration--by--parts reduction of the corresponding Feynman integrals. The relevant master integrals have been evaluated in \cite{Bianchi:2017afp,Bianchi:2017ujp} up to the required order in $\epsilon$.

\begin{figure}[h]
\centering
\includegraphics[scale=0.35]{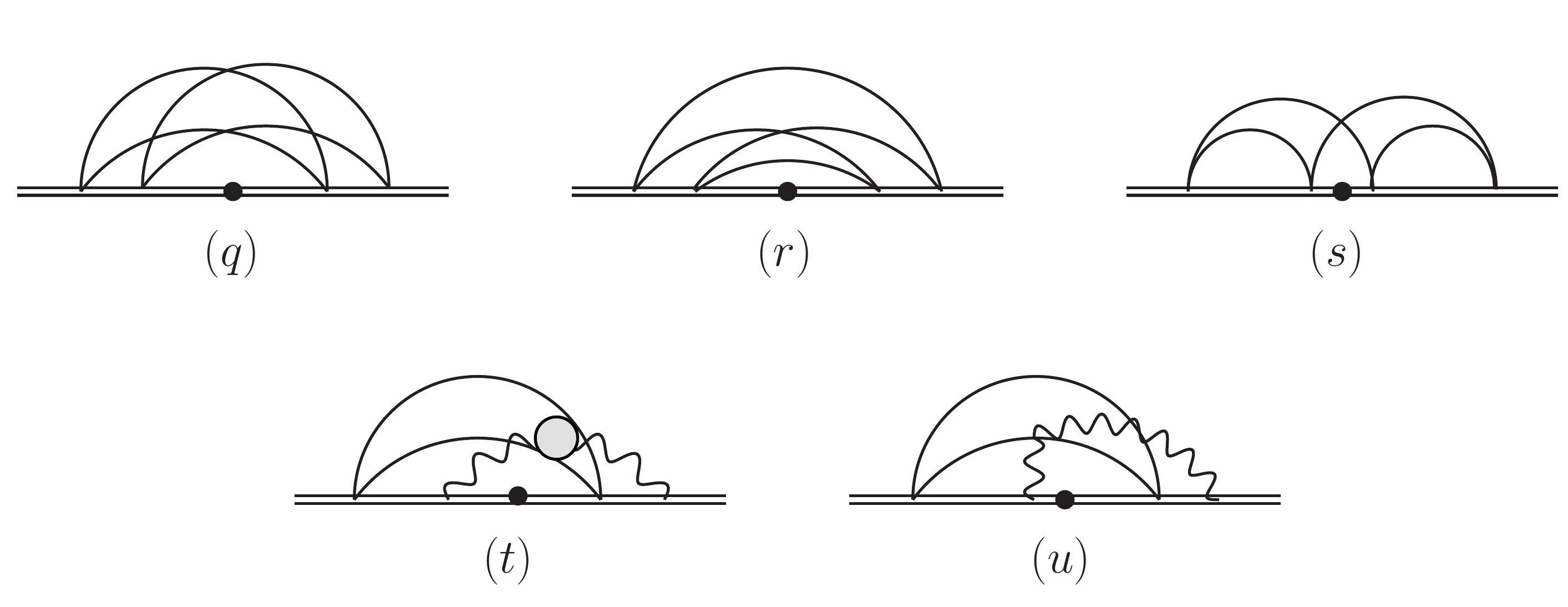}
\caption{List of non--planar diagrams contributing to the four--loop $\theta$--Bremsstrahlung function.} \label{fig:4loopsNP}
\end{figure}

\subsection{Results for the four--loop diagrams}\label{app:diagrams}

Here we report the results of the evaluation of the various diagrams. A common factor $\left(\frac{e^{-4\epsilon\gamma_E}}{k(4\pi)^{d/2}}\right)^4$ is understood.

The planar topologies of Figure \ref{fig:4loops}  yield  
\begin{align}
(a)= & N_2^2 C_2^2(R)\left( \frac{8 \pi ^2 C_{\theta }^2 \left(C_{\theta }^2-2\right)}{\epsilon ^2}+\frac{32 \pi ^2 C_{\theta }^2 \left((2 \log 2-1) C_{\theta }^2-4 \log 2\right)}{\epsilon }\right)+O\left(\epsilon ^0\right) \\
(b)= & N_2 \left(N_1^2 C_2(R)-N_1 \left(C_1^2(R)+2 C_3(R)\right)+2 \left(C_1(R) C_2(R)+C_4(R)\right)\right)\nonumber\\& \left(\frac{4 \pi ^2 C_{\theta }^2}{\epsilon ^3}+\frac{32 \pi ^2 \log 2 C_{\theta }^2}{\epsilon ^2}+\frac{4 \left(13 \pi ^4+96 \pi ^2 \log ^2(2)\right) C_{\theta }^2}{3 \epsilon }\right)+O\left(\epsilon ^0\right) \\
(c)= & N_2^2 C_2^2(R)\left(-\frac{16 \left(\pi ^2 C_{\theta }^2\right)}{\epsilon ^2}+\frac{16 \pi ^2 (7-8 \log 2) C_{\theta }^2}{\epsilon }\right)+O\left(\epsilon ^0\right) \\
(d)= & N_2^2 C_2^2(R)\left(\frac{16 \pi ^2 C_{\theta }^2}{\epsilon ^2}+\frac{16 \pi ^2 (1+8 \log 2) C_{\theta }^2}{\epsilon }\right)+O\left(\epsilon ^0\right) \\
(e)= & N_2^2 \left(C_1^2(R)-N_1 C_2(R)\right)\left(\frac{128 \pi ^2 C_{\theta }^2}{\epsilon }\right)+O\left(\epsilon ^0\right) \\
(f)= & N_2^2 \left(C_1^2(R)-N_1 C_2(R)\right)\left(-\frac{32 \left(\pi ^2 \left(\pi ^2-4\right) C_{\theta }^2\right)}{\epsilon }\right)+O\left(\epsilon ^0\right) \\
(g)= & N_2 \left(N_1^2 C_2(R)-N_1 \left(C_1^2(R)+4 C_3(R)\right)+4 \left(C_1(R) C_2(R)+C_4(R)\right)\right)\nonumber\\&
\left(-\frac{2 \left(\pi ^2 C_{\theta }^2\right)}{\epsilon ^3}-\frac{2 \left(\pi ^2 (8 \log 2-1) C_{\theta }^2\right)}{\epsilon ^2}\right.\nonumber\\&\left.-\frac{4 \left(\pi ^2 \left(-9+7 \pi ^2+12 \log 2 (4 \log 2-1)\right) C_{\theta }^2\right)}{3 \epsilon }\right)+O\left(\epsilon ^0\right) \\
(h)= & N_2 \left(N_1^2 C_2(R)-N_1 \left(C_1^2(R)+4 C_3(R)\right)+4 \left(C_1(R) C_2(R)+C_4(R)\right)\right)\nonumber\\& \left(-\frac{2 \left(\pi ^2C_{\theta }^2\right)}{\epsilon ^3}-\frac{2 \left(\pi ^2 (8 \log 2-1) C_{\theta }^2\right)}{\epsilon ^2}\right.\nonumber\\&\left.-\frac{2 \left(\pi ^2 \left(17 \pi ^2+6 (4 \log 2 (4 \log 2-1)-3)\right) C_{\theta }^2\right)}{3 \epsilon }\right)+O\left(\epsilon ^0\right) \\
(i)= & N_2 \left(N_1^2 C_2(R)-N_1 \left(C_1^2(R)+4 C_3(R)\right)+4 \left(C_1(R) C_2(R)+C_4(R)\right)\right)\nonumber\\&
\left(-\frac{2\pi ^4 C_{\theta }^2}{3 \epsilon }\right)+O\left(\epsilon ^0\right) 
\end{align}
Diagrams $(l)$-$(o)$ cancel pairwise and we have not shown their explicit result. The non-planar diagrams of Figure \ref{fig:4loopsNP} read
\begin{align}
(q)= & N_2^2 \left( C_1^2(R)-N_1 C_2(R) + C_2^2(R) \right) C_{\theta }^2\left(\frac{16 \pi ^2 }{\epsilon ^2}+\frac{32 \pi ^2  \left(C_{\theta }^2+4 \log 2\right)}{\epsilon }\right)+O\left(\epsilon ^0\right) \\
(t)= & N_2^2 \left(C_2(R) \left(C_2(R)-N_1\right)+C_1^2(R)\right) C_{\theta }^2\left(-\frac{32 \pi ^2}{\epsilon ^2}-\frac{32 \pi ^2  (1+8 \log 2)}{\epsilon }\right)+O\left(\epsilon ^0\right) \\
(r)= & N_2 \left(-N_1 C_3(R)+C_1(R) C_2(R)+C_4(R)\right) C_{\theta }^2\nonumber\\&
\left(\frac{4 \pi ^2 }{\epsilon ^3}+\frac{32 \pi ^2 \log 2}{\epsilon ^2}+\frac{4 \left(\pi ^4 \left(8 C_{\theta }^2+3\right)+96 \pi ^2 \log ^2 2\right)}{3 \epsilon }\right)+O\left(\epsilon ^0\right) \\
(s)= & N_2 \left(-N_1 C_3(R)+C_1(R) C_2(R)+C_4(R)\right) C_{\theta }^2\nonumber\\&
\left(\frac{4 \pi ^2 }{\epsilon ^3}+\frac{32 \pi ^2 \log 2}{\epsilon ^2}+\frac{4 \pi ^2 \left(19 \pi ^2+96 \log ^2 2\right)}{3 \epsilon }\right)+O\left(\epsilon ^0\right) \\
(u)= & N_2 \left(-N_1 C_3(R)+C_1(R) C_2(R)+C_4(R)\right) C_{\theta }^2\nonumber\\&
\left(-\frac{16 \left(\pi ^2\right)}{\epsilon ^2}+\frac{32 \pi ^2 \left(\pi ^2-3 (3+4 \log 2)\right)}{3 \epsilon }\right)+O\left(\epsilon ^0\right) 
\end{align}

\subsubsection*{Two-loop scalar propagator corrections}\label{app:prop}

As a by--product of this computation we present here the two--loop corrections to the scalar self--energy, including color subleading corrections.
Subleading corrections arise  from different contractions of the planar topologies of \eqref{eq:2loopscalar}
\begin{align}\label{eq:2loopscalar}
\raisebox{-2.5mm}{\includegraphics[scale=0.3]{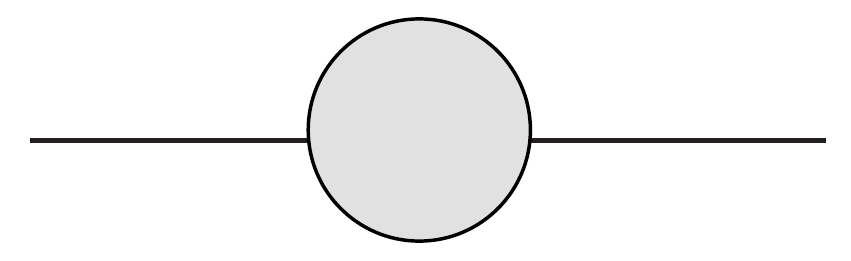}}\, =\, & \raisebox{-2.5mm}{\includegraphics[scale=0.3]{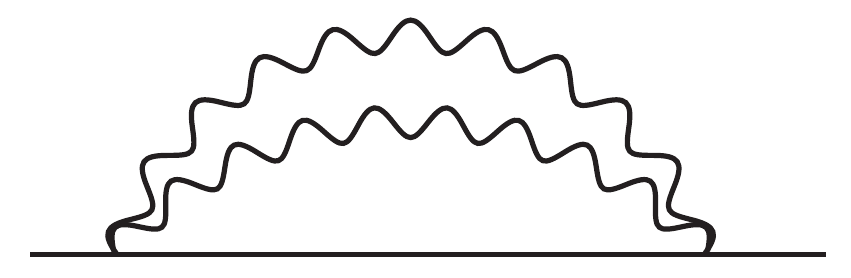}}\, +\, \raisebox{-3.5mm}{\includegraphics[scale=0.3]{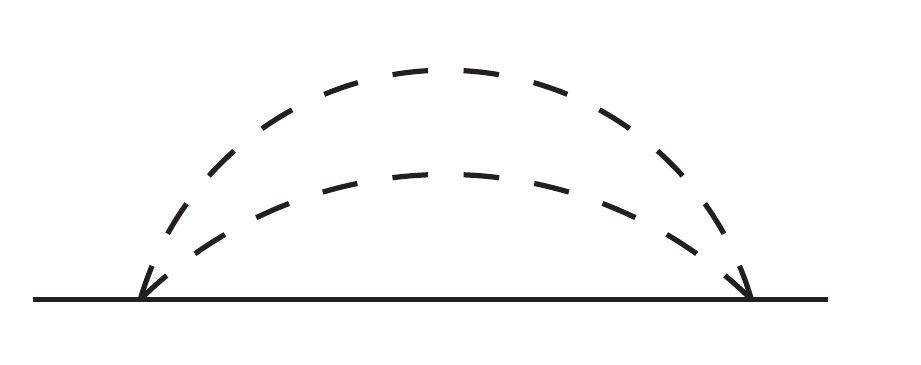}}\, +\, \raisebox{-4mm}{\includegraphics[scale=0.3]{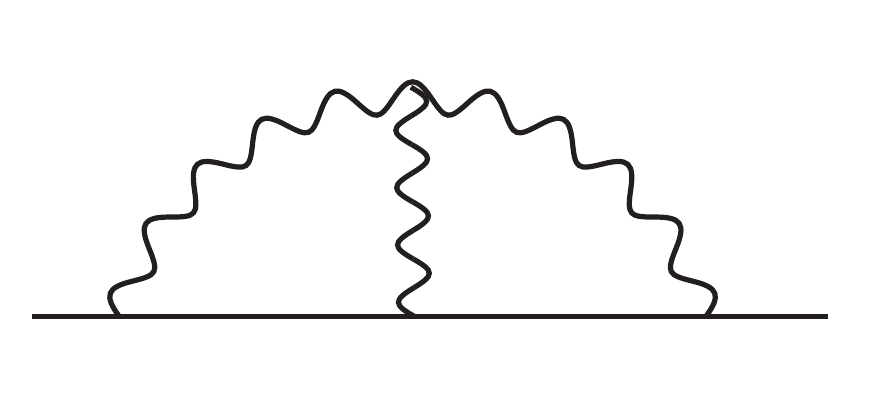}} \nonumber\\&
\raisebox{-6.5mm}{\includegraphics[scale=0.3]{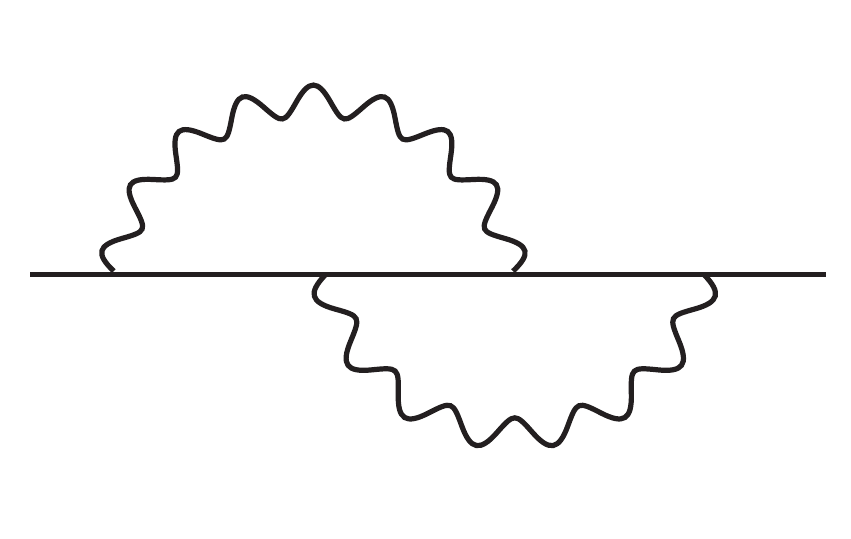}}\, +\, \raisebox{-5mm}{\includegraphics[scale=0.3]{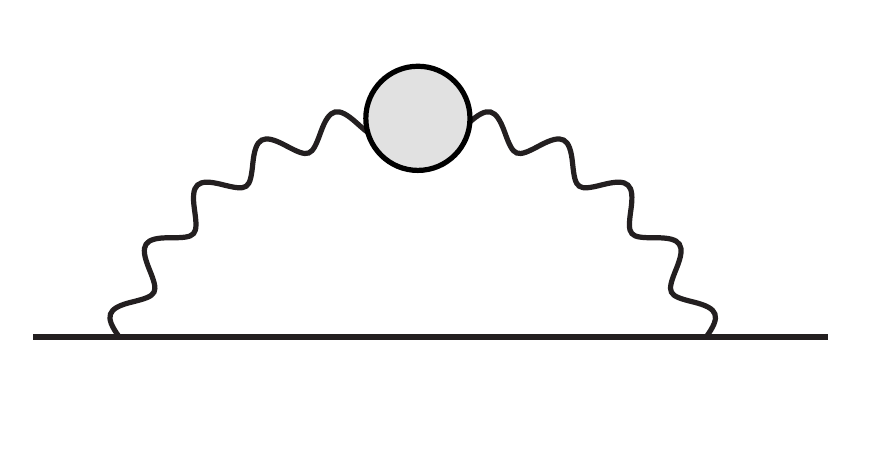}}
\end{align}

\begin{align}
&\raisebox{-2.5mm}{\includegraphics[scale=0.3]{scalarNP1}}\, =\,  C_2(R) N_2 \left(N_1^2-4 N_2 N_1+N_2^2+2\right)\left(\frac{\pi}{3 \epsilon }+2 \pi +O\left(\epsilon ^1\right)\right) \\
&\raisebox{-3.5mm}{\includegraphics[scale=0.3]{scalarNP5}}\,=\,  C_2(R) N_2 \left(N_1 N_2-1\right)\left(-\frac{56 \pi}{3 \epsilon }-112 \pi +O\left(\epsilon ^1\right)\right) \\
&\raisebox{-4mm}{\includegraphics[scale=0.3]{scalarNP4}}\, =\, C_2(R) N_2 \left(N_1^2+N_2^2-2\right)\left(-\frac{4\pi}{3 \epsilon }+\pi  \left(\pi ^2-8\right) +O\left(\epsilon ^1\right)\right) \\
&\raisebox{-6.5mm}{\includegraphics[scale=0.3]{scalarNP2}}\,= \, C_2(R) N_2 \left(N_1 N_2-1\right)\left(-\frac{16 \pi}{3 \epsilon }+4 \pi  \left(\pi ^2-8\right)+O\left(\epsilon ^1\right)\right) \\
&\raisebox{-5mm}{\includegraphics[scale=0.3]{scalarNP3}}\,=\,  C_2(R) N_2 \left(N_1 N_2-1\right)\left(\frac{64 \pi}{3 \epsilon }+64 \pi+O\left(\epsilon ^1\right) \right)
\end{align}
The corresponding contributions to diagram ($p_1$) in Figure \ref{diag_p} are obtained by multiplying these by $8\, B(1+2\epsilon,1)\, I(2,1/2+3\epsilon)$, where a factor of 2 stems from the two scalar propagators, a factor 4 comes from the normalization of HQET integrals and the indices of the bubble integrals are fixed by dimensional analysis.

\subsubsection*{Scalar bubble corrections}\label{app:vertex}

\begin{figure}[h]
\centering
\includegraphics[scale=0.3]{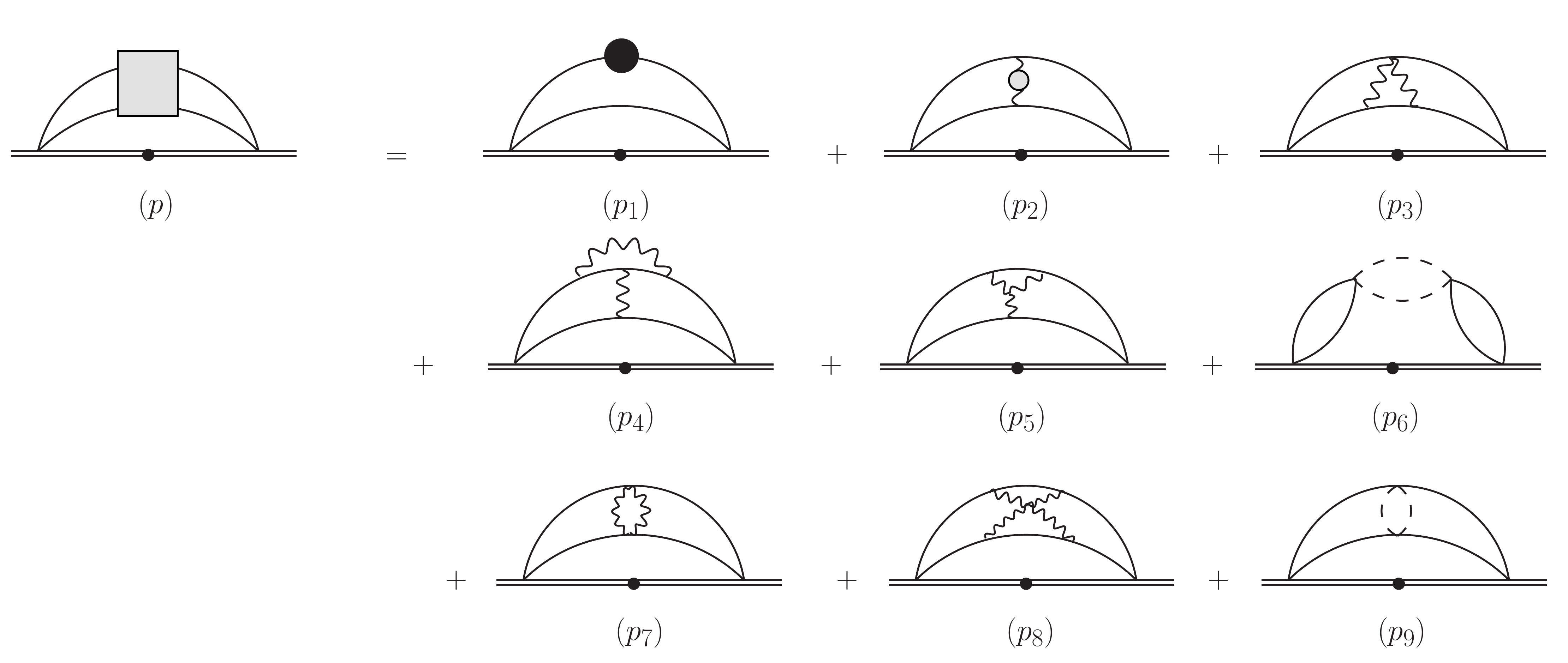}
\caption{Scalar bubble corrections} \label{diag_p}
\end{figure}

Diagram $(p)$ in Figure \ref{fig:4loops} comprises the corrections to the scalar bilinear two--point function. Its non--vanishing  contributions (some possible contractions generate for instance $\Tr M(\tau_{1,2}) = 0$), including color subleading ones are listed in Figure \ref{diag_p}.
In addition, diagram $(p_1)$ involves the two--loop correction to the scalar propagator, which we detailed above. 
Altogether, the various contributions from diagram ($p$) to the cusp expectation value read
\begin{align}
(p_1)= & C_2(R) N_2 \left(
-\frac{4 \left(\pi ^2 \left(N_1^2+4 N_2 N_1+N_2^2-6\right) C_{\theta }^2\right)}{\epsilon ^2} \right.\nonumber\\&
+\frac{4 \pi ^2 C_{\theta }^2}{\epsilon } \left((N_1^2+N_2^2) \left(-6+\pi ^2-8 \log 2\right)+4 N_2 N_1 \left(-22+\pi ^2-8 \log 2\right)\right.\nonumber\\&\left.\left.-6 \pi ^2+100+48 \log 2\right)\right)+O\left(\epsilon ^0\right) \\
(p_2)= & -\frac{16 \left(\pi ^2 \left(\pi ^2-12\right) N_2 \left(N_2+N_1 \left(N_1 N_2-2\right)\right) C_{\theta }^2\right)}{\epsilon }+O\left(\epsilon ^0\right) \\
(p_2)= & -\frac{16 \left(\pi ^2 \left(\pi ^2-12\right) N_2 \left(N_2 C_1^2(R)+\left(N_1 N_2-2\right) C_2(R)\right) C_{\theta }^2\right)}{\epsilon }+O\left(\epsilon ^0\right) \\
(p_3)= & -\frac{4 \left(\pi ^2 \left(\pi ^2-12\right) N_2 \left(\left(N_1-4 N_2\right) C_1^2(R)+\left(N_2^2+2\right) C_2(R)\right) C_{\theta }^2\right)}{\epsilon }+O\left(\epsilon ^0\right) \\
(p_4)= & \frac{16 \pi ^2 \left(\pi ^2-12\right) N_2 \left(N_2 C_1^2(R)+\left(N_1 N_2-2\right) C_2(R)\right) C_{\theta }^2}{3 \epsilon }+O\left(\epsilon ^0\right) \\
(p_5)= & \frac{8 \pi ^2 \left(\pi ^2-12\right) N_2 \left(N_1 C_1^2(R)+\left(N_2^2-2\right) C_2(R)\right) C_{\theta }^2}{3 \epsilon }+O\left(\epsilon ^0\right) \\
(p_6)= & -\frac{8 \left(\pi ^4 N_2 \left(N_2^2 C_2(R)+\left(N_1-2 N_2\right) C_1^2(R)\right) C_{\theta }^2\right)}{\epsilon }+O\left(\epsilon ^0\right) \\
(p_7)= & \frac{4 \pi ^2 N_2 \left(\left(N_1-4 N_2\right) C_1^2(R)+\left(N_2^2+2\right) C_2(R)\right) C_{\theta }^2}{\epsilon ^2}+\nonumber\\&\frac{8 \pi ^2 N_2 \left(\left(N_1-4 N_2\right) C_1^2(R)+\left(N_2^2+2\right) C_2(R)\right) (1+4 \log 2) C_{\theta }^2}{\epsilon }+O\left(\epsilon ^0\right) \\
(p_8)= & \frac{16 \pi ^2 \left(5 \pi ^2-48\right) N_2 \left(C_2(R)-N_2 C_1^2(R)\right) C_{\theta }^2}{3 \epsilon }+O\left(\epsilon ^0\right) \\
(p_9)= & \frac{32 \pi ^2 N_2 \left(C_2(R)-N_2 C_1^2(R)\right) C_{\theta }^2}{\epsilon ^2}-\frac{64 \left(\pi ^2 \left(N_1-N_2\right) N_2 (1+4 \log 2) C_{\theta }^2\right)}{\epsilon }+O\left(\epsilon ^0\right) 
\end{align}

\subsection{Bremsstrahlung function}

After summing up the diagrams, computing the cusp anomalous dimension and taking its small $\theta$ limit, the final result for the $\theta$--Bremsstrahlung function for the representation $R$ reads
\begin{align}\label{eq:B16result}
B_{1/6}^{\theta}(R) &= \frac{N_2 C_2(R)}{4 k^2}-\frac{\pi ^2 N_2}{24 k^4} \left(\left(N_1-5 N_2\right) C_1^2(R)+\left(N_2^2+5 N_1 N_2-2\right) C_2(R)\right.\nonumber\\&\left.+2 N_1 C_3(R)-2 C_2(R) C_1(R)-2 C_4(R)\right)+O\left(k^{-6}\right)
\end{align}
We can check that this result is in agreement with the conjecture
\begin{equation}\label{eq:B16conjecturegeneral}
B_{1/6}^{\theta}(R) = \frac{1}{8\pi^2}\, \partial_m\, |W_B^m(R)|\, \bigg|_{m=1}
\end{equation}
that generalizes \eqref{eq:B16conjecture} to generic representations.
In fact, plugging for instance \eqref{eq:Wn}--\eqref{eq:WnA3} in the right--hand--side of \eqref{eq:B16conjecturegeneral} we obtain
\begin{align}
&B_{1/6}^{\theta}(S_1) = \frac{N_1 N_2}{4 k^2}-\frac{\pi ^2 N_2 \left(5 N_1^2 N_2+N_1 N_2^2-3N_1-5 N_2\right)}{24 k^4}+{\cal O}\left(k^{-6}\right) \\&
B_{1/6}^{\theta}(S_2) = \frac{\left(N_1+1\right) N_2}{2k^2}\nonumber\\&~~~~-\frac{\pi ^2 N_2 \left(-2 \left(N_1+2\right){}^2+\left(N_1+1\right) N_2^2+5 \left(N_1^2+N_1-2\right) N_2\right)}{12 k^4}+{\cal O}\left(k^{-6}\right)\\&
B_{1/6}^{\theta}(S_3) = \frac{3 \left(N_1+2\right) N_2}{4 k^2}\nonumber\\&~~~~+\frac{\pi ^2 N_2 \left(4 N_1^2+21 N_1-\left(N_1+2\right) N_2^2-5 \left(N_1-1\right) \left(N_1+3\right) N_2+32\right)}{8 k^4}+{\cal O}\left(k^{-6}\right)\\&
B_{1/6}^{\theta}(A_2) = \frac{\left(N_1-1\right) N_2}{2k^2}\nonumber\\&~~~~+\frac{\pi ^2 N_2 \left(-2 \left(N_1-2\right)^2-\left(N_1-1\right) N_2^2+5 \left(-N_1^2+N_1+2\right) N_2\right)}{12 k^4}+{\cal O}\left(k^{-6}\right)\\&
B_{1/6}^{\theta}(A_3) = \frac{3 \left(N_1-2\right) N_2}{4 k^2}\nonumber\\&~~~~+\frac{\pi ^2 N_2 \left(-4 N_1^2+21 N_1-\left(N_1-2\right) N_2^2-5 \left(N_1-3\right) \left(N_1+1\right) N_2-32\right)}{8 k^4}+{\cal O}\left(k^{-6}\right)
\end{align}
These expressions agree with \eqref{eq:B16result}, upon using formulae \eqref{eq:casimir} for the corresponding representations.

\section{Perturbative expansion of the latitude matrix model}\label{app:pert}

The proposed matrix model \eqref{eq:matrixlat} for the latitude Wilson loop can be expanded at weak coupling in the same way as the one for 1/6--BPS Wilson loops given in Section \ref{sec:MMexp}.
We present here the expansion of the latitude expectation value at framing $f=\nu$, winding number $m$ and for the fundamental representation
\begin{align}
& \langle W_B^m(\nu) \rangle_\nu = 1+\frac{i \pi  \nu  m^2 N_1}{k}
\nonumber\\&~~~~
-\frac{\pi ^2 m^2}{6 k^2} \left( N_1^2 \left(\left(2 m^2+1\right) \nu ^2+1\right)-3 \left(\nu ^2+1\right) N_1 N_2 + \left(m^2-1\right) \nu ^2-1\right)
\nonumber\\&~~~~
-\frac{i \pi ^3 \nu  m^2}{18 k^3} \left(
N_1^3 \left(\nu ^2 m^4+2 \left(\nu ^2+1\right) m^2+1\right)
-6 N_1^2 N_2 \left(\left(\nu ^2+1\right) m^2+1\right)
\right.\nonumber\\&\left.~~~~~~~~
+9 N_1 N_2^2
+N_1 \left(2 \nu ^2 m^4-2 \left(\nu ^2+1\right) m^2-1\right)
-3 N_2 \left(\left(\nu ^2+1\right) m^2+1\right)\right)
\nonumber\\&~~~~
+ \frac{\pi ^4 m^2}{360k^4} \left(
m^2 N_1^4 \left(\left(2 \left(m^2+5\right) m^2+3\right) \nu ^4+10 \left(m^2+2\right) \nu ^2+3\right)
\right.\nonumber\\&~~~~~~~~
-15 N_1^3 N_2 m^2 \left(\nu ^2 \left(2 \left(\nu ^2+1\right) m^2+\nu ^2+8\right)+1\right)
\nonumber\\&~~~~~~~~
+30 N_1^2 N_2^2 \left(m^2 \left(\nu ^4+6 \nu ^2+1\right) -\nu ^2-1 \right)
\nonumber\\&~~~~~~~~
-30 N_1 N_2^3 \left(\nu ^2+1\right)
+10 N_1^2 \left(\nu ^4 m^6-m^2 \left(\left(\nu ^2+3\right) \nu ^2+1\right)\right)
\nonumber\\&~~~~~~~~
-15 N_1 N_2 \left(4 \nu ^2 \left(\nu ^2+1\right) m^4 - \left(3 \nu ^4-4 \nu ^2+3\right) m^2-2 \left(\nu ^2+1\right)\right)
\nonumber\\&~~~~~~~~
+ 15 N_2^2 \left( m^2 \left(\nu ^4+6 \nu ^2+1\right) + 2 \left(\nu ^2+1\right) \right)
\nonumber\\&\left.~~~~~~~~
+ 3 \nu ^4 m^6-10 \left(\nu ^4+\nu ^2\right) m^4+m^2 \left(7 \nu ^4+10 \nu ^2+7\right)
\right)
+ O\left(k^{-5}\right)
\end{align}
Remarkably, up to three loops this coincides with the perturbative computation \eqref{eq:Wlatm} at $f=\nu$. The four--loop term is a new prediction.

We report also the matrix model expansions for the totally symmetric and antisymmetric representations up to rank 3
\begin{align}
& \langle W_B^m(\nu,S_2) \rangle_\nu = 1+\frac{2 i \pi  \nu  \left(N_1+1\right)}{k}
\nonumber\\&~~~~
-\frac{\pi ^2}{3 k^2} \Big(6 \nu ^2+12 \nu ^2 N_1+\left(6 \nu ^2+1\right) N_1^2-3 \left(\nu ^2+1\right) N_2-3 \left(\nu ^2+1\right) N_1 N_2+N_1-2\Big)
\nonumber\\&~~~~
+\frac{i \pi ^3 \nu}{3 k^3}  \Big(-4 \nu ^2-2 \left(2 \nu ^2+1\right) N_1^3+N_1^2 \left(-12 \nu ^2+5 \nu ^2 N_2+7 N_2-4\right)
\nonumber\\&~~~~~~~~
+N_1 \left(-12 \nu ^2+\left(11 \nu ^2+13\right) N_2-3 N_2^2+2\right)+2 \left(4 \nu ^2+5\right) N_2-3 N_2^2+4\Big)
\nonumber\\&~~~~
+\frac{\pi ^4}{180 k^4} \Big(8 \left(15 \nu ^4-30 \nu ^2+7\right)-30 \left(\nu ^2+1\right) N_2^3+8 \left(15 \nu ^4+15 \nu ^2+1\right) N_1^4
\nonumber\\&~~~~~~~~
+N_1^3 \left(8 \left(60 \nu ^4+45 \nu ^2+2\right)-15 \left(17 \nu ^4+34 \nu ^2+3\right) N_2\right)
\nonumber\\&~~~~~~~~
+3 N_1^2 \left(4 \left(60 \nu ^4+10 \nu ^2-3\right)+5 \left(5 \nu ^4+28 \nu ^2+3\right) N_2^2-5 \left(55 \nu ^4+94 \nu ^2+5\right) N_2\right)
\nonumber\\&~~~~~~~~
+N_1 \big(480 \nu ^4-360 \nu ^2-30 \left(\nu ^2+1\right) N_2^3+15 \left(11 \nu ^4+64 \nu ^2+9\right) N_2^2
\nonumber\\&~~~~~~~~
-30 \left(35 \nu ^4+56 \nu ^2-5\right) N_2-44\big)+60 \left(2 \nu ^4+13 \nu ^2+3\right) N_2^2
\nonumber\\&~~~~~~~~
-30 \left(17 \nu ^4+22 \nu ^2-9\right) N_2\Big)
+ O\left(k^{-5}\right)
\end{align}
\begin{align}
& \langle W_B^m(\nu,A_2) \rangle_\nu = 1+\frac{2 i \pi  \nu  \left(N_1-1\right)}{k}
\nonumber\\&~~~~
+\frac{\pi ^2}{3 k^2} \left(12 \nu ^2 N_1-6 \nu ^2-\left(6 \nu ^2+1\right) N_1^2-3 \left(\nu ^2+1\right) N_2+3 \left(\nu ^2+1\right) N_1 N_2+N_1+2\right)
\nonumber\\&~~~~
+\frac{i \pi ^3 \nu}{3 k^3}  \Big(4 \nu ^2-2 \left(2 \nu ^2+1\right) N_1^3+N_1^2 \left(12 \nu ^2+\left(5 \nu ^2+7\right) N_2+4\right)
\nonumber\\&~~~~~~~~
-N_1 \left(12 \nu ^2+\left(11 \nu ^2+13\right) N_2+3 N_2^2-2\right)+8 \nu ^2 N_2+3 N_2^2+10 N_2-4\Big)
\nonumber\\&~~~~
+\frac{\pi ^4}{180 k^4} \Big(8 \left(15 \nu ^4-30 \nu ^2+7\right)+30 \left(\nu ^2+1\right) N_2^3+8 \left(15 \nu ^4+15 \nu ^2+1\right) N_1^4
\nonumber\\&~~~~~~~~
-N_1^3 \left(8 \left(60 \nu ^4+45 \nu ^2+2\right)+15 \left(17 \nu ^4+34 \nu ^2+3\right) N_2\right)
\nonumber\\&~~~~~~~~
+3 N_1^2 \left(4 \left(60 \nu ^4+10 \nu ^2-3\right)+5 \left(5 \nu ^4+28 \nu ^2+3\right) N_2^2+5 \left(55 \nu ^4+94 \nu ^2+5\right) N_2\right)
\nonumber\\&~~~~~~~~
-N_1 \big(480 \nu ^4-360 \nu ^2+30 \left(\nu ^2+1\right) N_2^3+15 \left(11 \nu ^4+64 \nu ^2+9\right) N_2^2
\nonumber\\&~~~~~~~~
+30 \left(35 \nu ^4+56 \nu ^2-5\right) N_2-44\big)+60 \left(2 \nu ^4+13 \nu ^2+3\right) N_2^2
\nonumber\\&~~~~~~~~
+30 \left(17 \nu ^4+22 \nu ^2-9\right) N_2\Big)
+ O\left(k^{-5}\right)
\end{align}
\begin{align}
& \langle W_B^m(\nu,S_3) \rangle_\nu = 1+\frac{3 i \pi  \nu  \left(N_1+2\right)}{k}
\nonumber\\&~~~~
-\frac{\pi ^2}{2 k^2} \left(36 \nu ^2+\left(9 \nu ^2+1\right) N_1^2+N_1 \left(36 \nu ^2-3 \left(\nu ^2+1\right) N_2+2\right)-6 \left(\nu ^2+1\right) N_2-3\right)
\nonumber\\&~~~~
-\frac{i \pi ^3 \nu}{2 k^3}  \Big(72 \nu ^2+\left(9 \nu ^2+3\right) N_1^3-2 N_1^2 \left(-27 \nu ^2+4 \nu ^2 N_2+5 N_2-6\right)
\nonumber\\&~~~~~~~~
+N_1 \left(108 \nu ^2-2 \left(17 \nu ^2+19\right) N_2+3 N_2^2+3\right)-3 \left(13 \nu ^2+14\right) N_2+6 N_2^2-18\Big)
\nonumber\\&~~~~
+ \frac{\pi ^4}{120 k^4} \Big(27 \left(240 \nu ^4-120 \nu ^2+7\right)-60 \left(\nu ^2+1\right) N_2^3+\left(405 \nu ^4+270 \nu ^2+13\right) N_1^4
\nonumber\\&~~~~~~~~
+N_1^3 \left(4 \left(810 \nu ^4+405 \nu ^2+13\right)-15 \left(43 \nu ^4+70 \nu ^2+5\right) N_2\right)
\nonumber\\&~~~~~~~~
+10 N_1^2 \left(972 \nu ^4+243 \nu ^2+3 \left(4 \nu ^4+23 \nu ^2+3\right) N_2^2-3 \left(137 \nu ^4+196 \nu ^2+9\right) N_2-5\right)
\nonumber\\&~~~~~~~~
-3 N_1 \big(-4320 \nu ^4+360 \nu ^2+10 \left(\nu ^2+1\right) N_2^3-10 \left(17 \nu ^4+100 \nu ^2+15\right) N_2^2
\nonumber\\&~~~~~~~~
+5 \left(611 \nu ^4+782 \nu ^2-11\right) N_2+68\big)+45 \left(13 \nu ^4+80 \nu ^2+15\right) N_2^2
\nonumber\\&~~~~~~~~
-60 \left(119 \nu ^4+133 \nu ^2-18\right) N_2\Big)
+ O\left(k^{-5}\right)
\end{align}
\begin{align}
& \langle W_B^m(\nu,A_3) \rangle_\nu = 1+\frac{3 i \pi  \nu  \left(N_1-2\right)}{k}
\nonumber\\&~~~~
-\frac{\pi ^2}{2 k^2} \Big(36 \nu ^2+\left(9 \nu ^2+1\right) N_1^2-N_1 \left(36 \nu ^2+3 \left(\nu ^2+1\right) N_2+2\right)+6 \left(\nu ^2+1\right) N_2-3\Big)
\nonumber\\&~~~~
-\frac{i \pi ^3 \nu}{2 k^3}  \Big(\left(9 \nu ^2+3\right) N_1^3-2 N_1^2 \left(27 \nu ^2+\left(4 \nu ^2+5\right) N_2+6\right)
\nonumber\\&~~~~~~~~
+N_1 \left(108 \nu ^2+\left(34 \nu ^2+38\right) N_2+3 N_2^2+3\right)
\nonumber\\&~~~~~~~~
-3 \left(24 \nu ^2+\left(13 \nu ^2+14\right) N_2+2 N_2^2-6\right)\Big)
\nonumber\\&~~~~
\frac{\pi ^4}{120 k^4} \Big(27 \left(240 \nu ^4-120 \nu ^2+7\right)+60 \left(\nu ^2+1\right) N_2^3+\left(405 \nu ^4+270 \nu ^2+13\right) N_1^4
\nonumber\\&~~~~~~~~
-N_1^3 \left(4 \left(810 \nu ^4+405 \nu ^2+13\right)+15 \left(43 \nu ^4+70 \nu ^2+5\right) N_2\right)
\nonumber\\&~~~~~~~~
+10 N_1^2 \left(972 \nu ^4+243 \nu ^2+3 \left(4 \nu ^4+23 \nu ^2+3\right) N_2^2+3 \left(137 \nu ^4+196 \nu ^2+9\right) N_2-5\right)
\nonumber\\&~~~~~~~~
-3 N_1 \big(4320 \nu ^4-360 \nu ^2+10 \left(\nu ^2+1\right) N_2^3+10 \left(17 \nu ^4+100 \nu ^2+15\right) N_2^2
\nonumber\\&~~~~~~~~
+5 \left(611 \nu ^4+782 \nu ^2-11\right) N_2-68\big)+45 \left(13 \nu ^4+80 \nu ^2+15\right) N_2^2
\nonumber\\&~~~~~~~~
+60 \left(119 \nu ^4+133 \nu ^2-18\right) N_2\Big)
+ O\left(k^{-5}\right)
\end{align}
Again these results show agreement with \eqref{eq:Wlatgeneric} at $f=\nu$.

\section{Checks on the strong coupling expansion}

\paragraph{Derivatives with respect to $\nu$}

As discussed in Section \ref{review}, for applications to the Bremsstrahlung function a simpler problem than computing the whole strong coupling expansion of the latitude Wilson loop consists in evaluating the expansion of its derivative with respect to $\nu$, at $\nu= 1$.
On the one hand, using prescription \eqref{eq:matrixlat} this reduces to the evaluation of the derivative of the ordinary 1/6--BPS Wilson loop with winding number $m$ (see discussion in Section \ref{sec:matrixlat}). Its strong coupling expansion is then inherited from the expansion of the derivative of the $m$--wound operator \cite{Klemm:2012ii} computed using the Fermi gas description. On the other hand, we can evaluate $\pa_\nu \langle W_B(\nu) \rangle |_{\nu=1}$ at strong coupling directly, using the prescription \eqref{eq:Wgrancanonical} with the $\nu$--derivative hitting the $n_{O_{1,\nu}}$ distribution.
The agreement between the two results will provide a test for the correctness of the expansion carried out in Section \ref{sec:Fermi}.

In order to evaluate the $\nu$--derivative in the Fermi gas approach it is convenient to use expression \eqref{eq:nOI} for the occupation number distribution expressed in terms of the $\tilde{I}_{1,\nu}$ (region I) and $\tilde{I}_{\nu,1}$ (region II) integrals defined in \eqref{eq:uint}. The $\nu \to 1$ limit makes the integrals over region III exponentially subleading and therefore we discard them. 

The effect of applying the $\nu$--derivative to the integrals in \eqref{eq:uint} (where for convenience we factor out $(-1)^{-\nu}$ in the integral for region I) is to produce the  following two new integrals 
\begin{equation}\label{eq:dernu2}
\tilde I_{1,\nu}^{(1)} = \int_0^{u_\ast} du\, \frac{u\, \log \frac{u}{1-u^2}}{1-u^2}
\end{equation}
and
\begin{equation}\label{eq:dernu1}
\tilde I_{\nu,1}^{(1)}  = \int_0^{u_\ast} du\, \frac{u\, \log u}{1-u^2}
\end{equation}
Integrating them and paying attention to the fact that the integration limits are complex, we find
\begin{equation}
\text{region I} \rightarrow \frac{e^{\frac{2 \mu }{k}} \left(i k \text{Li}_2(u^2)+\log \left(1-u^2\right) \left(i k \log u^2 -i k \log \left(1-u^2\right)+\pi  k-2 i k+4 i \mu \right)\right)}{16 \pi ^2}
\end{equation}
and
\begin{equation}
\text{region II} \rightarrow -\frac{i k e^{\frac{2 \mu }{k}} \left(-\text{Li}_2 (u^2) -\log  u^2 \log \left(1-u^2\right)\right)}{16 \pi ^2}\end{equation}
Summing these two expressions with the bulk contribution \eqref{eq:bulk} we obtain
\begin{align}\label{eq:dernusum}
\partial_\nu\, n_{O_{1,\nu}} \bigg|_{\nu=1} =& \frac{e^{\frac{2 \mu }{k}}}{16 \pi ^2} \Big[2 i k \text{Li}_2\left(u^2\right)+\log \left(1-u^2\right) \left(2 i k \log \left(u^2\right)+(\pi -2 i) k+4 i \mu \right) \nonumber\\&-i k \log ^2\left(1-u^2\right)+4 i k+\pi  k-4 i \mu \Big]
\end{align}
We now expand ${\rm Li}_2$ for asymptotically large values of its argument using the general identity
\begin{equation}
\text{Li}_s (z) = \sum_{k=0}^{\infty} (-1)^k(1-2^{1-2k})(2\pi)^{2k}\frac{B_{2k}}{(2k)!} \frac{\log^{s-2k}(-z)}{\Gamma(s+1-2k)}
\end{equation}
Note that for any given $s$ these expansions stop after a finite number of terms.

Plugging this expansion into \eqref{eq:dernusum} we finally obtain
\begin{equation}\label{eq:nderivative}
\partial_\nu\, n_{O_{1,\nu}} \bigg|_{\nu=1} = \frac{i e^{\frac{2 \mu }{k}} \left(\left(24+\pi ^2\right) k^2-48 k \mu +48 \mu ^2\right)}{96 \pi ^2 k}
\end{equation}
If we now take the same expression \eqref{eq:grandaverage2} for the $m$--wound 1/6--BPS Wilson loop, compute its derivative with respect to $m$ and set $m=1$ we indeed reproduce \eqref{eq:nderivative} correctly. This constitutes a successful test of our procedure.

\paragraph{The $n$-th derivative}
In principle further checks can be performed applying an arbitrarily large number of derivatives, provided there is enough computational power.
In particular, taking the $n^{th}$ derivative of the integrals in \eqref{eq:uint}, evaluating their large $\mu$ asymptotic behavior and plugging the result in \eqref{eq:nOI} we check that we obtain the same expression as from applying the $n^{th}$ derivative directly on the asymptotic expression \eqref{eq:nOnu}.
Here we report a sketch of this computation, primarily because the structure of the relevant integrals that get produced is particularly interesting: they belong to a class that can be entirely solved in terms of Harmonic polylogarithms \cite{Remiddi:1999ew}, for which a well established technology exists (and also the Mathematica package \cite{Maitre:2005uu,Maitre:2007kp}).

The relevant integral in region I, arising from taking the $n^{th}$ derivative of \eqref{eq:uint} evaluated at $a=1, b=\nu$, reads
\begin{equation}
\tilde I_{1,\nu}^{(n)}  \equiv \int_0^{u_\ast} du\, \frac{u\, \log^n \frac{u}{1-u^2}}{1-u^2}
\end{equation}
Upon a simple change of variables it can be reduced to an integral that can be immediately solved in terms of harmonic polylogarithms
\begin{equation} \label{eq:appF:result1}
\tilde I_{1,\nu}^{(n)} = \int_0^{u^2_\ast} du\, \frac{(\frac12 \log u - \log{(1-u)})^n}{2(1-u)}  \equiv \sum_{k=0}^n\, \left(-\frac12\right)^{k}\, \left(\begin{array}{c} n\\ k
\end{array}  \right)\, I(n-k,k)
\end{equation}
where  
\begin{equation}
I(a,b) = \int_0^{u^2_\ast} du\, \frac{\log^b u \log^a (1-u)}{2(1-u)}
= (-1)^a\frac{a!b!}{2}\, \sum_{r\in \{1^a\}\shuffle\{0^b\}}  {\rm H}_{1,r}(u^2) 
\end{equation}
 
In region II, applying the $n^{th}$ derivative to \eqref{eq:uint} this time evaluated at $a=\nu, b=1$, we obtain
\begin{equation}
\tilde I_{\nu,1}^{(n)} \equiv \int_0^{u_\ast} du\, \frac{u\, \log^n u}{1-u^2}
\end{equation}
Applying partial fractioning this can be reduced to an integral that enters the definition of the harmonic polylogarithms
\begin{equation} \label{eq:appF:result2}
\tilde I_{\nu,1}^{(n)}  = \frac{n!}{2^{n+1}}\, {\rm H}_{1,0^{n}}(u^2)
\end{equation}
Alternatively, this expression can be reduced to the following combination of ordinary polylogarithms 
\begin{equation}
\tilde I_{\nu,1}^{(n)}  = \frac12\, \sum_{k=1}^n \frac{(n-1)!}{(n-k)!}\, (-1)^{k+1}\, \log^{n-k}\, \left( {\rm Li}_k(u) + {\rm Li}_k(-u) \right)
\end{equation}
Solutions \eqref{eq:appF:result1} and \eqref{eq:appF:result2} have also been checked numerically.

\paragraph{Extraction of the asymptotic behavior}

We are interested in the large $\mu$ behavior of these integrals, that is when the argument of the polylogarithms grows exponentially.
To this end it is convenient to change their argument  as $u\to 1/t$, reduce the polylogarithms to have argument $t$ and finally extract their logarithmic divergence at $t=0$.
This procedure can be performed in a completely algorithmic (and recursive) manner, though it might take a long computing time for a large number $n$ of derivatives.

After extracting the leading behavior for the integrals (i.e. neglecting exponentially small corrections) and plugging them into \eqref{eq:nOI}, we have checked that the result coincides with taking the $n^{th}$ derivative directly on the general asypmptotic expression \eqref{eq:nOnu}. This provides a consistency check of our asymptotic expansions.

\bibliographystyle{JHEP}

\bibliography{biblio2}

\end{document}